\documentclass[preprint]{aastex61}

\usepackage{subfigure}
\usepackage{url}

\shorttitle{The RPA: $r$-Process Enhanced Metal-Poor Stars}
\shortauthors{Sakari et al.}

\begin{document}

\title{The $R$-Process Alliance: First Release from the Northern
  Search for $r$-Process Enhanced Metal-Poor Stars in the Galactic Halo}

\author{Charli M. Sakari}
\affil{Department of Astronomy, University of Washington, Seattle WA
98195-1580, USA}
\email{sakaricm@u.washington.edu}

\author{Vinicius M. Placco}
\affil{Department of Physics, University of Notre Dame, Notre Dame, IN 46556,  USA}
\affil{Joint Institute for Nuclear Astrophysics Center for the
  Evolution of the Elements (JINA-CEE), USA}

\author{Elizabeth M. Farrell}
\affil{Department of Astronomy, University of Washington, Seattle WA
98195-1580, USA}

\author{Ian U. Roederer}
\affil{Department of Astronomy, University of Michigan, 1085 S. University Ave.,
 Ann Arbor, MI 48109, USA}
\affil{Joint Institute for Nuclear Astrophysics Center for the Evolution of the 
Elements (JINA-CEE), USA}

\author{George Wallerstein}
\affil{Department of Astronomy, University of Washington, Seattle WA
98195-1580, USA}

\author{Timothy C. Beers}
\affil{Department of Physics, University of Notre Dame, Notre Dame, IN 46556,  USA}
\affil{Joint Institute for Nuclear Astrophysics Center for the Evolution of the 
Elements (JINA-CEE), USA}

\author{Rana Ezzeddine}
\affil{Department of Physics and Kavli Institute for Astrophysics and
  Space Research, Massachusetts Institute of Technology, Cambridge, MA
  02139, USA}

\author{Anna Frebel}
\affil{Department of Physics and Kavli Institute for Astrophysics and
  Space Research, Massachusetts Institute of Technology, Cambridge, MA
  02139, USA}

\author{Terese Hansen}
\affil{Observatories of the Carnegie Institution of Washington, 813
  Santa Barbara Street, Pasadena, CA 91101, USA}

\author{Erika M. Holmbeck}
\affil{Department of Physics, University of Notre Dame, Notre Dame, IN 46556,  USA}
\affil{Joint Institute for Nuclear Astrophysics Center for the
  Evolution of the Elements (JINA-CEE), USA}

\author{Christopher Sneden}
\affil{Department of Astronomy and McDonald Observatory, The University of Texas, Austin, TX 78712, USA}

\author{John J. Cowan}
\affil{Homer L. Dodge Department of Physics and Astronomy, University of Oklahoma, Norman, OK 73019, USA}

\author{Kim A. Venn}
\affil{Department of Physics and Astronomy, University of Victoria,
  Victoria, BC, Canada}

\author{Christopher Evan Davis}
\affil{Department of Astronomy, University of Washington, Seattle WA
98195-1580, USA}

\author{Gal Matijevi\v{c}}
\affil{Leibniz Institut f\"{u}r Astrophysik Potsdam (AIP), An der
Sterwarte 16, 14482 Potsdam, Germany}

\author{Rosemary F.G. Wyse}
\affil{Physics and Astronomy Department, Johns Hopkins University, 3400 North Charles Street, Baltimore, MD 21218, USA}

\author{Joss Bland-Hawthorn}
\affil{Sydney Institute for Astronomy, School of Physics A28, University of Sydney, NSW 2006, Australia}
\affil{ARC Centre of Excellence for All Sky Astrophysics (ASTRO-3D), Australia}

\author{Cristina Chiappini}
\affil{Leibniz Institut f\"{u}r Astrophysik Potsdam, An der Sternwarte 16, D-14482 Potsdam, Germany}

\author{Kenneth C. Freeman}
\affil{Research School of Astronomy \& Astrophysics, The Australian National University, Cotter Road, Canberra, ACT 2611}

\author{Brad K. Gibson}
\affil{E.A. Milne Centre for Astrophysics, University of Hull, Hull,
  HU6 7RX, United Kingdom}

\author{Eva K. Grebel}
\affil{Astronomisches Rechen-Institut, Zentrum f\"ur
Astronomie der Universit\"at Heidelberg, M\"onchhofstr.\ 12--14,
69120 Heidelberg, Germany}

\author{Amina Helmi}
\affil{Kapteyn Astronomical Institute, University of Groningen, P.O. Box 800,
NL-9700 AV Groningen, The Netherlands}

\author{Georges Kordopatis}
\affil{Universit\'e C\^ote d'Azur, Observatoire de la C\^ote d'Azur, CNRS, Laboratoire Lagrange, France}

\author{Andrea Kunder}
\affil{Saint Martin's University, 5000 Abbey Way SE, Lacey, WA 98503 USA}

\author{Julio Navarro}
\affil{Department of Physics and Astronomy, University of Victoria,
  Victoria, BC, Canada}

\author{Warren Reid}
\affil{Department of Physics and Astronomy, Macquarie University, Sydney, NSW 2109, Australia}
\affil{Western Sydney University, Locked bag 1797, Penrith South, NSW 2751, Australia}

\author{George Seabroke}
\affil{Mullard Space Science Laboratory, University College London, Holmbury St Mary, Dorking, RH5 6NT, UK}

\author{Matthias Steinmetz}
\affil{Leibniz Institut f\"{u}r Astrophysik Potsdam (AIP), An der
Sterwarte 16, 14482 Potsdam, Germany}

\author{Fred Watson}
\affil{Department of Industry, Innovation and Science, 105 Delhi Road, North Ryde, NSW 2113, Australia}



\begin{abstract}
This paper presents the detailed abundances and $r$-process
classifications of 126 newly identified metal-poor stars as part of an
ongoing collaboration, the $R$-Process Alliance.  The stars were
identified as metal-poor candidates from the RAdial Velocity
Experiment (RAVE) and were followed-up at high spectral resolution
($R~\sim~31,500$) with the 3.5~m telescope at
Apache Point Observatory. The atmospheric parameters were determined
spectroscopically from \ion{Fe}{1} lines, taking into account $<$3D$>$
non-LTE corrections and using differential abundances with respect
to a set of standards. Of the 126 new stars, 124 have
$[\rm{Fe/H}]<-1.5$, 105 have $[\rm{Fe/H}]<-2.0$, and 4 have
$[\rm{Fe/H}]<-3.0$. Nine new carbon-enhanced metal-poor stars have
been discovered, 3 of which are enhanced in $r$-process
elements. Abundances of neutron-capture elements reveal 60 new $r$-I
stars (with $+0.3~\le~[\rm{Eu/Fe}]~\le~+1.0$ and $[\rm{Ba/Eu}]<0$) and 4
new $r$-II stars (with $[\rm{Eu/Fe}]>+1.0$). Nineteen stars are found
to exhibit a ``limited-$r$'' signature ($[\rm{Sr/Ba}]>+0.5$,
$[\rm{Ba/Eu}]<0$). For the $r$-II stars, the second- and third-peak
main $r$-process patterns are consistent with the $r$-process
signature in other metal-poor stars and the Sun. The abundances of
the light, $\alpha$, and Fe-peak elements match those of typical Milky
Way halo stars, except for one $r$-I star which has high
Na and low Mg, characteristic of globular cluster stars. Parallaxes
and proper motions from the second {\it Gaia} data release yield $UVW$
space velocities for these stars which are consistent with membership
in the Milky Way halo. Intriguingly, all $r$-II and the majority of
$r$-I stars have retrograde orbits, which may indicate an accretion
origin.
\end{abstract}

\keywords{stars: abundances --- stars: atmospheres --- stars: fundamental parameters --- Galaxy: formation}

\section{Introduction}\label{sec:Intro}
Metal-poor stars ($[\rm{Fe/H}]\lesssim -1.0$) have received
significant attention in recent years, primarily because they are
believed to be some of the oldest remaining stars in the Galaxy
\citep{BeersChristlieb2005,FrebelNorris2015}. High-precision abundances
of a wide variety of elements, from lithium to uranium, provide
valuable information about the early conditions in the Milky Way (MW),
particularly the nucleosynthesis of rare elements, yields from early
neutron star mergers (NSMs) and supernovae, and the chemical evolution
of the MW. The low iron content of the most metal-poor stars suggests
that their natal gas clouds were polluted by very few stars, in some
cases by only a single star (e.g., \citealt{Ito2009,Placco2014a}).
Observations of the most metal-poor stars therefore provide valuable
clues to the formation, nucleosynthetic yields, and evolutionary fates
of the first stars and the early assembly history of the MW and its
neighboring galaxies.

The stars that are enhanced in elements that form via the rapid
($r$-) neutron-capture process are particularly useful for
investigating the nature of the first stars and early galaxy assembly
(e.g.,
\citealt{Sneden1996,Hill2002,Christlieb2004,Frebel2007,Roederer2014b,Placco2017,Hansen2018,Holmbeck2018}).
The primary nucleosynthetic site of the $r$-process is still
under consideration. Photometric and spectroscopic follow-up of
GW~170817 \citep{Abbott2017} detected signatures of $r$-process
nucleosynthesis (e.g., \citealt{Chornock2017,Drout2017,Shappee2017}),
strongly supporting the NSM paradigm (e.g.,
\citealt{LattimerSchramm1974,Rosswog2014,Lippuner2017}).  This
paradigm is also supported by chemical evolution arguments
(e.g., \citealt{Cescutti2015,Cote2017}), comparisons with other
abundances (e.g., Mg; \citealt{MaciasRamirezRuiz2016}), and detections
of $r$-process enrichment in the ultra faint dwarf galaxy Reticulum~II
\citep{Ji2016,Roederer2016,Beniamini2018}.

However, the {\it ubiquity} of the $r$-process \citep{Roederer2010},
particularly in a variety of ultra faint dwarf galaxies, suggests that
NSMs may not be the only site of the $r$-process
\citep{TsujimotoNishimura2015,Tsujimoto2017}.  Standard core-collapse
supernovae are unlikely to create the main $r$-process elements
\citep{ArconesThielemann2013}; instead, the most likely
candidate for a second site of $r$-process formation may be the ``jet
supernovae,'' the resulting core collapse supernovae from strongly
magnetic stars (e.g., \citealt{Winteler2012,Cescutti2015}).  The physical
conditions (electron fraction, temperature, density), occurrence
rates, and timescales for jet supernovae may differ from
NSMs---naively, this could lead to different abundance patterns
(particularly between the $r$-process peaks) and different levels of
enrichment (e.g., see \citealt{Mosta2017}).  This then raises several
questions. Why is the relative abundance pattern for the main
$r$-process (barium and above) so robust across $\sim 3$ dex in
metallicity (e.g., \citealt{Sakari2018})? (In other words, why don't
the $r$-process yields vary?) Why is $r$-process contamination so
ubiquitous, even in low-mass systems where $r$-process events should
be rare? Finally, how can such low-mass systems like the ultra faint
dwarf galaxies retain the ejecta from such energetic events? (See
\citealt{BlandHawthorn2015} and \citealt{Beniamini2018} for
discussions of the mass limits of dwarfs that can retain ejecta for
subsequent star formation.) Addressing these questions requires
collaboration between theorists, experimentalists, modelers, and
observers.

Observationally, the $r$-process-enhanced, metal-poor stars may
provide the most useful information for identifying the site(s) of the 
$r$-process.  There are two main reasons for this:  1) The enhancement
in $r$-process elements ensures that spectral lines from a wide
variety of $r$-process elements are sufficiently strong to be
measured, while the (relative) lack of metal lines (compared to more
metal-rich stars) reduces the severe blending typically seen in the
blue spectral region; and 2) These stars are selected to have
little-to-no contamination from the slow ($s$-) process, simplifying
comparisons with models of $r$-process yields. If the enhancement in
radioactive elements like Th and U is sufficiently high,
cosmo-chronometric ages can also be determined (see, e.g.,
\citealt{Holmbeck2018} and references therein).

The $r$-process-enhanced, metal-poor stars have historically been
divided into two main categories \citep{BeersChristlieb2005}: the
$r$-I stars have $+0.3\le [\rm{Eu/Fe}]\le +1.0$, while $r$-II stars
have $[\rm{Eu/Fe}]~>~+1.0$; both require $[\rm{Ba/Eu}]<0$ to avoid
contamination from the $s$-process.  Prior to 2015, there were
$\sim30$ $r$-II and $\sim 75$ $r$-I stars known, according to the
JINAbase compilation \citep{Abohalima2017}.  Observations of
these $r$-process-enhanced stars have found a common pattern among the
main $r$-process elements, which is in agreement with the Solar
$r$-process residual.  Despite the consistency of the main $r$-process
patterns, $r$-process-enhanced stars are known to have deviations from
the Solar pattern for the lightest and heaviest neutron-capture
elements.  Variations in the lighter neutron-capture elements, such as
Sr, Y, and Zr have been observed in several stars (e.g.,
\citealt{SiqueiraMello2014,Placco2017,Spite2018}).  A new limited-$r$
designation \citep{Frebel2018}, with $[\rm{Sr/Ba}]~>~+0.5$, has been
created to classify stars with enhancements in these lighter elements.
(Though note that fast rotating massive stars can create some light
elements via the $s$-process;
\citealt{Chiappini2011,Frischknecht2012,Cescutti2013,Frischknecht2016}. In
highly $r$-process-enhanced stars, however, this signal may be swamped
by the larger contribution from the $r$-process; \citealt{Spite2018}.)
A subset of $r$-II stars ($\sim30$\%) also exhibit an enhancement in
Th and U that is referred to as an ``actinide boost'' (e.g.,
\citealt{Hill2002,Mashonkina2014,Holmbeck2018})---a complete
explanation for this phenomenon remains elusive (though
\citealt{Holmbeck2018b} propose one possible model), but it may prove 
critical for constraining the $r$-process site(s).

The numbers of stars in these categories will be important for
understanding the source(s) of the $r$-process.  If NSMs are the
dominant site of the $r$-process, they may be responsible for the
enhancement in both $r$-I and $r$-II stars---if so, the relative
frequencies of $r$-I and $r$-II stars can be compared with NSM rates.
Finally, there has been speculation that $r$-process-enhanced stars
may form in dwarf galaxies (e.g., Reticulum II; \citealt{Ji2016}),
which are later accreted into the MW.  The combination of abundance
information from high-resolution spectroscopy and proper motions and
parallaxes from {\it Gaia} DR2 \citep{GaiaDR2REF} will enable the
birth sites of the $r$-process-enhanced stars to be assessed, as has
already been done for several halo $r$-II stars
\citep{Sakari2018,Roederer2018}.

These are the observational goals of the $R$-Process Alliance (RPA), a
collaboration with the aim of identifying the site(s) of the
$r$-process.  This paper presents the first data set from the Northern
Hemisphere component of the RPA's search for $r$-process-enhanced
stars in the MW; the first Southern Hemisphere data set is presented
in \citet{Hansen2018}.  The observations and data reduction for this
sample are outlined in Section \ref{sec:Observations}.  Section
\ref{sec:AtmParams} presents the atmospheric parameters (temperature,
surface gravity, and microturbulence) and Fe and C abundances of a set
of standard stars, utilizing Local Thermodynamic Equilibrium (LTE)
\ion{Fe}{1} abundances both with and without non-LTE (NLTE)
corrections.  The parameters for the targets are then determined
differentially with respect to the set of standards.  The detailed
abundances are given in Section \ref{sec:Abunds};
Section~\ref{sec:Discussion} then discusses the $r$-process
classifications, the derived $r$-process patterns, implications for
the site(s) of the $r$-process, and comparisons with other MW halo
stars.  The choice of NLTE corrections is justified by comparisons
with other techniques for deriving atmospheric parameters, e.g.,
photometric temperatures, in Appendix \ref{appendix:Comp}.  LTE
parameters and abundances are also provided in Appendix
\ref{appendix:LTE}, and a detailed analysis of systematic errors is
given in Appendix \ref{appendix:Errors}.  Future papers from the RPA
will present additional discoveries of $r$-I and $r$-II stars.

\section{Observations and Data Reduction}\label{sec:Observations}
The metal-poor targets in this study were selected from two sources.  
Roughly half of the stars were selected from the fourth
\citep{RAVEDR4ref} and fifth \citep{RAVEDR5ref} data releases from the
RAdial Velocity Experiment \citep[RAVE]{RAVEref} and the
\citet{BaBref} sample.  These stars had their atmospheric parameters
($T_{\rm{eff}}$, $\log g$, and [Fe/H]) and [C/Fe] ratios validated
through optical ($3500-5500$~\AA), medium-resolution ($R~\sim~2000$)
spectroscopy \citep{Placco2018}.  The other half were part of a
re-analysis of RAVE data by \citet{Matijevic2017}.  The stars that
were targeted for high-resolution follow-up all had metallicity
estimates $[\rm{Fe/H}] \la -1.8$ and (in the case of the Placco et
al. subsample) were not carbon enhanced. Additionally, twenty
previously observed metal-poor stars were included to serve as
standard stars. Altogether, 131 stars with $V$-band magnitudes between
9 and 13 were observed, as shown in Table \ref{table:Targets1}, where
IDs, coordinates, and magnitudes are listed.

All targets were observed in 2015-2017 with the Astrophysical Research
Consortium (ARC) 3.5~-~m telescope at Apache Point Observatory (APO).
The seeing ranged from $0.6-2\arcsec$, with a median value of
$1.15\arcsec$.  The ARC Echelle Spectrograph (ARCES) was utilized in
its default setting, with a $1.6\arcsec \times 3.2\arcsec$ slit,
providing a spectral resolution of $R\sim 31,500$.  The spectra cover
the entire optical range, from $3800-10400$ \AA , though the S/N is
often prohibitively low below 4000 \AA.  Initial ``snapshot'' spectra
were taken to determine $r$-process enhancement; exposure times were
typically adjusted to obtain S/N ratios $>30$ (per pixel) in the blue,
which leads to S/N ratios $\ga 60$ near 6500 \AA.  Any interesting
targets were then observed again to obtain higher S/N.  Observation
dates, exposure times, and S/N ratios are reported in Table
\ref{table:Targets1}.

\begin{figure}[h!]
\begin{center}
\centering
\includegraphics[scale=0.65,trim=0 0.75in 0in 0in]{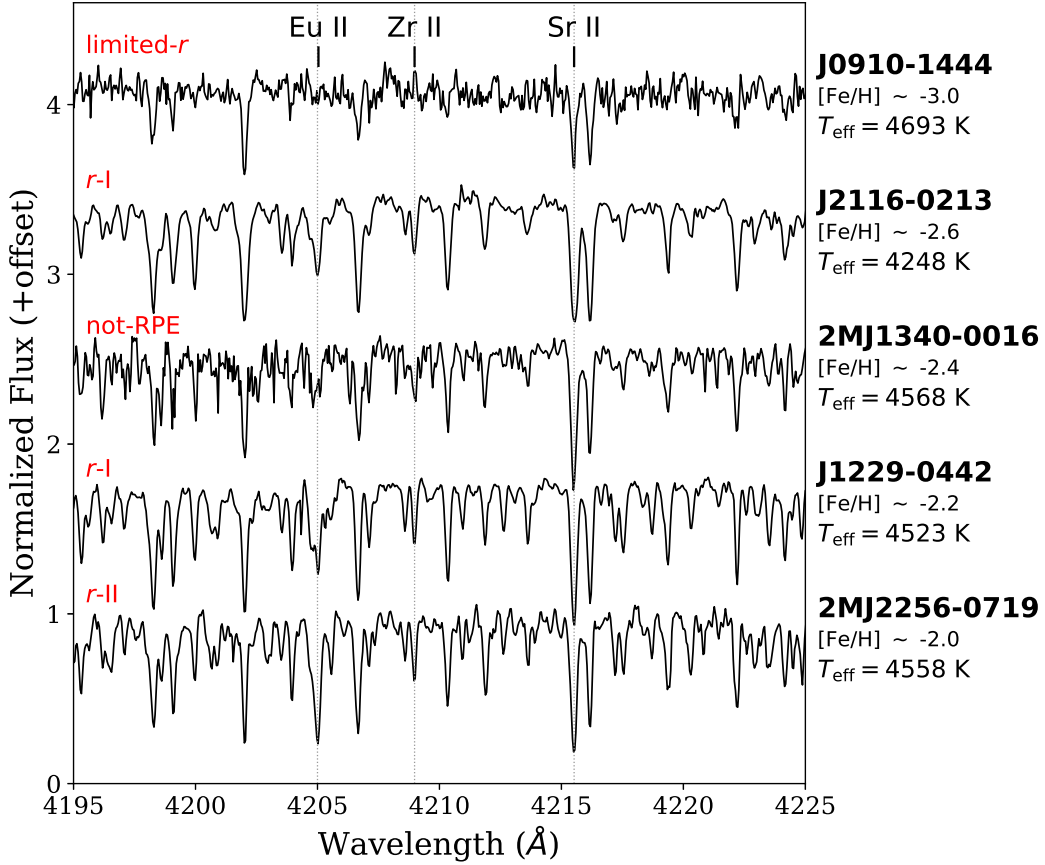}
\caption{Sample spectra for stars with a range of S/N, metallicity,
  temperature, and $r$-process enhancement.  ``Not-RPE'' indicates
  that the stars is not enhanced in $r$-process elements.  Three \ion{Sr}{2}, \ion{Zr}{2}, and \ion{Eu}{2} lines that were used in this analysis are identified.\label{fig:SampleSpectra}}
\end{center}
\end{figure}

The data were reduced in the Image Reduction and Analysis Facility
program (IRAF)\footnote{IRAF is distributed by the National Optical
  Astronomy Observatory, which is operated by the Association of
  Universities for Research in Astronomy, Inc., under cooperative
  agreement with the National Science Foundation.} with the standard
ARCES reduction recipe (see the manual by
J. Thorburn\footnote{\url{http://astronomy.nmsu.edu:8000/apo-wiki/attachment/wiki/ARCES/Thorburn_ARCES_manual.pdf}}),
yielding non-normalized spectra with 107 orders each.  The blaze
function was determined empirically through Legendre polynomial fits
to high S/N, extremely metal-poor stars.  The spectra of the other
targets were divided by these blaze function fits and refit with
low-order (5-7) polynomials (with strong lines, molecular bands, and
telluric features masked out).  All spectra were shifted to the
rest-frame through cross-correlations with a very high-resolution,
high S/N spectrum of Arcturus (from the \citealt{Hinkle2003}
atlas). The individual observations were then combined with average
$\sigma$-clipping techniques, weighting the individual spectra by
their flux near 4150 \AA.  Sample spectra around the 4205
\AA \hspace{0.02in} \ion{Eu}{2} line are shown in Figure
\ref{fig:SampleSpectra}.

\begin{figure}[h!]
\begin{center}
\centering
\subfigure{\includegraphics[scale=0.55,trim=0.3in 0in 0.45in 0.3in,clip]{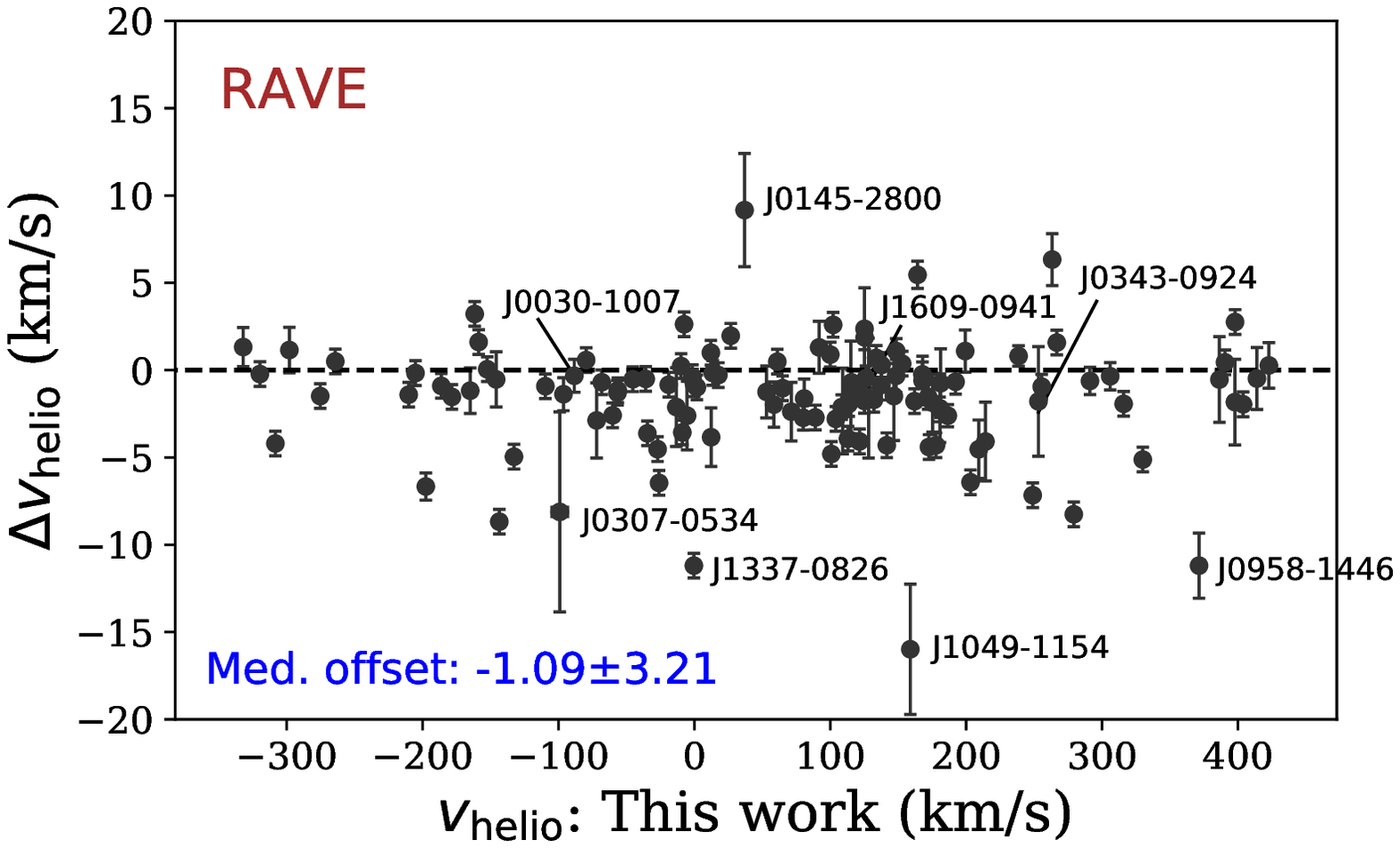}}\label{fig:RV_RAVE}
\subfigure{\includegraphics[scale=0.55,trim=0.3in 0in 0.6in 0.3in,clip]{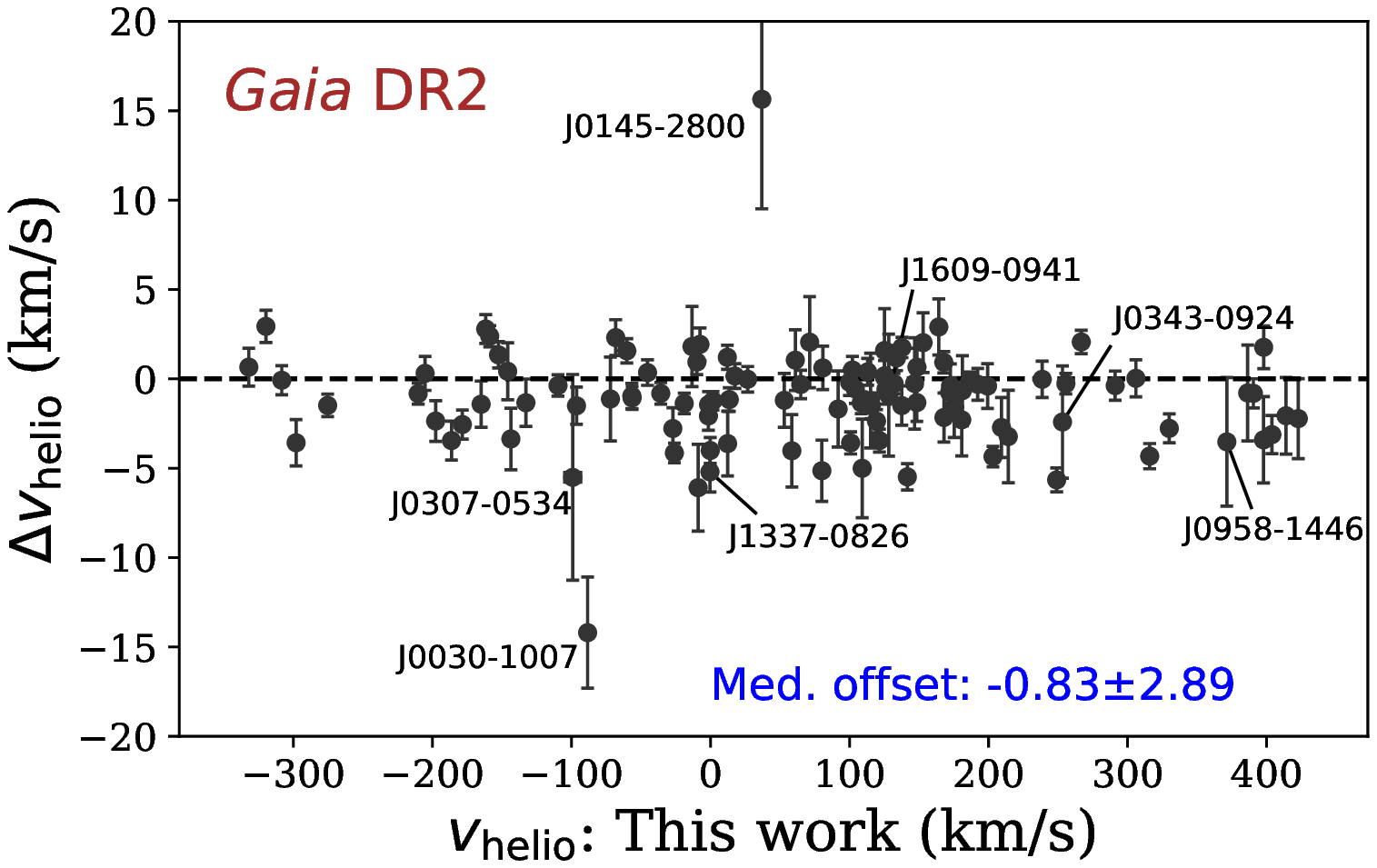}}\label{fig:RV_Gaia}
\caption{A comparison of the average heliocentric radial velocities in
  this work with those from RAVE (left) and {\it Gaia} DR2 (right).
  There are 122 stars with RAVE velocities, and 111 with {\it Gaia}
  DR2 velocities. The labeled outliers have offsets $>1\sigma$ from
  the median and/or large dispersions in velocity, and may be
  binaries.\label{fig:Velocities}}
\end{center}
\end{figure}

The final S/N ratios and heliocentric radial velocities are given in
Tables \ref{table:Targets1}, while Figure \ref{fig:Velocities} shows a
comparison with the radial velocities from RAVE and {\it Gaia} DR2
\citep{GaiaREF,GaiaDR2REF}.  The agreement is generally excellent,
with a small median offset and standard deviation of $-1.1\pm 3.2$ km
s$^{-1}$ from RAVE and $-0.8\pm2.9$~km~s$^{-1}$ from {\it Gaia}.
There are several outliers with offsets $1\sigma$ from the mean, which
may be binaries.\footnote{Note that the radial velocity for
  J2325$-$0815 is in agreement with {\it Gaia}, but in RAVE has been
  marked as unreliable owing to the low S/N ratio. The RAVE value for
  this star has been disregarded in this discussion.} In the case of
J0145$-$2800, J0307$-$0534, and J0958$-$1446, multi-epoch observations
in this paper show large radial velocity variations; in these cases,
the RAVE and {\it Gaia} radial velocities also differ.  Even if these
stars are unresolved binaries, none of the spectra show any signs of
contamination from a companion.

\clearpage

\startlongtable
\begin{deluxetable}{@{}ccccccccDr}
\tabletypesize{\scriptsize}
\tablecolumns{11}
\tablewidth{0pt}
\tablecaption{Targets\tablenotemark{a}\label{table:Targets1}}
\hspace*{-2in}
\tablehead{
Star\tablenotemark{b} & RA & Dec & $V$                & Observation & Exposure & \multicolumn{2}{c}{S/N\tablenotemark{c}} & \multicolumn{2}{c}{$v_{\rm{helio}}$\tablenotemark{d}} & Note\tablenotemark{e}\\
     & \multicolumn{2}{c}{(J2000)} & & Dates       & Time (s) & 4400 \AA        & 6500 \AA   & \multicolumn{2}{c}{(km s$^{-1}$)} & 
}
\decimals
\startdata
J000738.2$-$034551     & 00:07:38.16 & $-$03:45:50.4 & 11.52 & 9, 11 Sep 2016      & 2700 & 60 & 156 & -145.9\pm1.5 & P18 \\ 
J001236.5$-$181631     & 00:12:36.47 & $-$18:16:31.0 & 10.95 & 22 Jan, 28 Sep 2016 & 1500 & 80 & 150 & -96.4\pm0.8 &  \\ 
J002244.9$-$172429     & 00:22:44.86 & $-$17:24:29.1 & 12.89 & 22 Jan, 28 Sep 2016 & 3600 & 18 &  62 & 91.8\pm1.4 &  \\ 
J003052.7$-$100704     & 00:30:52.67 & $-$10:07:04.2 & 12.77 & 28 Sep 2016,        & 2700 & 25 &  60 & -88.4\pm3.0 &  \\ 
                       &             &             &       & 2 Feb 2017          & & & & \multicolumn{2}{c}{} & \\
J005327.8$-$025317     & 00:53:27.84 & $-$02:53:16.8 & 10.34 & 20 Jan 2016         & 2400 & 53 & 220 & -197.7\pm0.6 & P18\\ 
                       &             &             &       & 31 Jan 2017         &      &    &     & \multicolumn{2}{c}{} & \\
J005419.7$-$061155     & 00:54:19.65 & $-$06:11:55.4 & 13.06 & 28 Sep 2016         & 1800 & 20 &  75 & -132.8\pm0.5 & \\ 
J010727.4$-$052401     & 01:07:27.37 & $-$05:24:00.9 & 11.88 & 28 Sep 2016         & 1800 & 58 &  98 & -1.4\pm0.5 & \\ 
J012042.2$-$262205     & 01:20:42.20 & $-$26:22:04.7 & 10.21 & 22 Jan 2016         & 1200 & 43 & 100 & 15.2\pm0.5 &  \\ 
CS 31082-0001          & 01:29:31.14 & $-$16:00:45.5 & 11.32 & 22 Jan 2016         & 1440 & 30 & 106 & 137.6\pm0.7 & Std\\
J014519.5$-$280058     & 01:45:19.52 & $-$28:00:58.4 & 11.55 & 2 Feb, 28 Dec 2017  & 3000 & 20 &  75 &  36.9\pm3.2 & \\ 
\enddata
\tablenotetext{a}{Only a portion of this table is shown here to
  demonstrate its form and content. A machine-readable version of the
  full table is available.}
\tablenotetext{b}{The standard stars are identified by their names in
  SIMBAD.  Otherwise, the target stars are identified by their RAVE IDs, unless
  preceded by ``2M'', in which case their IDs from the Two Micron All
  Sky Survey (2MASS) are given \citep{2MASSref}.}
\tablenotetext{c}{S/N is per pixel; there are 2.5 pixels per
  resolution element.}
\tablenotetext{d}{The quoted errors are based on the uncertainty in
  the mean, with an adopted minimum of 0.5 km s$^{-1}$.}
\tablenotetext{e}{``P18'' indicates that the target was included in
  the medium-resolution follow-up of \citet{Placco2018}, while ``Std''
  indicates that the star was previously observed by others.}
\tablenotetext{f}{Based on radial velocity variations, this object is
  a suspected or confirmed binary.}
\end{deluxetable}

\normalsize

\section{Atmospheric Parameters, Metallicities, and Carbon Abundances}\label{sec:AtmParams}
High-resolution analyses utilize a variety of techniques to refine the
stellar temperatures, surface gravities, microturbulent velocities,
and metallicities, each with varying strengths and weaknesses.
The most common way to determine atmospheric parameters is from
the strengths of Fe lines, under assumptions of LTE.  
  Note that the atmospheric parameters are all somewhat
  degenerate---the assumption of LTE therefore can systematically
  affect all the parameters.  In a typical high-resolution analysis,
temperatures and microturbulent velocities are found by removing any
trends in the \ion{Fe}{1} abundance with line excitation potential
(EP) and reduced EW (REW),\footnote{REW = $\log$(EW/$\lambda$), where
  $\lambda$ is the   wavelength of the transition.} respectively.
However, each \ion{Fe}{1} line will have a different sensitivity to
NLTE effects. Similarly, surface gravities are sometimes determined
by requiring agreement between the \ion{Fe}{1} and \ion{Fe}{2}
abundances; however, the abundances derived from \ion{Fe}{1} lines
more sensitive to NLTE effects than those from \ion{Fe}{2} lines
\citep{KraftIvans2003}.  There are ways to determine the stellar
parameters that will not be as affected by NLTE effects, e.g., using
colors \citep{RamirezMelendez2005,Casagrande2010} to determine
temperatures or isochrones to determine surface gravities (e.g.,
\citealt{Sakari2017}), but these techniques require some {\it a
  priori} knowledge of the reddening, distance, etc.  Some groups also
utilize empirical corrections to LTE spectroscopic temperatures to more
closely match the photometric temperatures (e.g.,
\citealt{Frebel2013}).  Recently, it has become
  possible to apply NLTE corrections directly to the LTE abundances
  \citep{Lind2012,Ruchti2013,Amarsi2016,Ezzeddine2017}.  This
technique has the benefit of enabling the atmospheric parameters to be
determined solely from the spectra.

An ideal approach should provide the most accurate abundances for
future use, while maintaining compatibility with other samples of
metal-poor stars.  Sections \ref{subsec:StandardAtmParams} and Appendix
\ref{appendix:Comp} demonstrate that adopting spatially- and
temporally-averaged three-dimensional ($<$3D$>$), NLTE corrections (in
this case from \citealt{Amarsi2016}) provide parameters that are in
better agreement with independent methods, compared to purely
spectroscopic LTE parameters.  Although NLTE-corrected parameters from
$<$3D$>$ models are ultimately selected as the preferred values in
this paper, LTE parameters and abundances are provided in Appendix
\ref{appendix:LTE} to facilitate comparisons with LTE studies. Section
\ref{subsec:TargetAtmParams} presents the adopted parameters for the
target stars, Section \ref{subsec:CFe} discusses the [C/Fe] ratios,
and Section \ref{subsec:AtmParamErrors} then discusses the
uncertainties in these parameters.

In the analyses that follow, Fe abundances are determined from
equivalent widths (EWs), which are measured using the program {\tt
  DAOSPEC} \citep{DAOSPECref}.  Only lines with $\rm{REW}<-4.7$ were
used, to avoid uncertainties that arise from, e.g., uncertain damping
constants \citep{McWilliam1995}. All abundances are determined with
the 2017 version of {\tt MOOG} \citep{Sneden}, including an
appropriate treatment for scattering
\citep{Sobeck2011}.\footnote{\url{https://github.com/alexji/moog17scat}}
Kurucz model atmospheres were used \citep{KuruczModelAtmRef}.
For all cases below, the final atmospheric parameters are determined
entirely from the spectra.  Surface gravities are determined by
enforcing ionization equilibrium in iron (i.e., the surface gravities
are adjusted so that the average \ion{Fe}{1} abundance is equal to the
average \ion{Fe}{2} abundance).  Temperatures and microturbulent
velocities are determined by flattening trends in \ion{Fe}{1} line
abundances with EP and REW.  For the NLTE cases, corrrections were
applied to LTE abundance from each \ion{Fe}{1} line, according to the
current atmospheric parameters in that iteration.  The corrections are
determined with the interpolation grid from
\citet{Amarsi2016}.\footnote{\url{http://www.mso.anu.edu.au/~ama51/data/}}

\subsection{Standard Stars}\label{subsec:StandardAtmParams}
The parameters of the previously observed standard stars are first
presented, to 1) establish the effects of the NLTE corrections on the
atmospheric parameters and 2) demonstrate agreement with results from
the literature.

\subsubsection{LTE vs. NLTE}\label{subsubsec:LTEvNLTE}
The LTE and NLTE atmospheric parameters for the standard stars are
shown in Table \ref{table:NLTEStandardAtmParams}.  The naming
convention of \citet{Amarsi2016} is adopted: the 1D, NLTE corrections
are labeled ``NMARCS'' while the $<$3D$>$, NLTE corrections are
``NMTD'' (i.e., NMARCS 3D).  These corrections were applied as in
\citet{Ruchti2013}, using the 1D and $<$3D$>$ NLTE grids from
\citet{Amarsi2016}.  The interpolation scheme from \citet{Lind2012}
and \citet{Amarsi2016} is used to determine the appropriate
corrections for each set of atmospheric parameters; these corrections
are then applied on-the-fly to the LTE abundance from
each \ion{Fe}{1} line (note that the NLTE corrections for the
\ion{Fe}{2} lines are negligible; \citealt{Ruchti2013}).

\begin{figure}[h!]
\begin{center}
\centering
\includegraphics[scale=0.75,trim=0.5in 0.8in 1.0in 0.5in,clip]{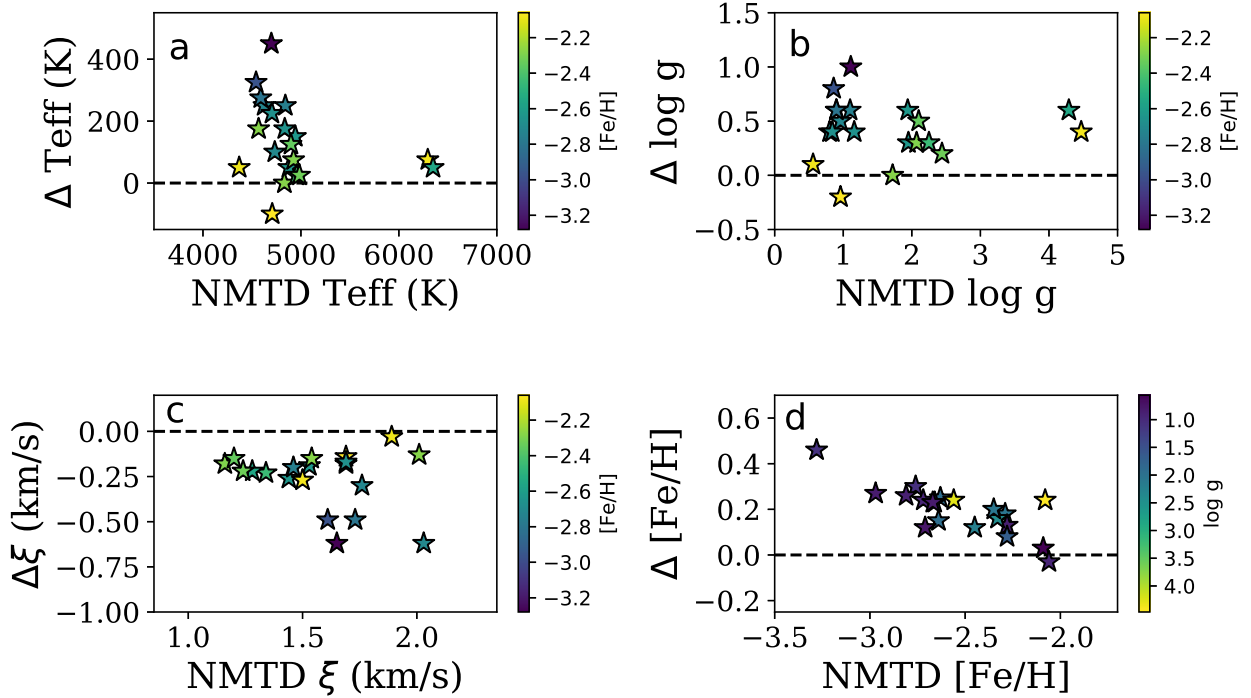}
\caption{Offsets in the atmospheric parameters (NMTD $-$ LTE), as a
  function of the NMTD parameters, for the standard stars.  In panels
  a, b, and c, the points are color-coded according to their [Fe/H]
  ratios, while in panel d they are color-coded according to surface
  gravities.\label{fig:NLTEvsLTE}}
\end{center}
\end{figure}

A qualitative trend is evident from Table
\ref{table:NLTEStandardAtmParams}, and is demonstrated in Figure
\ref{fig:NLTEvsLTE}.  Compared to the LTE values, the NMARCS
corrections moderately affect $T_{\rm{eff}}$, while the NMTD
corrections increase $T_{\rm{eff}}$.  The surface gravities
and metallicities are also generally increased when the NLTE
corrections are applied, while the microturbulent velocities decrease.
These changes are most  severe at the metal-poor end and for the
cooler giants.  It is worth noting that these changes qualitatively
agree with the known problems that occur in purely spectroscopic LTE
analyses, where the temperatures, surface gravities, and metallicities
that are derived from \ion{Fe}{1} lines are known to be
under-estimated, while the microturbulent velocities are
over-estimated.  Appendix \ref{appendix:Comp} more completely
validates the choice of the NMTD parameters through comparisons with
photometric temperatures and parallax-based distances.

\begin{deluxetable}{@{}lcccccccccccDD}
\tabletypesize{\scriptsize}
\tablecolumns{14}
\tablewidth{0pt}
\tablecaption{Atmospheric Parameters and [C/Fe]: Standard Stars\label{table:NLTEStandardAtmParams}}
\hspace*{-4in}
\tablehead{
 & \multicolumn{4}{l}{LTE} & \multicolumn{4}{l}{NMARCS} & \multicolumn{7}{l}{NMTD} \\
     & $T_{\rm{eff}}$ & $\log g$ & $\xi\;\;\;\;$ & [Fe/H] &
  $T_{\rm{eff}}$ & $\log g$ & $\xi\;\;\;\;$  & [Fe/H] & $T_{\rm{eff}}$
  & $\log g$ & $\xi\;\;\;\;$  & \multicolumn{2}{r}{[Fe I/H] ($N$)\tablenotemark{a}$\;\;$} & \multicolumn{2}{r}{[C/Fe]\tablenotemark{b}$\;$} \\
Star & (K)          & \multicolumn{3}{c}{(km s$^{-1}$)$\;\;$} & (K)          & \multicolumn{3}{c}{$\;$(km s$^{-1}$)$\;$}  & (K)          & \multicolumn{3}{c}{$\;$(km s$^{-1}$)}  & \multicolumn{1}{r}{ } & \multicolumn{2}{r}{     }
}
\decimals
\startdata
CS 31082-001    & 4827 & 1.65 & 1.70 & $-2.79$ & 4827 & 1.95 & 1.59 & $-2.68$ & 4877 & 1.95 & 1.44 & -2.64\pm0.01\;\;(87)  & 0.04\pm0.10  \\
TYC 5861-1732-1 & 4850 & 1.77 & 1.34 & $-2.47$ & 4825 & 1.87 & 1.23 & $-2.39$ & 4925 & 2.07 & 1.16 & -2.29\pm0.02\;\;(109) & -0.29\pm0.11 \\
CS 22169-035    & 4483 & 0.50 & 2.01 & $-3.03$ & 4458 & 0.50 & 2.03 & $-3.02$ & 4683 & 0.70 & 1.75 & -2.80\pm0.02\;\;(86)  & 0.58\pm0.10  \\
TYC 75-1185-1   & 4793 & 1.34 & 1.72 & $-2.88$ & 4793 & 1.54 & 1.63 & $-2.79$ & 4943 & 1.94 & 1.53 & -2.63\pm0.02\;\;(89)  & 0.05\pm0.10  \\
TYC 5911-452-1  & 6220 & 4.07 & 1.77 & $-2.32$ & 6195 & 4.27 & 1.60 & $-2.19$ & 6295 & 4.47 & 1.50 & -2.08\pm0.02\;\;(39)  & -0.15\pm0.20 \\
TYC 5329-1927-1 & 4393 & 0.30 & 2.14 & $-2.41$ & 4368 & 0.20 & 2.12 & $-2.41$ & 4568 & 0.90 & 2.01 & -2.28\pm0.02\;\;(101) & 0.43\pm0.11  \\
TYC 6535-3183-1 & 4320 & 0.46 & 1.92 & $-2.12$ & 4295 & 0.36 & 1.91 & $-2.15$ & 4370 & 0.56 & 1.89 & -2.09\pm0.02\;\;(103) & 0.23\pm0.10  \\
TYC 4924-33-1   & 4831 & 1.72 & 1.69 & $-2.36$ & 4806 & 1.82 & 1.62 & $-2.30$ & 4831 & 1.72 & 1.54 & -2.28\pm0.01\;\;(112) & 0.27\pm0.10  \\
HE 1116$-$0634  & 4248 & 0.01 & 2.17 & $-3.72$ & 4198 & 0.01 & 2.28 & $-3.75$ & 4698 & 1.11 & 1.65 & -3.28\pm0.02\;\;(58)  & 0.54\pm0.20  \\
TYC 6088-1943-1 & 4931 & 1.95 & 1.57 & $-2.54$ & 4931 & 2.25 & 1.50 & $-2.43$ & 4956 & 2.25 & 1.34 & -2.45\pm0.01\;\;(96)  & -0.14\pm0.11 \\
BD-13 3442      & 6299 & 3.69 & 1.50 & $-2.80$ & 6299 & 4.09 & 1.35 & $-2.64$ & 6349 & 4.29 & 1.28 & -2.56\pm0.02\;\;(14)  & <0.55        \\
BD-01 2582      & 4960 & 2.24 & 1.46 & $-2.49$ & 4960 & 2.54 & 1.40 & $-2.37$ & 4985 & 2.44 & 1.24 & -2.33\pm0.01\;\;(100) & 0.71\pm0.10  \\
HE 1317$-$0407  & 4660 & 0.76 & 1.87 & $-2.89$ & 4660 & 0.86 & 1.79 & $-2.83$ & 4835 & 1.16 & 1.69 & -2.66\pm0.02\;\;(86)  & 0.15\pm0.20  \\
HE 1320$-$1339  & 4591 & 0.50 & 1.66 & $-3.06$ & 4591 & 0.60 & 1.60 & $-3.02$ & 4841 & 1.10 & 1.46 & -2.76\pm0.02\;\;(81)  & 0.0\pm0.20   \\
HD 122563       & 4374 & 0.46 & 2.06 & $-2.96$ & 4324 & 0.26 & 2.09 & $-2.97$ & 4624 & 0.96 & 1.76 & -2.71\pm0.01\;\;(96)  & 0.49\pm0.13  \\
TYC 4995-333-1  & 4807 & 1.16 & 1.83 & $-2.02$ & 4707 & 0.96 & 1.75 & $-2.07$ & 4707 & 0.96 & 1.71 & -2.06\pm0.02\;\;(107) & 0.14\pm0.10  \\
HE 1523-0901    & 4290 & 0.20 & 2.13 & $-3.09$ & 4315 & 0.40 & 2.16 & $-3.06$ & 4590 & 0.90 & 1.73 & -2.81\pm0.02\;\;(79)  & 0.39\pm0.15  \\
TYC 6900-414-1  & 4798 & 1.50 & 1.24 & $-2.45$ & 4823 & 1.80 & 1.17 & $-2.35$ & 4898 & 2.00 & 1.10 & -2.28\pm0.02\;\;(108) & -0.04\pm0.10 \\
J2038-0023      & 4579 & 0.84 & 2.03 & $-2.89$ & 4579 & 0.94 & 1.97 & $-2.84$ & 4704 & 0.94 & 1.77 & -2.71\pm0.02\;\;(88)  & 0.59\pm0.10  \\
BD-02 5957      & 4217 & 0.06 & 2.05 & $-3.22$ & 4192 & 0.06 & 2.10 & $-3.23$ & 4567 & 0.96 & 1.57 & -2.91\pm0.02\;\;(78)  & 0.54\pm0.10  \\
\enddata
\medskip
\tablenotetext{a}{Note that the NLTE \ion{Fe}{2} abundances are
  required to be equal to the \ion{Fe}{1} abundances.  The quoted
  uncertainty is the random error in the mean, and is the
  line-to-line dispersion divided by $\sqrt{N}$, where $N$ is the
  number of spectral lines.} 
\tablenotetext{b}{The [C/Fe] ratios have been corrected for
  evolutionary effects \citep{Placco2014}.}
\end{deluxetable}

The NMARCS parameters were also compared with parameters derived
using the 1D NLTE corrections following \citet{Ezzeddine2017}.  
Similar to the process for the \citet{Amarsi2016} corrections, the 
NLTE corrections for each \ion{Fe}{1} line were found by interpolating
the measured EWs over a calculated grid of NLTE EWs over a dense
parameter space in effective temperature, surface gravity,
metallicity, and microturbulent velocity.  The 1D MARCS model
atmospheres \citep{Gustafsson2008} were used with the NLTE radiative
transfer code \texttt{MULTI2.3} \citep{Carlsson1986,Carlsson1992} to
calculate the EW grid.  A comprehensive \ion{Fe}{1}/\ion{Fe}{2} model
atom is used in the calculations, with up-to-date inelastic collisions
with hydrogen implemented from \citet{Barklem2018}; see
\citet{Ezzeddine2016} for more details on the atomic model and data.  
Compared to the NMARCS values, the Ezzeddine et al. corrections lead
to agreement in temperature within 50 K, surface gravities within 0.5
dex, microturbulent velocities within 0.5 km~s$^{-1}$, and
metallicities within 0.1 dex.

\begin{figure}[h!]
\begin{center}
\centering
\includegraphics[scale=0.75,trim=0.5in 0.8in 1.0in 0.5in,clip]{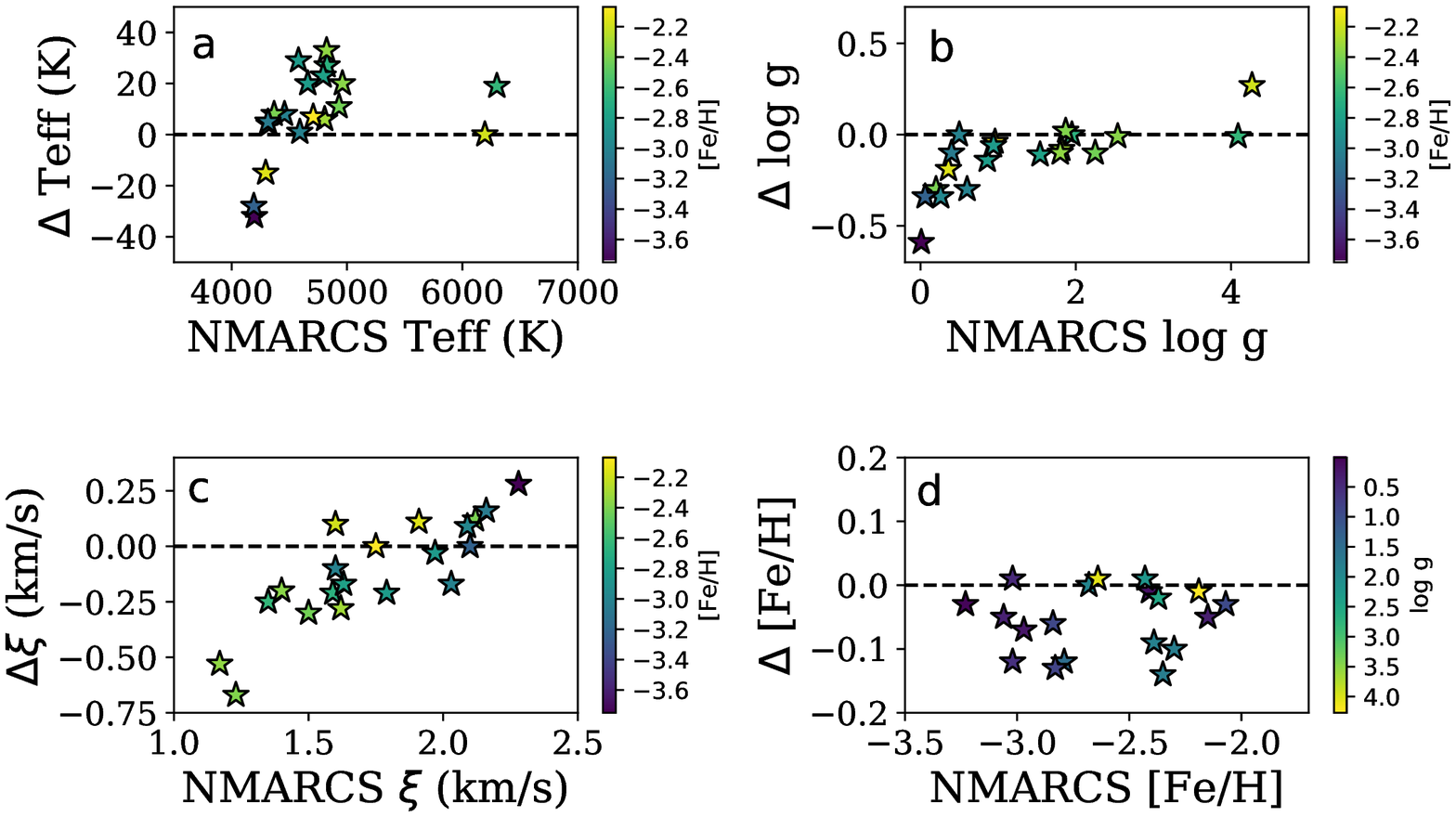}
\caption{Offsets in the atmospheric parameters derived with 1D NLTE
  corrections (Amarsi et al. $-$ Ezzeddine et al.) for the standard
  stars.  Panels are color-coded as in Figure
  \ref{fig:NLTEvsLTE}.\label{fig:NMARCS_vs_Ezzeddine}}
\end{center}
\end{figure}

\subsubsection{Comparisons with Literature Values}\label{subsubsec:StdAtmParamLitComp}
\begin{figure}[h!]
\begin{center}
\centering
\includegraphics[scale=0.75,trim=0.5in 0.4in 1.5in 0.35in,clip]{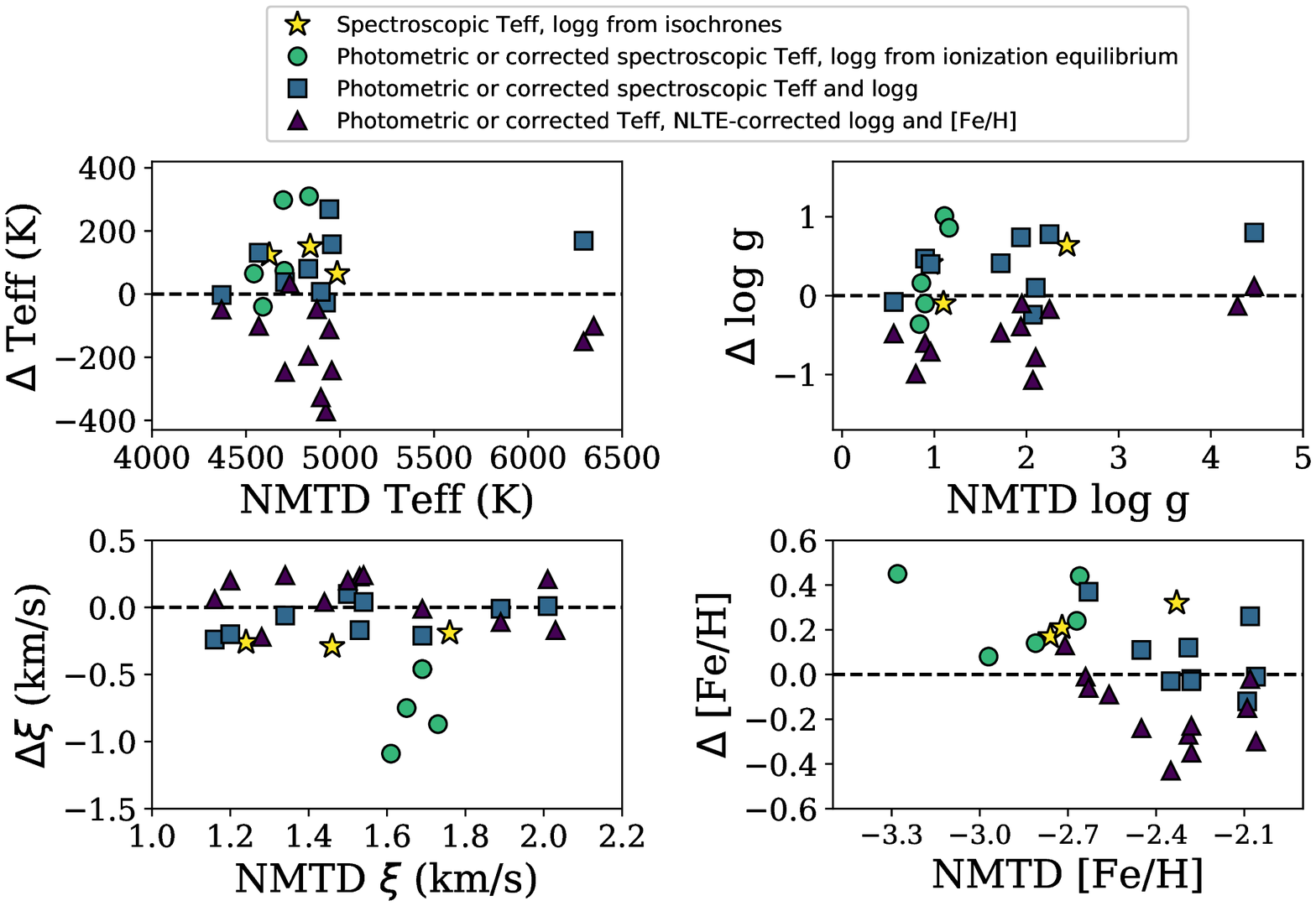}
\caption{Differences between the $<$3D$>$, NLTE (NMTD) atmospheric
  parameters and parameters from the literature for the standard
  stars (NMTD$-$literature).  Some stars are shown multiple times from
  different studies. The yellow stars show comparisons with
  spectroscopic LTE temperatures and isochrone-based surface gravities
  from \citet{Roederer2014}.  The green circles show comparisons with
  \citet{Frebel2007}, \citet{Hollek2011}, \citet{Thanathibodee2016},
  and \citet{Placco2017}, LTE analyses that utilized either
  photometric temperatures or spectroscopic temperatures with
  corrections to match photometric temperatures, and surface gravities
  derived by requiring ionization equilibrium.  The blue squares
  compare with \citet{Ruchti2011}, who utilized photometric or
  corrected LTE spectroscopic temperatures and surface gravities
  derived from photometry.  Finally, the purple triangles show
  comparisons with \citet{Hansen2013} and \citet{Ruchti2013}, who used
  photometric or corrected spectroscopic temperatures and 1D NLTE
  corrections to determine the surface gravity and metallicity.
\label{fig:CompAbundParamLit}}
\end{center}
\end{figure}

The NMTD parameters are compared to LTE and NLTE literature
values in Figure \ref{fig:CompAbundParamLit}.  As with any set of
spectroscopic analyses, the techniques used to derive the atmospheric
parameters vary significantly between groups; the points in Figure
\ref{fig:CompAbundParamLit} are therefore grouped roughly by
technique.  Again, the results qualitatively make sense when compared
with the LTE results from the literature (from
\citealt{Frebel2007,Hollek2011,Roederer2014,Thanathibodee2016,Placco2017}):
the NMTD temperatures are slightly higher than values derived
spectroscopically, occasionally even when empirical corrections are
included to raise the temperature.  The surface gravities are
typically higher than the values derived with LTE ionization
equilibrium and isochrones, while the microturbulent velocities are
much lower than the studies that utilize LTE ionization equilibrium to
derive surface gravities. Finally, the [Fe/H] ratios agree reasonably
well at the metal-rich end, but become increasingly discrepant with
lower [Fe/H].  These findings are all consistent with those from
\citet{Amarsi2016}.

\citet{Hansen2013} and \citet{Ruchti2013} adopted NLTE corrections of
some sort in previous analyses of standard stars in this paper, albeit
with slightly different techniques for deriving the final atmospheric
parameters.  \citet{Hansen2013} adopted photometric temperatures and
then applied 1D NLTE corrections to $\log g$ and [Fe/H]; the agreement
with those points is generally good. \citet{Ruchti2013} applied 1D
NLTE corrections to LTE abundances, as in this paper; a key
difference, however, is that Ruchti et al. did not use \ion{Fe}{1}
lines with $\rm{EP}<2$ eV, which they argue are more sensitive to the
NLTE effects.  As a result, Ruchti et al. find even higher
temperatures, surface gravities, and metallicities, values which would
no longer agree with the previous LTE analyses, even when photometric
temperatures and parallax-based surface gravities are adopted.

Given that the spectroscopic NMTD-corrected parameters in this paper
agree well with the photometric temperatures and gravities from the
literature (also see Appendix \ref{appendix:Comp}), the NMTD
parameters are adopted for the rest of the paper.

\subsubsection{The Case of HD 122563}\label{subsubsec:HD122563}
The standard HD 122563 was one of the stars in
\citet{Amarsi2016}, the paper which provides the $<$3D$>$, NLTE
corrections that are used in this analysis.  Amarsi et
al. were able to achieve ionization equilibrium with NMTD corrections
for all of their target stars {\it except} for HD~122563.  They
suggested that the parallax-based surface gravity from the
literature was too high, and that $\log g\approx 1.1$ was more
appropriate. Naturally, with the Amarsi et al. corrections the NMTD
spectroscopic gravity in Table \ref{table:NLTEStandardAtmParams},
$\log g = 0.96$, is indeed lower than the parallax-based value used in
\citet{Hansen2013}. \citet{Roederer2014} also find a lower value using
isochrones. Indeed, {\it Gaia} DR2 provides a smaller parallax and
error than the {\it Hipparcos} value: {\it Gaia} finds a parallax of
$3.44\pm0.06$, while {\it Hipparcos} found $4.22\pm0.35$
\citep{HIPref}.  This suggests that the surface gravity is indeed
lower (i.e., the star is farther away and intrinsically brighter) than
previously predicted (also see Section \ref{subsec:CompGAIA}).

\subsection{Atmospheric Parameters: Target Stars}\label{subsec:TargetAtmParams}
Beyond the choice of LTE or NLTE, stellar abundance analyses suffer
from a variety of other systematic errors as a result of, e.g., atomic
data, choice of model atmospheres, etc.  These effects have been
mitigated in the past by performing {\it differential} analyses with
respect to a set of standard stars.  A differential
  analysis reduces the systematic offsets relative to the standard
  star, enabling higher precision parameters and abundances to be
  determined.  This type of analysis has been performed on both
metal-rich
\citep{Fulbright2006,Fulbright2007,KochMcWilliam2008,McWilliam2013,Sakari2017}
and metal-poor stars \citep{OMalley2017,Reggiani2016,Reggiani2017} and
is the approach that is chosen for the target stars.  The stars
identified in Table \ref{table:FinalAtmParams} are used as the
differential standards.

Each target is matched up with a standard star based on its initial
atmospheric parameters, and $\Delta \log \epsilon$(\ion{Fe}{1})
abundances are calculated for each line with respect to the standard,
again using NLTE $<$3D$>$ corrections.  Flattening the slopes in 
$\Delta \log \epsilon$(\ion{Fe}{1}) with EP and REW provide the
relative temperature and microturbulent velocity offsets for the
target, while the offset between the $\Delta
\log\epsilon$(\ion{Fe}{1}) and $\Delta \log \epsilon$(\ion{Fe}{2})
abundances is then used to determine the relative $\log g$.  These
relative offsets are then applied to the NLTE atmospheric parameters
of the standard stars.  If the atmospheric parameters are
  in better agreement with another standard, the more appropriate
  standard is selected and the process is redone.  Note that the
  choice of standard does not significantly affect the final
  atmospheric parameters, unless the two stars have very different
  parameters (and therefore few lines in common); in this case, the
  final atmospheric parameters indicate that another standard would be
  more appropriate. This process is very similar to that of
\citet{OMalley2017}, except that this analysis utilizes $<$3D$>$ NLTE
corrections.

The final NMTD atmospheric parameters are shown in Table
\ref{table:FinalAtmParams}.  Because LTE parameters are still widely
used in the community, LTE parameters are also provided in Appendix
\ref{appendix:LTE}.  However, it is worth noting that the NMTD values
in this paper produce similar results to the photometric temperatures
and gravities, and the LTE values may not be the best choice for
comparisons with literature values.

\begin{figure}[h!]
\begin{center}
\centering
\includegraphics[scale=0.65,trim=0.8in 0.4in 0.5in 0.35in,clip]{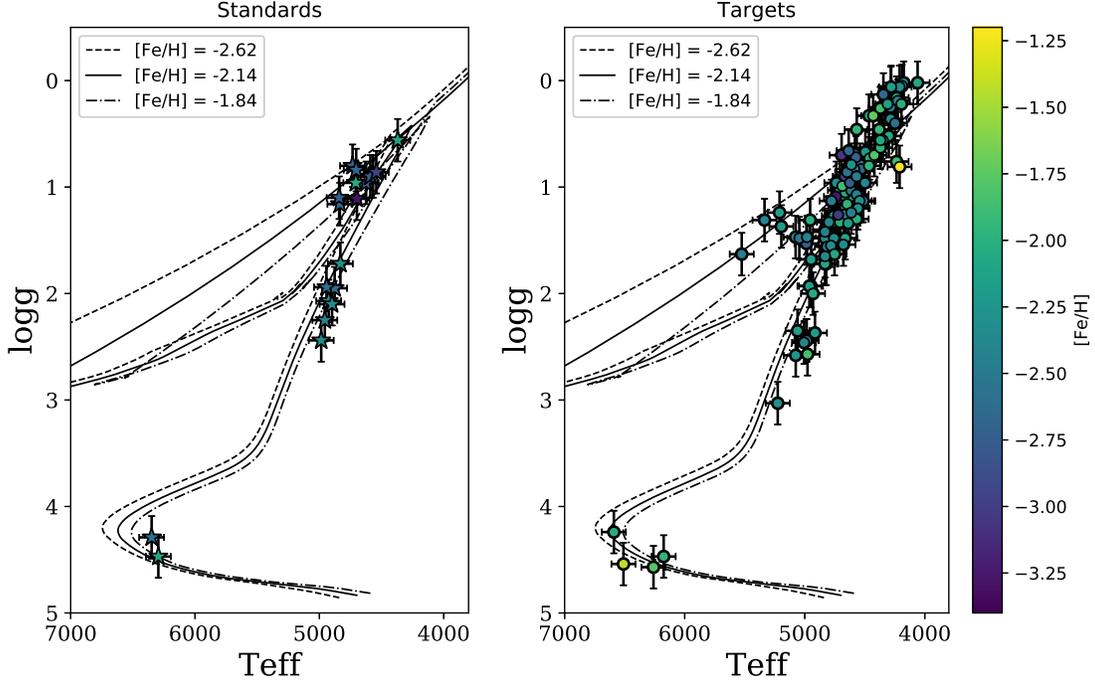}
\caption{A Hertzsprung-Russell diagram showing surface gravity versus
  effective temperature.  The standard stars are shown in the left
  panel, while the targets are shown in the right panel; both are
  color-coded by [Fe/H].  Three BaSTI isochrones are shown, with
  $[\rm{Fe/H}] = -1.84$, $-2.14$, and $-2.62$ (both with
  $[\alpha/\rm{Fe}]=+0.4$ and ages of 14 Gyr).\label{fig:HRD}}
\end{center}
\end{figure}

The spectroscopic temperatures, gravities, and metallicities can be
directly compared to stellar isochrones, e.g., the BaSTI/Teramo models
\citep{BaSTIref}.  Figure \ref{fig:HRD} shows a spectroscopic
HR-Diagram with the standard and target stars color-coded by [Fe/H].
Overplotted are 14 Gyr, $\alpha$-enhanced BaSTI isochrones at
$[\rm{Fe/H}] = -1.84$, $-2.14$ and $-2.62$.  The BaSTI isochrones
persist through the AGB phase; extended AGBs with a mass-loss
parameter of $\eta=-0.2$ are shown. Some of the brightest stars are
slightly hotter than the RGB for their [Fe/H], indicating that they
may be AGB stars.  Four of the targets are main-sequence stars.

\startlongtable
\begin{deluxetable}{@{}cccccDDD}
\tabletypesize{\scriptsize}
\tablecolumns{8}
\tablewidth{0pt}
\tablecaption{Atmospheric Parameters and [C/Fe]: Target Stars\tablenotemark{a}\label{table:FinalAtmParams}}
\hspace*{-4in}
\tablehead{
Star & Reference Standard & $T_{\rm{eff}}$ (K)\tablenotemark{b} &
$\log g$\tablenotemark{b} & $\xi$ (km/s)\tablenotemark{b} &
\multicolumn{2}{c}{[\ion{Fe}{1}/H] ($N$)\tablenotemark{c}} & \multicolumn{2}{c}{[\ion{Fe}{2}/H] ($N$)\tablenotemark{c}} & \multicolumn{2}{c}{[C/Fe]\tablenotemark{d}}
}
\decimals
\startdata
J0007$-$0345   & TYC 5329-1927-1  & 4663 & 1.48 & 2.07 & -2.09\pm0.01\;\;(91)  & -2.10\pm0.03\;\;(24) &  0.17\pm0.07 \\ 
J0012$-$1816   & BD$-$01 2582       & 4985 & 2.44 & 1.27 & -2.28\pm0.01\;\;(94)  & -2.27\pm0.02\;\;(17) & -0.26\pm0.15 \\ 
J0022$-$1724   & HE 1116-0634     & 4718 & 1.11 & 1.29 & -3.38\pm0.03\;\;(30)  & -3.44\pm0.11\;\;(3)  &  1.87\pm0.13\tablenotemark{e} \\ 
J0030$-$1007   & TYC 4924-33-1    & 4831 & 1.48 & 1.97 & -2.35\pm0.02\;\;(90)  & -2.34\pm0.06\;\;(14) &  0.50\pm0.20 \\ 
J0053$-$0253   & TYC 6535-3183-1  & 4370 & 0.56 & 1.81 & -2.16\pm0.01\;\;(93)  & -2.16\pm0.04\;\;(25) &  0.40\pm0.07 \\ 
J0054$-$0611   & TYC 4995-333-1   & 4707 & 1.03 & 1.74 & -2.32\pm0.02\;\;(89)  & -2.37\pm0.08\;\;(15) &  0.50\pm0.14 \\ 
J0107$-$0524   & BD$-$01 2582       & 5225 & 3.03 & 1.20 & -2.32\pm0.01\;\;(82)  & -2.36\pm0.03\;\;(13) & -0.09\pm0.07 \\ 
J0145$-$2800   & TYC 4995-333-1   & 4582 & 0.69 & 1.57 & -2.60\pm0.02\;\;(79)  & -2.58\pm0.05\;\;(12) &  0.34\pm0.15 \\ 
J0156$-$1402   & TYC 4995-333-1   & 4622 & 1.09 & 2.27 & -2.08\pm0.02\;\;(86)  & -2.07\pm0.05\;\;(20) &  0.37\pm0.13 \\ 
2MJ0213$-$0005 & TYC 5911-452-1   & 6225 & 4.54 & 2.33 & -1.88\pm0.02\;\;(38)  & -1.93\pm0.08\;\;(5)  & -0.38\pm0.07 \\ 
\enddata
\tablenotetext{a}{Only a portion of this table is shown here to
  demonstrate its form and content. A machine-readable version of the
  full table is available.}
\tablenotetext{b}{Errors in the atmospheric parameters are discussed
  in Section \ref{subsec:AtmParamErrors}.}
\tablenotetext{c}{The quoted uncertainty is the random error in the
  mean, and is the line-to-line dispersion divided by $\sqrt{N}$,
  where $N$ is the number of spectral lines.}
\tablenotetext{d}{The [C/Fe] ratios have been corrected for
  evolutionary effects \citep{Placco2014}.}
\tablenotetext{e}{The star's high [C/Fe] makes it a CEMP star,
  according to the $[\rm{C/Fe}]>+0.7$ criterion.}
\end{deluxetable}

A small number of stars were also erroneously flagged as metal-poor
($[\rm{Fe/H}]<-1$) in the moderate-resolution observations.  These
stars are shown in Table \ref{table:BadGuys}, and include hot,
metal-rich stars and cool M dwarfs.

\begin{deluxetable}{@{}cc}
\tabletypesize{\scriptsize}
\tablecolumns{2}
\tablewidth{0pt}
\tablecaption{Stars that are Likely not Metal-Poor}\label{table:BadGuys}
\hspace*{-4in}
\tablehead{
 & Type
}
\startdata
J0120$-$2622       & Hot, metal-rich star\\  
J0958$-$0323       & Hot, modestly metal-rich star ($[\rm{Fe/H}] \sim -0.8$)\\ 
J1555$-$0359       & M star\\ 
\enddata
\end{deluxetable}

\subsection{Carbon}\label{subsec:CFe}
Carbon abundances were determined from syntheses of the CH $G$-band at
4312 \AA \hspace{0.025in} and the neighboring feature at 4323 \AA.  In
some stars, particularly the hotter ones, only upper limits are
available.  The evolutionary corrections of \citet{Placco2014} were
applied to account for C depletion after the first dredge up.  Most of
the stars have [C/Fe] ratios that are consistent with typical
metal-poor MW halo stars, though there are a few carbon-enhanced
metal-poor (CEMP) stars with $[\rm{C/Fe}]~>~+~0.7$.  One of the standards,
BD$-$01 2582, is a CEMP star, in agreement with \citet{Roederer2014}.
Of the targets, eight are found to be CEMP stars---these stars will be
further classified according to their $r$- and $s$-process enrichment
in Section \ref{subsec:TargetAbunds}.

\subsection{Uncertainties in Atmospheric Parameters}\label{subsec:AtmParamErrors}
Uncertainties in the atmospheric parameters are calculated for
seven standard stars covering a range in [Fe/H], temperature, and
surface gravity.  The full details are given in Appendix
\ref{appendix:Errors}.  Briefly, because the parameters are determined
from Fe lines, the uncertainties increase with decreasing [Fe/H] and
increasing temperature, a natural result of having fewer \ion{Fe}{1}
and \ion{Fe}{2} lines.  The detailed analysis in Appendix
\ref{appendix:Errors} demonstrates that the typical uncertainties in
temperature range from 20 to 200 K, in $\log g$ from 0.05 to 0.3 dex,
and in microturbulence from $0.10$ to $0.35$ km s$^{-1}$.  These
parameters are not independent, as demonstrated by the
covariances in Table \ref{table:AppendixErrors}---however, the
covariances are generally fairly small.


\section{Chemical Abundances}\label{sec:Abunds}
All abundances are determined in
{\texttt{MOOG}}.  In general, lines with $\rm{REW}>-4.7$ are not
utilized because of issues with damping and treatment of the outer
layers of the atmosphere \citep{McWilliam1995}; some exceptions are
made, and are noted below.  The line lists were generated with the
{\tt linemake}
code\footnote{\vspace{0.5in}\url{https://github.com/vmplacco/linemake}},
and include hyperfine structure, isotopic splitting, and molecular
lines from CH, C$_{2}$, and CN.  Abundances of Mg, Si, K, Ca, Sc, Ti,
V, Cr, Mn, and Ni were determined from EWs (see Table
\ref{table:EWs}), while abundances of Li, O, Na, Al, Cu, Zn, Sr, Y,
Zr, Ba, La, Ce, Pr, Nd, Sm, Eu, Dy, Os, and Th were determined from
spectrum syntheses (see Table \ref{table:SSs}), whenever the lines are
sufficiently strong. Note that most of the stars will only have
detectable lines from a handful of these latter elements.


All [X/H] ratios are calculated line-by-line with respect to the Sun
when the Solar line is sufficiently weak (REW$<-4.7$; see Table
\ref{table:SSs}); otherwise, the Solar abundance from
\citet{Asplund2009} is adopted. The Solar EWs from
\citet{Fulbright2006,Fulbright2007} are adopted when EW analyses are
used.  The use of ionization equilibrium to derive $\log g$ ensures
that [\ion{Fe}{1}/H] and [\ion{Fe}{2}/H] are equal within the errors;
regardless, [X/Fe] ratios for singly ionized species utilize
\ion{Fe}{2}, while neutral species utilize \ion{Fe}{1}.  Systematic
errors that occur as a result of uncertainties in the atmospheric
parameters are discussed in Appendix \ref{appendix:Errors}.

Table \ref{table:SrBaEuAbunds} shows the abundances of Sr, Ba, and Eu
and the corresponding classifications, while the other abundances are
given in Table \ref{table:Abunds}.  The stars are classified according
to their $r$-process enhancement, where $[\rm{Ba/Eu}]<0$ defines stars
without significant $s$-process contamination.  The $r$-I and $r$-II
definitions ($+0.3\le[\rm{Eu/Fe}]\le+1$ and $[\rm{Eu/Fe}]>+1$,
respectively) are from \citet{BeersChristlieb2005}, and the
limited-$r$ definition ($[\rm{Eu/Fe}] < +0.3$, $[\rm{Sr/Ba}]>+0.5$) is
from \citet{Frebel2018}.  The CEMP-$r$ definition has been expanded to
include $r$-I stars, as in \citet{Hansen2018}.  Stars with
$0~<~[\rm{Ba/Eu}]~<~+~0.5$ are classified as $r/s$, following the
scheme from \citet{BeersChristlieb2005}. However, recent work by
\citet{Hampel2016}  attributes the heavy-element abundance patterns in
these stars to the $i$-process, a form of neutron-capture
nucleosynthesis with neutron densities \textit{intermediate} between
the $r$- and $s$-processes \citep{CowanRose1977,Herwig2011}.
The stars with $[\rm{Eu/Fe}] < +0.3$, $[\rm{Ba/Eu}]~<~0$, and
$[\rm{Sr/Ba}]<+0.5$ are not $r$-process-enhanced, and are classified
as ``not-RPE.'' 

Below, the abundances of the standard stars are compared with the
literature values, the abundances of the target stars are introduced,
and the abundances and $r$-process classifications of the target stars
are presented.

\subsection{Standard Stars: Comparison with Literature Values}\label{subsec:StandardAbunds}
\begin{figure}[h!]
\begin{center}
\centering
\includegraphics[scale=0.65]{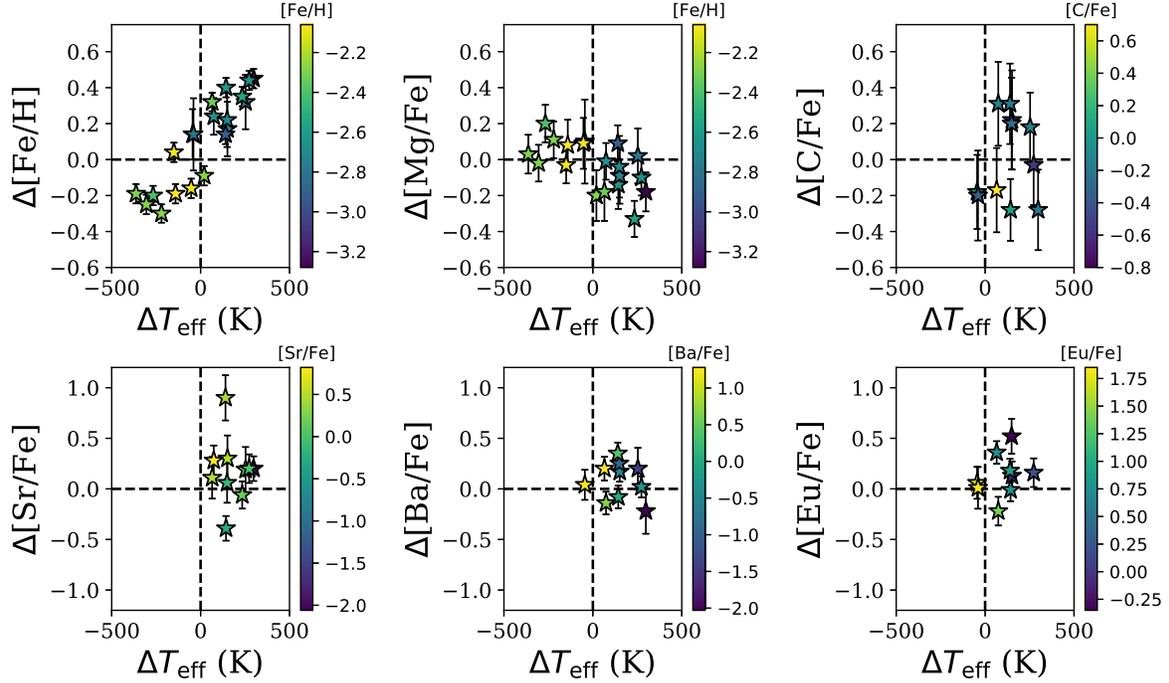}
\caption{Offsets between the abundances in this paper and those from
  the literature, as a function of offsets in the adopted effective
  temperature.  Note that the literature atmospheric parameters are
  all derived in slightly different ways.  With the
    exception of some [Fe/H] ratios, all literature abundances were
    determined under assumptions of LTE.  References for literature abundances are
  given in the text.\label{fig:CompAbunds_Teff}} 
\end{center}
\end{figure}

With the exception of Fe (for some stars), all literature abundances
were determined only under assumptions of LTE; any offsets from
previous analyses are thus likely driven by the differences in
the atmospheric parameters (see Appendix \ref{appendix:Errors}).  The
abundance offsets between this study and those in the literature are
shown in Figure \ref{fig:CompAbunds_Teff}, utilizing the LTE
abundances from \citet{Barklem2005}, \citet{Boesgaard2011},
\citet{Hollek2011}, \citet{Ruchti2011}, \citet{Roederer2014}, and
\citet{Thanathibodee2016}.  The abundances are given as a function of
the difference in temperature, and are color-coded according to their
[Fe/H] or [X/Fe] ratios.  Only the most important elements for this
paper are shown: Fe, the proxy for metallicity; C, which is necessary
to identify CEMP stars; Mg, a representative for the
$\alpha$-abundance; and Sr, Ba, and Eu, which are used to characterize
the $r$- and $s$-process enrichment.  Figure \ref{fig:CompAbunds_Teff}
shows that there is a strong dependence on temperature for [Fe/H],
with good agreement when the temperatures are similar.  There are
fewer data points for the other elements, yet they show decent
agreement even with large temperature offsets except for a few
outliers.

Despite slight differences in the abundance ratios, the Sr, Ba, and Eu
ratios lead to $r$-process classifications (Table
\ref{table:SrBaEuAbunds}) that agree with those from the literature:
CS 31082-001, HE 1523$-$0901, and J2038$-$0023 are correctly identified as
$r$-II stars, while TYC 75-1185-1 and BD$-$02 5957 are identified as
$r$-I stars. Some of these stars have not had previous analyses of the
neutron-capture elements, since \citet{Ruchti2011} only examined the
$\alpha$-elements. This paper has therefore discovered three new
$r$-I stars in the standard sample: TYC 5329-1927-1, TYC 6535-3183-1,
and TYC 6900-414-1.  CS 22169$-$035, HE 1320$-$1339, and HD 122563 were
correctly found to have ``limited-$r$'' signatures (see
\citealt{Frebel2018}); BD$-$13 3442's abundances hint at a possible
limited-$r$ signature as well, based on its [Sr/Ba] ratio.  This
analysis has also re-identified a CEMP-$s$ star, BD$-$01 2582, and a
number of metal-poor stars with $[\rm{Eu/Fe}]<+0.3$.

\clearpage

\subsection{Abundances of Target Stars}\label{subsec:TargetAbunds}

\subsubsection{$r$-Process Enhancement}\label{subsubsec:RPE}
\begin{figure}[h!]
\begin{center}
\centering
\includegraphics[scale=0.7,trim=0.3in 0.55in 0.2in 0.7in,clip]{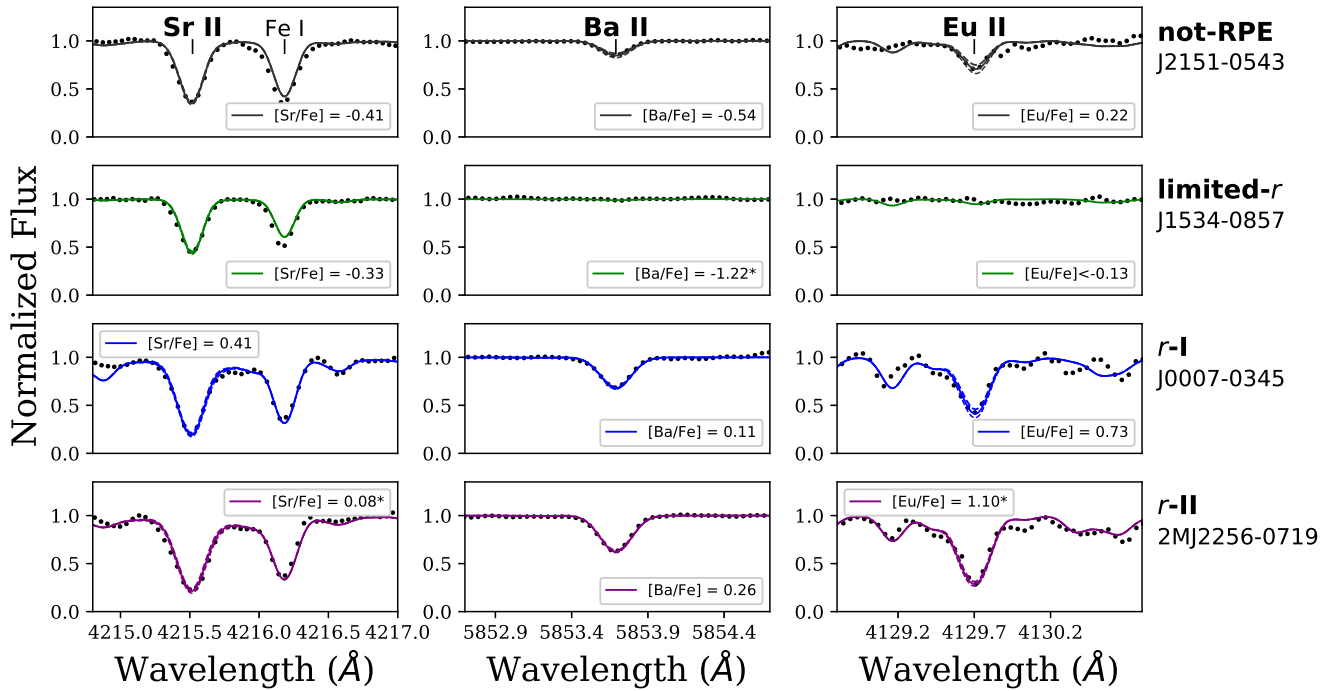}
\caption{Syntheses of Sr, Ba, and Eu lines in four different stars,
  one not-RPE, one limited-$r$, one $r$-I, and one $r$-II star.  The
  dashed lines show the $\pm1\sigma$ errors for a single line.  The
  lines marked with asterisks were not used to determine the
  abundances, either because they were too strong or too weak in
  that star; in this case, they are merely shown for illustrative
  purposes.
\label{fig:Synths}}
\end{center}
\end{figure}

The ultimate goal of this paper is to identify $r$-process-enhanced
metal-poor stars; particular emphasis is therefore placed on the
elements used for this classification, Sr, Ba, and Eu, which are all
determined via spectrum syntheses (see Figure \ref{fig:Synths}).  The
\ion{Sr}{2} line at 4077 \AA \hspace{0.025in} is frequently too strong
for a reliable abundance; conversely, the line at 4161
\AA \hspace{0.025in} is frequently too weak.  The line at
4215~\AA \hspace{0.025in} is generally the best of the three lines,
though it is occasionally slightly stronger than the $\rm{REW}=-4.7$
limit. In this case, the Y abundances provide additional constraints
on the lighter neutron-capture elements.  Ba abundances are determined
for all of the stars in the sample, from the \ion{Ba}{2} 4554, 5853,
6141, and 6496 \AA \hspace{0.025in} lines.  The 4554
\AA \hspace{0.025in} line is really only sufficiently weak in the
hottest ($T\ga 6000$ K) or most barium-poor ($[\rm{Ba/H}] \la -3$)
stars.  Note that the strong 4554 \AA \hspace{0.025in} \ion{Ba}{2} and
4077 and 4215 \AA \hspace{0.025in} \ion{Sr}{2} lines may be affected
by NLTE effects; however, \citet{ShortHauschildt2006} quote an offset
in Ba of only $+0.14$ dex in red giant stars, with smaller effects on
Sr.

Eu abundances or upper limits are also provided for all stars,
from the \ion{Eu}{2} 4129, 4205, 4435, and (only in certain cases)
6645~\AA \hspace{0.025in} lines.  In some cases, the Eu upper limits
may not be sufficient to determine if the star is
$r$-process-enhanced, particularly if the star is hotter than 
$\sim 5500$ K. Occasionally, the lower limits in [Ba/Eu] lie below the
lower limit for the Solar $r$-process residual; in this case, a second
set of limits is also provided in parentheses in Table
\ref{table:SrBaEuAbunds}, assuming that $[\rm{Ba/Eu}]~>~-~0.89$
\citep{Burris2000}.  Table \ref{table:SrBaEuAbunds} shows the
classifications for the 20 standards and the 126 new targets.

\begin{figure}[h!]
\begin{center}
\centering
\includegraphics[scale=0.65]{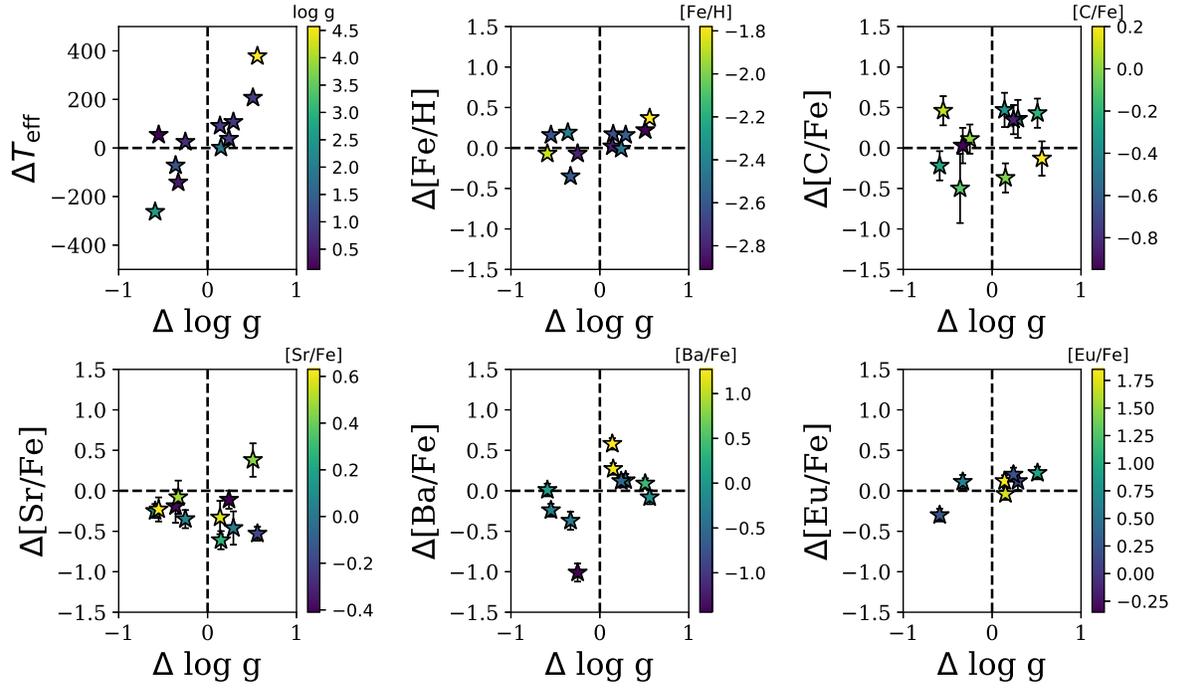}
\caption{Offsets (this paper $-$ Hansen et al.) between the abundances
  in this paper and those from \citet{Hansen2018}, as a function of
  the surface gravity (the parameter which varies most between the
  studies).\label{fig:CompAbunds_TH}}
\end{center}
\end{figure}

Seven of the target stars and three of the standards overlap with the
Southern Hemisphere sample from \citet{Hansen2018}---Figure
\ref{fig:CompAbunds_TH} shows the parameter and abundance comparison.
The temperatures, [Fe/H], and [Eu/Fe] ratios are generally in good
agreement; though Hansen et al. did not employ NLTE corrections, they
did use the \citet{Frebel2013} correction to their spectroscopic
temperatures.  The Sr abundances in this paper are slightly lower, on
average, than Hansen et al., and there are occasional disagreements in
[Ba/Fe]. Still, the $r$-I and $r$-II classifications match, with one
exception: Hansen et al. classify CS 22169-035 as an $r$-I star while
here it is classified as limited-$r$.

\startlongtable
\begin{deluxetable}{@{}lcDDDDD}
\tabletypesize{\scriptsize}
\tablewidth{0pt}
\tablecaption{$r$-Process-Enhancement Classifications and Sr, Ba, and Eu Abundance Ratios\label{table:SrBaEuAbunds}}
\tablehead{
Star       & Class      & \multicolumn{2}{c}{[Sr/Fe]} & \multicolumn{2}{c}{[Ba/Fe]} & \multicolumn{2}{c}{[Eu/Fe]} & \multicolumn{2}{c}{[Ba/Eu]} & \multicolumn{2}{c}{[Sr/Ba]} }
\decimals
\startdata
{\bf Standards} & & \multicolumn{2}{c}{ } & \multicolumn{2}{c}{ } & \multicolumn{2}{c}{ } & \multicolumn{2}{c}{ } & \multicolumn{2}{c}{ } \\
CS 31082-001 & $r$-II   &  0.27\pm0.10\;\;(1) &  1.22\pm0.05\;\;(3) & 1.72\pm0.05\;\;(4) & -0.50\pm0.07 & -0.95\pm0.11 \\ 
TYC 5861-1732-1 & not-RPE  & -0.48\pm0.10\;\;(1) & -0.45\pm0.05\;\;(3) & \multicolumn{2}{c}{$<0.29$} & \multicolumn{2}{c}{$>-0.16$} & -0.03\pm0.11 \\ 
CS 22169-035 & limited-$r$ & -0.07\pm0.20\;\;(1) & -1.44\pm0.10\;\;(2) & \multicolumn{2}{c}{$<0.01\;\; (<-0.55\tablenotemark{a})$} & \multicolumn{2}{c}{$>-1.45\;\;(>-0.89)$} & 1.51\pm0.22 \\ 
TYC 75-1185-1   & $r$-I    & -0.28\pm0.07\;\;(2) &  0.0\pm0.05\;\;(3)  & 0.78\pm0.05\;\;(2) & -0.78\pm0.07 & -0.28\pm0.09 \\ 
TYC 5911-452-1  & not-RPE  & -0.23\pm0.07\;\;(2) & -0.68\pm0.10\;\;(1) & \multicolumn{2}{c}{$<0.72\;\; (<-0.21\tablenotemark{a})$} & \multicolumn{2}{c}{$>-1.40\;\;(>-0.89)$} & 0.45\pm0.12 \\ 
TYC 5329-1927-1 & $r$-I\tablenotemark{b}& -0.07\pm0.10\;\;(1) &  0.13\pm0.10\;\;(1) & 0.89\pm0.05\;\;(2) & -0.76\pm0.11 & -0.20\pm0.14 \\ 
TYC 6535-3183-1 & $r$-I\tablenotemark{b}& -0.19\pm0.20\;\;(1) & -0.19\pm0.05\;\;(1) & 0.31\pm0.04\;\;(2) & -0.50\pm0.06 & 0.00\pm0.21   \\ 
TYC 4924-33-1   & not-RPE  & -0.21\pm0.10\;\;(1) & -0.44\pm0.05\;\;(3) & 0.20\pm0.14\;\;(2) & -0.64\pm0.15 &  0.23\pm0.11 \\ 
HE 1116$-$0634  & not-RPE  & -2.06\pm0.07\;\;(2) & -2.03\pm0.20\;\;(1) & \multicolumn{2}{c}{$<0.67\;\; (<-1.14\tablenotemark{a})$} & \multicolumn{2}{c}{$>-2.70\;\;(>-0.89)$} & -0.03\pm0.21 \\ 
TYC 6088-1943-1 & not-RPE  & -0.20\pm0.20\;\;(1) & -0.48\pm0.06\;\;(3) & \multicolumn{2}{c}{$<0.06$} & \multicolumn{2}{c}{$>-0.54$} & 0.28\pm0.21 \\ 
BD$-$13 3442   & limited-$r$?& 0.15\pm0.09\;\;(2) & -0.60\pm0.20\;\;(1) & \multicolumn{2}{c}{$<1.70\;\; (<0.29\tablenotemark{a})$} & \multicolumn{2}{c}{$>-2.30\;\;(>-0.89)$} & 0.75\pm0.22 \\ 
BD$-$01 2582   & CEMP-$s$ &  0.48\pm0.15\;\;(1) &  1.28\pm0.05\;\;(3) & 0.74\pm0.05\;\;(3) &  0.54\pm0.06 & -0.80\pm0.16  \\ 
HE1317$-$0407  & not-RPE   & -0.02\pm0.10\;\;(1) & -0.33\pm0.03\;\;(3) & 0.18\pm0.10\;\;(1) & -0.51\pm0.10 &  0.31\pm0.10  \\ 
HE1320$-$1339  & limited-$r$ &  0.50\pm0.14\;\;(2) & -0.51\pm0.04\;\;(2) & -0.08\pm0.10\;\;(1) & -0.43\pm0.11 & 1.01\pm0.15 \\ 
HD 122563     & limited-$r$ & -0.13\pm0.10\;\;(1) & -0.92\pm0.03\;\;(3) & -0.32\pm0.05\;\;(2) & -0.60\pm0.06 & 0.79\pm0.10 \\ 
TYC 4995-333-1  & not-RPE   & -0.24\pm0.20\;\;(1) & -0.19\pm0.05\;\;(3) & 0.18\pm0.05\;\;(1) & -0.37\pm0.07 & -0.05\pm0.21  \\ 
HE 1523$-$0901  & $r$-II   &  0.57\pm0.20\;\;(1) &  1.27\pm0.05\;\;(1) & 1.82\pm0.05\;\;(1) & -0.55\pm0.07 & -0.70\pm0.21  \\ 
TYC 6900-414-1  & $r$-I\tablenotemark{b}& -0.68\pm0.10\;\;(1) &  0.08\pm0.07\;\;(2) & 0.49\pm0.07\;\;(2) & -0.41\pm0.10 & -0.76\pm0.12  \\ 
J2038$-$0023   & $r$-II   &  0.82\pm0.10\;\;(1) &  0.69\pm0.05\;\;(1) & 1.42\pm0.10\;\;(1) & -0.73\pm0.11 &  0.13\pm0.11  \\ 
BD$-$02 5957   & $r$-I    &  0.45\pm0.20\;\;(1) &  0.40\pm0.04\;\;(3) &
0.91\pm0.06\;\;(2) & -0.51\pm0.07 &  0.05\pm0.20  \\ 
 & & \multicolumn{2}{c}{ } & \multicolumn{2}{c}{ } & \multicolumn{2}{c}{ } & \multicolumn{2}{c}{ } & \multicolumn{2}{c}{ } \\
{\bf Targets} & & \multicolumn{2}{c}{ } & \multicolumn{2}{c}{ } & \multicolumn{2}{c}{ } & \multicolumn{2}{c}{ } & \multicolumn{2}{c}{ } \\
J0007$-$0345 & $r$-I      &  0.41\pm0.20\;\;(1) &  0.11\pm0.07\;\;(2) & 0.73\pm0.04\;\;(3) & -0.62\pm0.08 & 0.41\pm0.22 \\  
J0012$-$1816 & not-RPE    & -0.51\pm0.10\;\;(1) & -0.63\pm0.05\;\;(3) & \multicolumn{2}{c}{$<-0.12$} & \multicolumn{2}{c}{$>-0.51$} &  0.12\pm0.12 \\  
J0022$-$1724 & CEMP-no    & -0.83\pm0.10\;\;(1) & -0.73\pm0.10\;\;(2) & \multicolumn{2}{c}{$<2.12\;\; (<0.16\tablenotemark{a})$} & \multicolumn{2}{c}{$>-2.85\;\;(>-0.89)$} & -0.10\pm0.14  \\  
J0030$-$1007 & limited-$r$   &  0.50\pm0.20\;\;(1) & -0.71\pm0.03\;\;(2) & 0.0\pm0.10\;\;(2)  & -0.71\pm0.10 & 1.21\pm0.20  \\  
J0053$-$0253 & $r$-I      & -0.05\pm0.10\;\;(1) & -0.24\pm0.03\;\;(2) & 0.39\pm0.02\;\;(3) & -0.63\pm0.04 & 0.19\pm0.10  \\  
J0054$-$0611 & $r$-I      &  0.26\pm0.20\;\;(1) & -0.21\pm0.05\;\;(3) & 0.59\pm0.11\;\;(2) & -0.80\pm0.12 & 0.47\pm0.21 \\  
J0107$-$0524 & limited-$r$&  0.14\pm0.10\;\;(1) & -0.61\pm0.06\;\;(3) & \multicolumn{2}{c}{$<0.16$} & \multicolumn{2}{c}{$>-0.77$} & 0.75\pm0.12 \\  
J0145$-$2800 & limited-$r$& -0.02\pm0.20\;\;(1) & -1.05\pm0.06\;\;(2) & \multicolumn{2}{c}{$<0.10\;\; (<-0.16\tablenotemark{a})$} & \multicolumn{2}{c}{$>-1.15\;\;(>-0.89)$} & 1.03\pm0.21  \\  
J0156$-$1402 & $r$-I      &  0.10\pm0.20\;\;(1) & -0.11\pm0.10\;\;(1) & 0.76\pm0.06\;\;(3) & -0.87\pm0.12 & 0.21\pm0.22 \\  
J0213$-$0005 & not-RPE    & -0.54\pm0.06\;\;(2) &  0.05\pm0.07\;\;(2) & \multicolumn{2}{c}{$<0.16$} & \multicolumn{2}{c}{$>-0.11$} & -0.59\pm0.09 \\  
J0227$-$0519 & $r$-I      &  0.72\pm0.10\;\;(1) & -0.18\pm0.10\;\;(1) & 0.42\pm0.06\;\;(3) & -0.60\pm0.12 & 0.90\pm0.14 \\  
J0229$-$1307 & ?          & -0.37\pm0.14\;\;(2) & -0.32\pm0.07\;\;(2) & \multicolumn{2}{c}{$<0.95\;\; (<0.57\tablenotemark{a})$} & \multicolumn{2}{c}{$>-1.27\;\;(>-0.89)$} & -0.05\pm0.16  \\ 
J0236$-$1202 & not-RPE    & -0.41\pm0.10\;\;(1) & -0.29\pm0.08\;\;(3) & <0.30              & >-0.59       & -0.12\pm0.13 \\ 
J0241$-$0427 & $r$-I      &  0.24\pm0.20\;\;(1) & -0.26\pm0.06\;\;(3) & 0.48\pm0.07\;\;(2) & -0.74\pm0.09 & 0.50\pm0.21 \\  
J0242$-$0707 & ?          &  0.37\pm0.13\;\;(2) & -0.08\pm0.10\;\;(1) & \multicolumn{2}{c}{$<1.04$ ($<0.82$\tablenotemark{a})} & \multicolumn{2}{c}{$>-1.12$ ($>-0.89$)} & 0.45\pm0.16 \\  
J0243$-$3249 & not-RPE?   & \multicolumn{2}{c}{$<-0.59$} & -0.95\pm0.09\;\;(4) & \multicolumn{2}{c}{$<0.93$ ($<0.05$\tablenotemark{a})} & \multicolumn{2}{c}{$>-1.88$ ($>-0.89$)} & \multicolumn{2}{c}{$<0.36$} \\  
J0246$-$1518 & $r$-II     &  0.33\pm0.20\;\;(1) &  0.65\pm0.06\;\;(3) & 1.29\pm0.07\;\;(2) & -0.64\pm0.09 & -0.42\pm0.21 \\ 
J0307$-$0534 & $r$-I      &  0.38\pm0.20\;\;(1) &  0.17\pm0.06\;\;(3) & 0.50\pm0.07\;\;(2) & -0.33\pm0.09 & 0.21\pm0.21 \\ 
J0313$-$1020 & $r$-I      & -0.17\pm0.20\;\;(1) & -0.12\pm0.06\;\;(3) & 0.42\pm0.07\;\;(2) & -0.54\pm0.09 & -0.05\pm0.21 \\ 
J0343$-$0924 & $r$-I      & -0.02\pm0.20\;\;(1) & -0.07\pm0.10\;\;(1) & 0.38\pm0.07\;\;(2) & -0.45\pm0.12 &  0.05\pm0.22 \\ 
J0346$-$0730 & not-RPE    &  0.11\pm0.10\;\;(1) & -0.19\pm0.06\;\;(3) & 0.16\pm0.06\;\;(2) & -0.35\pm0.08 & 0.30\pm0.12 \\ 
J0355$-$0637 & limited-$r$&  0.50\pm0.15\;\;(1) & -0.28\pm0.07\;\;(2) & 0.25\pm0.07\;\;(2) & -0.53\pm0.10 &  0.78\pm0.17 \\ 
J0419$-$0517 & $r$-I      &  0.23\pm0.20\;\;(1) &  0.0\pm0.10\;\;(1)  & 0.40\pm0.07\;\;(2) & -0.40\pm0.12 & 0.23\pm0.22 \\ 
J0423$-$1315 & not-RPE    & -0.24\pm0.20\;\;(1) & -0.29\pm0.10\;\;(1) & 0.08\pm0.15\;\;(1) & -0.37\pm0.18 & 0.05\pm0.22 \\ 
J0434$-$2325 & limited-$r$?& -0.42\pm0.07\;\;(2) & -2.27\pm0.11\;\;(2) & \multicolumn{2}{c}{$<-0.53$ ($<-1.38$\tablenotemark{a})} & \multicolumn{2}{c}{$>-1.74$ ($>-0.89$)} & 1.85\pm0.13 \\ 
J0441$-$2303 & ?          & -0.22\pm0.20\;\;(1) & -0.41\pm0.13\;\;(2) & \multicolumn{2}{c}{$<0.55$ ($<0.48$\tablenotemark{a})} & \multicolumn{2}{c}{$>-0.96$ ($>-0.89$)} & 0.19\pm0.24 \\ 
J0453$-$2437 & $r$-I      & -0.21\pm0.10\;\;(1) & -0.04\pm0.07\;\;(3) & 0.59\pm0.05\;\;(3) & -0.63\pm0.09 & -0.17\pm0.12 \\ 
J0456$-$3115 & $r$-I      &  0.02\pm0.20\;\;(2) & -0.33\pm0.10\;\;(1) & 0.34\pm0.10\;\;(1) & -0.67\pm0.14 & 0.35\pm0.22 \\ 
J0505$-$2145 & not-RPE    & -0.22\pm0.20\;\;(1) & -0.32\pm0.07\;\;(2) & 0.15\pm0.08\;\;(2) & -0.47\pm0.11 & 0.10\pm0.21 \\ 
J0517$-$1342 & not-RPE    & -0.43\pm0.11\;\;(2) & -0.43\pm0.06\;\;(3) & 0.21\pm0.07\;\;(2) & -0.64\pm0.09 & 0.0\pm0.13 \\ 
J0525$-$3049 & not-RPE    &  0.40\pm0.15\;\;(1) &  0.02\pm0.07\;\;(2) & 0.12\pm0.20\;\;(1) & -0.10\pm0.21 & 0.38\pm0.17 \\ 
J0610$-$3141 & limited-$r$?& -0.37\pm0.20\;\;(1) & -1.57\pm0.10\;\;(1) & \multicolumn{2}{c}{$<1.30$ ($<-0.68$\tablenotemark{a})} & \multicolumn{2}{c}{$>-2.87$ ($>-0.89$)} & 1.20\pm0.22 \\ 
J0705$-$3343 & $r$-I      &  0.03\pm0.15\;\;(1) & -0.17\pm0.06\;\;(3) & 0.62\pm0.07\;\;(2) & -0.79\pm0.09 & 0.20\pm0.16 \\ 
J0711$-$3432 & $r$-II     & $<0.24$             &  0.50\pm0.06\;\;(3) & 1.30\pm0.10\;\;(1) & -0.80\pm0.12 & $<-0.26$ \\ 
J0910$-$1444 & limited-$r$   & -0.20\pm0.14\;\;(2) & -1.64\pm0.09\;\;(2) & \multicolumn{2}{c}{$<0.03$ ($<-0.78$\tablenotemark{a})} & \multicolumn{2}{c}{$>-1.67$ ($>-0.89$)} & 1.44\pm0.17 \\ 
J0918$-$2311 & $r$-I      & -0.51\pm0.10\;\;(1) & -0.06\pm0.10\;\;(1) & 0.71\pm0.08\;\;(3) & -0.77\pm0.13 & -0.45\pm0.14 \\ 
J0929$-$2905 & not-RPE    & -0.36\pm0.20\;\;(1) & -0.37\pm0.06\;\;(3) & 0.14\pm0.08\;\;(2) & -0.51\pm0.10 & 0.01\pm0.21 \\ 
J0946$-$0626 & $r$-I      &  0.03\pm0.10\;\;(1) & -0.07\pm0.07\;\;(2) & 0.35\pm0.08\;\;(2) & -0.42\pm0.11 &  0.10\pm0.12 \\ 
J0949$-$1617 & CEMP-$r$/$s$\tablenotemark{c} &  0.16\pm0.15\;\;(1) &  0.61\pm0.10\;\;(1) & 0.36\pm0.07\;\;(2) &  0.25\pm0.12 & -0.45\pm0.18 \\ 
J0950$-$2506 & not-RPE    & -0.42\pm0.20\;\;(1) & -0.57\pm0.07\;\;(2) & <0.10              & >-0.67       & 0.15\pm0.21 \\ 
J0952$-$0855 & limited-$r$&  0.00\pm0.20\;\;(1)  & -1.05\pm0.05\;\;(3) & \multicolumn{2}{c}{$<0.24$ ($<-0.16$\tablenotemark{a})} & \multicolumn{2}{c}{$>-1.29$ ($>-0.89$)} & 1.05\pm0.21 \\ 
J0958$-$1446 & $r$-I      &  0.59\pm0.20\;\;(1) &  0.20\pm0.15\;\;(2) & 0.59\pm0.05\;\;(3) & -0.39\pm0.16 & 0.39\pm0.25 \\ 
J1004$-$2706 & $r$-I      &  0.0\pm0.15\;\;(1)  & -0.38\pm0.06\;\;(3) & 0.41\pm0.07\;\;(2) & -0.79\pm0.09 & 0.38\pm0.16 \\ 
J1022$-$3400 & $r$-I      &  0.35\pm0.20\;\;(1) & -0.29\pm0.06\;\;(3) & 0.37\pm0.06\;\;(3) & -0.66\pm0.08 & 0.64\pm0.21 \\ 
J1031$-$0827 & not-RPE    &  0.24\pm0.20\;\;(1) & -0.23\pm0.06\;\;(3) & 0.26\pm0.22\;\;(2) & -0.49\pm0.23 & 0.47\pm0.21 \\ 
J1036$-$1934 & limited-$r$&  0.22\pm0.20\;\;(1) & -0.38\pm0.06\;\;(3) & 0.26\pm0.06\;\;(3) & -0.64\pm0.08 & 0.60\pm0.12 \\ 
J1049$-$1154 & $r$-I      & -0.06\pm0.20\;\;(1) & -0.16\pm0.06\;\;(3) & 0.33\pm0.07\;\;(2) & -0.49\pm0.09 & 0.10\pm0.21 \\ 
J1051$-$2115 & $r$-I      &  0.03\pm0.20\;\;(1) & -0.27\pm0.07\;\;(2) & 0.32\pm0.07\;\;(2) & -0.59\pm0.10 & 0.30\pm0.21 \\ 
J1059$-$2052 & $r$-I      &  0.26\pm0.07\;\;(2) & -0.07\pm0.06\;\;(3) & 0.35\pm0.06\;\;(3) & -0.42\pm0.08 & 0.33\pm0.09 \\ 
J1120$-$2406 & not-RPE    & -0.16\pm0.20\;\;(1) & -0.17\pm0.06\;\;(3) & \multicolumn{2}{c}{$<0.16$} & \multicolumn{2}{c}{$>-0.33$} & 0.01\pm0.21 \\ 
J1124$-$2155 & not-RPE    &  0.20\pm0.10\;\;(1) & -0.17\pm0.06\;\;(3) & 0.22\pm0.07\;\;(2) & -0.39\pm0.09 & 0.37\pm0.12 \\ 
J1130$-$1449 & $r$-I      &  0.08\pm0.07\;\;(2) & -0.12\pm0.06\;\;(3) & 0.50\pm0.07\;\;(1) & -0.62\pm0.09 & 0.20\pm0.09 \\ 
J1139$-$0558 & not-RPE    & -0.10\pm0.20\;\;(1) & -0.30\pm0.06\;\;(3) & 0.29\pm0.07\;\;(2) & -0.59\pm0.09 & 0.20\pm0.21 \\ 
J1144$-$0409 & $r$-I      & -0.01\pm0.10\;\;(1) & -0.26\pm0.07\;\;(2) & 0.58\pm0.06\;\;(3) & -0.84\pm0.09 & 0.25\pm0.12 \\ 
2MJ1144$-$1128 & $r$-I    &  0.03\pm0.07\;\;(2) & -0.29\pm0.06\;\;(3) & 0.35\pm0.07\;\;(2) & -0.64\pm0.09 &  0.32\pm0.09 \\ 
J1146$-$0422 & CEMP-$r$   & -0.28\pm0.25\;\;(1) &  0.32\pm0.10\;\;(1) & 0.62\pm0.06\;\;(3) & -0.30\pm0.12 & -0.60\pm0.27 \\ 
J1147$-$0521 & $r$-I      &  0.0\pm0.20\;\;(1)  & -0.22\pm0.06\;\;(3) & 0.31\pm0.06\;\;(3) & -0.53\pm0.08 & 0.22\pm0.21 \\ 
J1158$-$1522 & limited-$r$& -0.37\pm0.20\;\;(1) & -1.07\pm0.14\;\;(2) & \multicolumn{2}{c}{$<0.15$ ($<-0.18$\tablenotemark{a})} & \multicolumn{2}{c}{$>-1.22$ ($>-0.89$)} & 0.70\pm0.24 \\ 
J1204$-$0759 & $r$-I      & -0.29\pm0.10\;\;(1) & -0.11\pm0.06\;\;(3) & 0.33\pm0.20\;\;(1) & -0.44\pm0.21 & -0.18\pm0.12 \\ 
2MJ1209$-$1415 & $r$-I    & -0.01\pm0.20\;\;(1) &  0.11\pm0.13\;\;(2) & 0.81\pm0.06\;\;(3) & -0.70\pm0.14 & -0.12\pm0.21 \\ 
J1218$-$1610 & limited-$r$& -0.20\pm0.11\;\;(2) & -1.50\pm0.20\;\;(1) & \multicolumn{2}{c}{$<0.17$ ($<-0.61$\tablenotemark{a})} & \multicolumn{2}{c}{$>-1.67$ ($>-0.89$)} & 1.30\pm0.23 \\
J1229$-$0442 & $r$-I      &  0.0\pm0.20\;\;(1)  & -0.22\pm0.06\;\;(3) & 0.46\pm0.04\;\;(4) & -0.68\pm0.07 & 0.22\pm0.21 \\ 
J1237$-$0949 & not-RPE    &  0.22\pm0.20\;\;(1) & -0.27\pm0.07\;\;(2) & 0.19\pm0.06\;\;(3) & -0.46\pm0.09 & 0.49\pm0.21 \\ 
J1250$-$0307 & $r$-I      & -0.57\pm0.14\;\;(2) &  0.10\pm0.06\;\;(3) & 0.45\pm0.12\;\;(2) & -0.35\pm0.13 & -0.67\pm0.15 \\ 
J1256$-$0834 & $r$-I      &  0.32\pm0.15\;\;(1) & -0.28\pm0.07\;\;(2) & 0.45\pm0.06\;\;(3) & -0.73\pm0.09 & 0.60\pm0.17 \\ 
J1302$-$0843 & $r$/$s$\tablenotemark{c} & $<0.73$             &  0.55\pm0.07\;\;(1) & 0.41\pm0.07\;\;(2) &  0.14\pm0.09 & <0.18 \\ 
J1306$-$0947 & not-RPE    & -0.21\pm0.11\;\;(2) & -0.12\pm0.04\;\;(3) & 0.12\pm0.07\;\;(3) & -0.24\pm0.08 & -0.09\pm0.12 \\ 
2MJ1307$-$0931 & not-RPE  &  0.02\pm0.20\;\;(1) & -0.38\pm0.05\;\;(3) & 0.10\pm0.06\;\;(3) & -0.48\pm0.08 & 0.40\pm0.21 \\ 
J1321$-$1138 & not-RPE    & -0.03\pm0.15\;\;(1) & -0.36\pm0.06\;\;(3) & 0.08\pm0.07\;\;(2) & -0.44\pm0.09 & 0.33\pm0.16 \\ 
2MJ1325$-$1747 & $r$-I    & -0.02\pm0.20\;\;(1) & -0.44\pm0.07\;\;(2) & 0.40\pm0.06\;\;(3) & -0.84\pm0.09 & 0.42\pm0.21 \\ 
J1326$-$1525 & limited-$r$& -0.10\pm0.07\;\;(2) & -0.67\pm0.06\;\;(3) & -0.28\pm0.10\;\;(2) & -0.39\pm0.12 & 0.57\pm0.09 \\ 
J1328$-$1731 & not-RPE    & -0.02\pm0.20\;\;(1) & -0.08\pm0.06\;\;(3) & 0.20\pm0.11\;\;(1) & -0.28\pm0.13 & 0.06\pm0.21 \\ 
J1333$-$2623 & limited-$r$&  0.11\pm0.12\;\;(2) & -0.55\pm0.06\;\;(3) & 0.20\pm0.08\;\;(3) & -0.75\pm0.10 & 0.66\pm0.13 \\ 
J1335$-$0110 & $r$-I      & -0.39\pm0.20\;\;(1) & -0.22\pm0.05\;\;(3) & 0.53\pm0.07\;\;(2) & -0.75\pm0.09 & -0.17\pm0.21 \\ 
J1337$-$0826 & $r$-I      &  0.17\pm0.20\;\;(1) &  0.02\pm0.02\;\;(3) & 0.93\pm0.11\;\;(2) & -0.91\pm0.11 & 0.15\pm0.20 \\ 
J1339$-$1257 & not-RPE    &  0.08\pm0.20\;\;(1) & -0.42\pm0.06\;\;(3) & 0.10\pm0.20\;\;(1) & -0.52\pm0.21 & 0.27\pm0.21 \\  
2MJ1340$-$0016 & not-RPE  &  0.05\pm0.20\;\;(1) & -0.30\pm0.06\;\;(3) & 0.29\pm0.11\;\;(2) & -0.59\pm0.13 & 0.35\pm0.21 \\ 
J1342$-$0717 & $r$-I      &  0.04\pm0.20\;\;(1) & -0.26\pm0.06\;\;(3) & 0.44\pm0.06\;\;(3) & -0.70\pm0.08 & 0.30\pm0.21 \\ 
2MJ1343$-$2358 & CEMP-no  & -0.37\pm0.20\;\;(2) & -0.77\pm0.07\;\;(2) & \multicolumn{2}{c}{$<0.15$ ($<0.12$\tablenotemark{a})} & \multicolumn{2}{c}{$>-0.92$ ($>-0.89$)} & 0.40\pm0.21 \\ 
J1403$-$3214 & not-RPE    & -0.60\pm0.20\;\;(1) & -0.08\pm0.06\;\;(2) & 0.12\pm0.10\;\;(1) & -0.20\pm0.12 & -0.52\pm0.21 \\ 
2MJ1404$+$0011 & CEMP-$r$ &  0.43\pm0.20\;\;(1) &  0.38\pm0.07\;\;(2) & 0.58\pm0.06\;\;(3) & -0.28\pm0.09 & 0.05\pm0.21 \\ 
J1410$-$0343 & $r$-I      & -0.15\pm0.14\;\;(2) & -0.12\pm0.06\;\;(3) & 0.67\pm0.07\;\;(2) & -0.79\pm0.09 & -0.03\pm0.15 \\ 
J1416$-$2422 & not-RPE    &  0.02\pm0.20\;\;(1) & -0.31\pm0.06\;\;(3) & 0.14\pm0.10\;\;(1) & -0.45\pm0.12 & 0.33\pm0.21 \\ 
J1418$-$2842 & $r$-I      & -0.41\pm0.20\;\;(1) & -0.11\pm0.06\;\;(3) & 0.43\pm0.12\;\;(2) & -0.54\pm0.13 & -0.30\pm0.21 \\ 
J1419$-$0844 & $r$-I      &  0.34\pm0.20\;\;(1) & -0.15\pm0.06\;\;(3) & 0.34\pm0.06\;\;(3) & -0.49\pm0.08 & 0.49\pm0.21 \\ 
J1500$-$0613 & $r$-I      &  0.12\pm0.09\;\;(2) & -0.10\pm0.06\;\;(3) & 0.39\pm0.06\;\;(3) & -0.49\pm0.08 & 0.32\pm0.11 \\ 
J1502$-$0528 & not-RPE    &  0.02\pm0.09\;\;(2) &  0.00\pm0.06\;\;(3) & 0.24\pm0.06\;\;(3) & -0.24\pm0.08 & 0.02\pm0.11 \\ 
J1507$-$0659 & $r$-I      &  0.12\pm0.07\;\;(2) & -0.10\pm0.06\;\;(3) & 0.36\pm0.06\;\;(3) & -0.46\pm0.08 & 0.22\pm0.09 \\ 
J1508$-$1459 & $r$-I      &  0.0\pm0.10\;\;(1)  & -0.10\pm0.06\;\;(3) & 0.49\pm0.07\;\;(3) & -0.59\pm0.09 & 0.10\pm0.12 \\ 
J1511$+$0025 & $r$-I      &  0.02\pm0.20\;\;(1) & -0.18\pm0.06\;\;(3) & 0.41\pm0.06\;\;(3) & -0.59\pm0.08 & 0.20\pm0.21 \\ 
J1516$-$2122 & CEMP-no    & -0.03\pm0.20\;\;(1) & -0.48\pm0.06\;\;(3) & 0.09\pm0.07\;\;(2) & -0.59\pm0.09 &  0.45\pm0.09 \\ 
2MJ1521$-$0607 & $r$-I    & -0.18\pm0.20\;\;(1) &  0.10\pm0.07\;\;(2) & 0.93\pm0.07\;\;(2) & -0.83\pm0.10 & -0.28\pm0.21 \\ 
J1527$-$2336 & ?          & -0.18\pm0.07\;\;(1) & -0.11\pm0.07\;\;(2) & <0.74             & >-0.85       & -0.07\pm0.10 \\ 
J1534$-$0857 & limited-$r$& -0.33\pm0.07\;\;(2) & -1.22\pm0.05\;\;(3) & \multicolumn{2}{c}{$<-0.13$ ($<-0.33$\tablenotemark{a})} & \multicolumn{2}{c}{$>-1.09$ ($>-0.89$)} & 0.89\pm0.09 \\ 
J1538$-$1804 & $r$-II     &  0.44\pm0.20\;\;(1) &  0.62\pm0.07\;\;(2) & 1.27\pm0.05\;\;(5) & -0.65\pm0.09 & -0.18\pm0.21 \\ 
J1542$-$0131 & not-RPE    &  0.02\pm0.20\;\;(1) & -0.35\pm0.06\;\;(3) & 0.26\pm0.07\;\;(2) & -0.61\pm0.09 & 0.37\pm0.21 \\ 
J1547$-$0837 & limited-$r$&  0.78\pm0.20\;\;(1) & -0.50\pm0.06\;\;(3) & -0.10\pm0.14\;\;(2) & -0.40\pm0.15 & 1.28\pm0.21 \\ 
J1554$+$0021 & not-RPE    &  0.19\pm0.20\;\;(1) & -0.26\pm0.06\;\;(3) & -0.09\pm0.07\;\;(2) & -0.17\pm0.09 & 0.45\pm0.21 \\ 
J1602$-$1521 & not-RPE    &  0.10\pm0.07\;\;(2) &  0.09\pm0.06\;\;(3) & 0.25\pm0.06\;\;(3) & -0.16\pm0.08 & 0.01\pm0.09 \\ 
J1606$-$0400 & not-RPE    & -0.02\pm0.20\;\;(1) & -0.17\pm0.07\;\;(2) & 0.23\pm0.09\;\;(2) & -0.40\pm0.11 & 0.15\pm0.21 \\ 
J1606$-$1632 & limited-$r$&  0.01\pm0.20\;\;(1) & -0.57\pm0.07\;\;(2) & -0.27\pm0.10\;\;(1) & -0.30\pm0.12 & 0.58\pm0.21 \\ 
J1609$-$0941 & $r$-I      & -0.06\pm0.15\;\;(1) & -0.30\pm0.05\;\;(3) & 0.41\pm0.06\;\;(3) & -0.71\pm0.08 & 0.24\pm0.16 \\ 
J1612$-$0541 & not-RPE    &  0.07\pm0.20\;\;(1) &  0.03\pm0.06\;\;(3) & 0.20\pm0.07\;\;(2) & -0.17\pm0.09 & 0.00\pm0.21 \\ 
J1612$-$0848 & $r$-I      &  0.29\pm0.20\;\;(1) &  0.04\pm0.06\;\;(3) & 0.58\pm0.05\;\;(4) & -0.54\pm0.08 & 0.25\pm0.21 \\ 
J1616$-$0401 & $r$-I      &  0.08\pm0.14\;\;(2) & -0.19\pm0.07\;\;(3) & 0.52\pm0.06\;\;(3) & -0.71\pm0.09 & 0.27\pm0.16 \\ 
J1618$-$0630 & not-RPE?   &  0.01\pm0.20\;\;(1) & -0.59\pm0.10\;\;(1) & <-0.27 & >-0.32 & 0.58\pm0.22 \\ 
J1627$-$0848 & not-RPE    &  0.00\pm0.20\;\;(1) &  0.10\pm0.06\;\;(3) & 0.12\pm0.20\;\;(1) & -0.02\pm0.21 & -0.10\pm0.21 \\ 
J1628$-$1014 & $r$-I      & -0.26\pm0.10\;\;(1) & -0.02\pm0.06\;\;(3) & 0.36\pm0.06\;\;(3) & -0.38\pm0.08 & -0.24\pm0.12 \\ 
J1639$-$0522 & limited-$r$&  0.36\pm0.20\;\;(1) & -0.26\pm0.06\;\;(3) & -0.07\pm0.20\;\;(1) & -0.19\pm0.21 & 0.62\pm0.21 \\ 
J1645$-$0429 & limited-$r$&  0.38\pm0.30\;\;(1) & -0.37\pm0.06\;\;(3) & -0.15\pm0.10\;\;(1) & -0.22\pm0.12 & 0.75\pm0.31 \\ 
J1811$-$2126 & not-RPE    & -0.09\pm0.20\;\;(1) &  0.18\pm0.10\;\;(1) & 0.28\pm0.10\;\;(1) & -0.10\pm0.14 & -0.27\pm0.22 \\ 
J1905$-$1949 & $r$-I      & -0.01\pm0.20\;\;(1) & -0.08\pm0.03\;\;(3) & 0.36\pm0.04\;\;(3) & -0.44\pm0.05 &  0.07\pm0.20 \\ 
J2005$-$3057 & $r$-I      & -0.16\pm0.20\;\;(1) &  0.36\pm0.07\;\;(2) & 0.86\pm0.07\;\;(2) & -0.50\pm0.10 & -0.52\pm0.22 \\ 
J2010$-$0826 & $r$-I      &  0.04\pm0.14\;\;(2) & -0.39\pm0.04\;\;(3) & 0.42\pm0.07\;\;(3) & -0.81\pm0.08 &  0.43\pm0.15 \\ 
J2032$+$0000 & not-RPE    &  0.16\pm0.20\;\;(1) & -0.29\pm0.07\;\;(2) & 0.26\pm0.06\;\;(3) & -0.55\pm0.10 & 0.45\pm0.21 \\ 
J2036$-$0714 & CEMP-$r$   &  0.02\pm0.20\;\;(1) & -0.57\pm0.10\;\;(1) & 0.48\pm0.10\;\;(1) & -0.87\pm0.12 & 0.59\pm0.22 \\ 
J2038$-$0252 & $r$-I      &  0.39\pm0.10\;\;(1) & -0.26\pm0.10\;\;(1) & 0.59\pm0.06\;\;(3) & -0.85\pm0.12 & 0.65\pm0.22 \\ 
J2054$-$0033 & CEMP-no/lim-$r$ & 0.63\pm0.14\;\;(2) & -0.27\pm0.06\;\;(3) & <-0.18        & >-0.14       &  0.90\pm0.15 \\ 
J2058$-$0354 & $r$-I      & -0.24\pm0.07\;\;(2) & -0.09\pm0.06\;\;(3) & 0.36\pm0.06\;\;(3) & -0.45\pm0.08 & -0.15\pm0.09 \\ 
J2116$-$0213 & $r$-I      & -0.41\pm0.20\;\;(1) & -0.31\pm0.10\;\;(1) & 0.60\pm0.07\;\;(2) & -0.91\pm0.12 & -0.10\pm0.22 \\ 
J2151$-$0543 & not-RPE    & -0.41\pm0.10\;\;(1) & -0.54\pm0.06\;\;(3) & 0.22\pm0.07\;\;(2) & -0.76\pm0.09 & 0.13\pm0.12 \\ 
2MJ2256$-$0719 & $r$-II   &  0.08\pm0.20\;\;(1) &  0.26\pm0.04\;\;(3) & 1.10\pm0.07\;\;(2) & -0.84\pm0.08 & -0.18\pm0.20 \\ 
J2256$-$0500 & not-RPE    & -0.10\pm0.20\;\;(1) & -0.46\pm0.06\;\;(3) & -0.06\pm0.07\;\;(2) & -0.40\pm0.09 & 0.36\pm0.21 \\ 
J2304$+$0155 & not-RPE    &  0.01\pm0.20\;\;(1) & -0.20\pm0.07\;\;(2) & 0.26\pm0.07\;\;(2) & -0.45\pm0.10 & 0.21\pm0.21 \\ 
J2325$-$0815 & $r$-I      & -0.42\pm0.20\;\;(1) & -0.33\pm0.07\;\;(2) & 0.55\pm0.07\;\;(2) & -0.88\pm0.10 & -0.09\pm0.10 \\ 
\enddata
\raggedright
\tablenotetext{a}{This Eu upper limit can be lowered by assuming $[\rm{Ba/Eu}] > -0.89$, as required by the Solar $r$-process residual \citep{Burris2000}.}
\tablenotetext{b}{\citet{Ruchti2011} did not determine abundances of
  neutron-capture elements, and therefore did not detect the
  $r$-process enhancement in these stars.}
\tablenotetext{c}{The $r$/$s$ designation is based on the criteria
  from \citet{BeersChristlieb2005}, though note that this category may
  also contain stars with signatures of an intermediate, or $i$-,
  process (e.g., \citealt{CowanRose1977,Hampel2016}).}
\end{deluxetable}

\clearpage
\subsubsection{Other Neutron-Capture Abundances}\label{subsec:NeutronCapture}
Abundances of other neutron-capture elements are given in Table
\ref{table:Abunds}.  Abundances of Y, La, Ce, and Nd are available for
most of the stars, while Zr, Pr, Sm, Dy, and Os are only available in
the stars with high S/N, higher [Fe/H], and/or high $r$-process
enhancement.  Th is heavily blended, and was only detectable in a
handful of stars.  Abundances of all these elements were determined
with spectrum syntheses.


\begin{splitdeluxetable}{lccccccccBcccccccccBcccccccccc}
\tabletypesize{\scriptsize}
\tablecaption{Elemental Abundances\tablenotemark{a}\label{table:Abunds}}
\tablehead{
Star     & [O/Fe] & [Na/Fe]    & [Mg/Fe] & [Si/Fe] & [K/Fe]  & [Ca/Fe] & [Sc/Fe] & [Ti I/Fe] & [Ti II/Fe] & 
 [V/Fe]  & [Cr II/Fe] & [Mn/Fe] & [Co/Fe] & [Ni/Fe] & [Cu/Fe] & [Zn/Fe] & [Y/Fe]    & [Zr/Fe] &
 [La/Fe] & [Ce/Fe]    & [Pr/Fe] & [Nd/Fe] & [Sm/Fe] & [Dy/Fe] & [Os/Fe]   & [Th/Fe]
}
\decimals
\startdata
CS 31082-001     & --- & ---               & $0.46\pm0.04$ (4) &  ---              & $0.17\pm0.10$ (1) & $0.44\pm0.01$ (23) & $-0.03\pm0.04$ (5) & $0.20\pm0.01$ (14) & $0.38\pm0.01$ (32) & 
   ---            & $0.37\pm0.10$ (1) & ---               & $0.03\pm0.10$ (1) & $0.03\pm0.05$ (5) & ---                & $0.15\pm0.10$ (1)  & $0.45\pm0.05$ (6)  & $0.62\pm0.10$ (1) & 
$1.23\pm0.05$ (3) & $1.04\pm0.05$ (6) & $1.24\pm0.06$ (4) & $1.29\pm0.05$ (8) & $1.42\pm0.05$ (1) & $1.22\pm0.10$ (1)  & $1.65\pm0.07$ (2)  & ---  \\  
T5861-1732-1     & ---  & $0.50\pm0.05$ (2) & $0.42\pm0.04$ (3) & $0.52\pm0.10$ (1) & $0.34\pm0.10$ (1) & $0.31\pm0.01$ (24) & $-0.15\pm0.02$ (10)& $0.07\pm0.01$ (17) & $0.21\pm0.01$ (29) & 
   ---            &$-0.01\pm0.05$ (4) & ---               &$-0.20\pm0.11$ (2) &$-0.11\pm0.01$ (12)& ---                & $0.03\pm0.13$ (2)  &$-0.41\pm0.08$ (2)  & --- & 
   ---            &  ---              &  ---              & ---               & ---               & ---                &  ---               & ---  \\  
CS 22169-035    & ---   &  ---              & $0.32\pm0.03$ (2) & ---               & $0.31\pm0.10$ (1) & $0.18\pm0.01$ (12) & $-0.25\pm0.03$ (5) & $-0.12\pm0.01$ (6) & $-0.17\pm0.02$ (18) & 
   ---            &  ---              &$-0.27\pm0.10$ (1) & $0.22\pm0.06$ (2) & $0.10\pm0.03$ (5) & ---                & ---                & $-0.39\pm0.05$ (2) & --- &
   ---            &  ---              &  ---              & ---               & ---               & ---                &  ---               & ---  \\  
T75-1185-1      & ---   &  ---              & $0.30\pm0.09$ (2) & ---               & $0.30\pm0.10$ (1) & $0.35\pm0.01$ (16) & $-0.10\pm0.04$ (5) & $0.27\pm0.02$ (15) & $0.27\pm0.01$ (30) & 
   ---            & $0.22\pm0.10$ (1) &$-0.29\pm0.02$ (2) &$-0.04\pm0.02$ (3) & $0.17\pm0.02$ (5) & ---                & $0.10\pm0.10$ (1)  & $-0.20\pm0.07$ (3) & --- & 
   ---            &  ---              &  ---              & ---               & ---               & ---                &  ---               & ---  \\  
T5911-452-1     & ---   &  ---              & $0.32\pm0.03$ (2) & ---               & $0.45\pm0.10$ (1) & $0.26\pm0.02$ (14) & $0.04\pm0.02$ (3)  & $0.39\pm0.04$ (5)  & $0.44\pm0.02$ (15) & 
   ---            &  ---              &  ---              & ---               & $0.09\pm0.10$ (1) & ---                & ---                &  ---               &  --- & 
   ---            &  ---              &  ---              & ---               & ---               & ---                &  ---               & ---  \\  
\enddata
\tablenotetext{a}{Table \ref{table:Abunds} is published in
its entirety in the electronic edition of \textit{The Astrophysical Journal}. A portion is shown here for
guidance regarding its form and content.}
\end{splitdeluxetable}


\subsubsection{The $\alpha$-Elements and K}\label{subsec:Alpha}
In most of the stars there are many clear \ion{Ca}{1}, \ion{Ti}{1},
and \ion{Ti}{2} lines; the Ca and Ti abundances were therefore determined
differentially with respect to a standard, similar to \ion{Fe}{1} and
\ion{Fe}{2}.  Note that the Ti lines follow similar trends as the Fe
lines when NLTE corrections are not applied, i.e., the \ion{Ti}{1}
lines yield lower Ti abundances than the \ion{Ti}{2} abundances.
Because the [\ion{Ti}{1}/H] ratios are likely to be too low, the
average differential offsets in [\ion{Ti}{1}/H] and [\ion{Ti}{2}/H]
are both applied relative to the [\ion{Ti}{2}/H] ratios in the
standard stars.

The other elements were not determined differentially. The \ion{Mg}{1}
lines at 4057, 4167, 4703, 5528, and 5711 \AA $\;$ are generally
detectable, though at the metal-rich end some become prohibitively
strong.  The \ion{Si}{1} lines are generally very weak in metal-poor
stars, and are occasionally difficult to detect even in high S/N
spectra.  The \ion{K}{1} line at 7699 \AA $\;$ lies at the edge of a
series of telluric absorption lines; when the K line is distinct from the
telluric features a measurement is provided.  In a handful of stars,
the O abundance can be determined from the 6300 and 6363 \AA $\;$
forbidden lines.

\subsubsection{Iron-peak Elements, Cu, and Zn}\label{subsec:FePeak}
Abundances of \ion{Sc}{2}, \ion{V}{1}, \ion{Cr}{2}, \ion{Mn}{1},
\ion{Co}{1}, and \ion{Ni}{1} were all determined from EWs,
considering HFS when necessary.  Each species has a multitude of
available lines.  Note that \ion{Cr}{1} lines are not included, as
they are expected to suffer from NLTE effects
\citep{BergemannCescutti2010}.   The Mn lines in these metal-poor
stars may require NLTE corrections $\sim0.5-0.7$ dex
\citep{BergemannGehren2008}, but they have not been applied here.

Cu and Zn were determined via spectrum syntheses, using the 5105 and
5782 \AA $\;$ \ion{Cu}{1} lines and the 4722 and 4810 \AA $\;$
\ion{Zn}{1} lines.  Note that the \ion{Cu}{1} lines are likely to
suffer from NLTE issues (e.g., \citealt{Shi2018}); these corrections
are also not applied here.

\subsubsection{Light Elements: Li and Na}\label{subsec:Light}
In some stars, Na abundances can be determined from the \ion{Na}{1}
doublet at 5682/5688~\AA.  In the most metal-poor stars, the
\ion{Na}{1} doublet at 5889 and 5895 \AA $\;$ is weak enough for an
abundance determination, but is only used if the interstellar
contamination is either insignificant or is sufficiently offset from
the stellar lines.  Note that the NaD lines may suffer from NLTE
effects (e.g., \citealt{Andrievsky2007}), but the 5682/5688
\AA \hspace{0.025in} lines are not likely to have significant NLTE
corrections in this metallicity range \citep{Lind2011}.

The \ion{Li}{1} line at 6707 \AA \hspace{0.025in} is detectable in
nine stars, as listed in Table \ref{table:LiAbunds}.  These Li 
abundances are typical for the evolutionary state of the stars; the
main sequence stars have values that are consistent with the Spite
plateau, while the giants show signs of Li depletion.  Two $r$-II,
three $r$-I, and one limited-$r$ stars have Li detections.

\begin{deluxetable}{@{}lcc}
\tabletypesize{\scriptsize}
\tablecolumns{3}
\tablewidth{0pt}
\tablecaption{Stars with Li Measurements\label{table:LiAbunds}}
\hspace*{-4in}
\tablehead{
Star       & $\log \epsilon$ (Li) & $T_{\rm{eff}}$ (K)}
\startdata
J0107$-$0524   & $1.20\pm0.05$ & 5225 \\ 
2MJ0213$-$0005 & $2.42\pm0.05$ & 6225 \\
J0517$-$1342   & $0.91\pm0.10$ & 4961 \\
J0705$-$3343   & $0.81\pm0.10$ & 4757 \\
J0711$-$3432   & $0.94\pm0.10$ & 4767 \\
J1022$-$3400   & $0.79\pm0.05$ & 4831 \\
J1333$-$2623   & $0.94\pm0.05$ & 4821 \\
J1527$-$2336   & $2.46\pm0.10$ & 6260 \\
J1538$-$1804   & $0.81\pm0.05$ & 4752 \\
J2058$-$0354   & $0.89\pm0.05$ & 4831 \\
\enddata
\end{deluxetable}


\clearpage
\section{Discussion}\label{sec:Discussion}

\subsection{The $r$-process-Enhanced Stars}\label{subsec:RPEstars}
\begin{figure*}
\begin{center}
\centering
\subfigure[]{\includegraphics[scale=0.53,trim=0.0in 0.1in 0.5in 0.45in,clip]{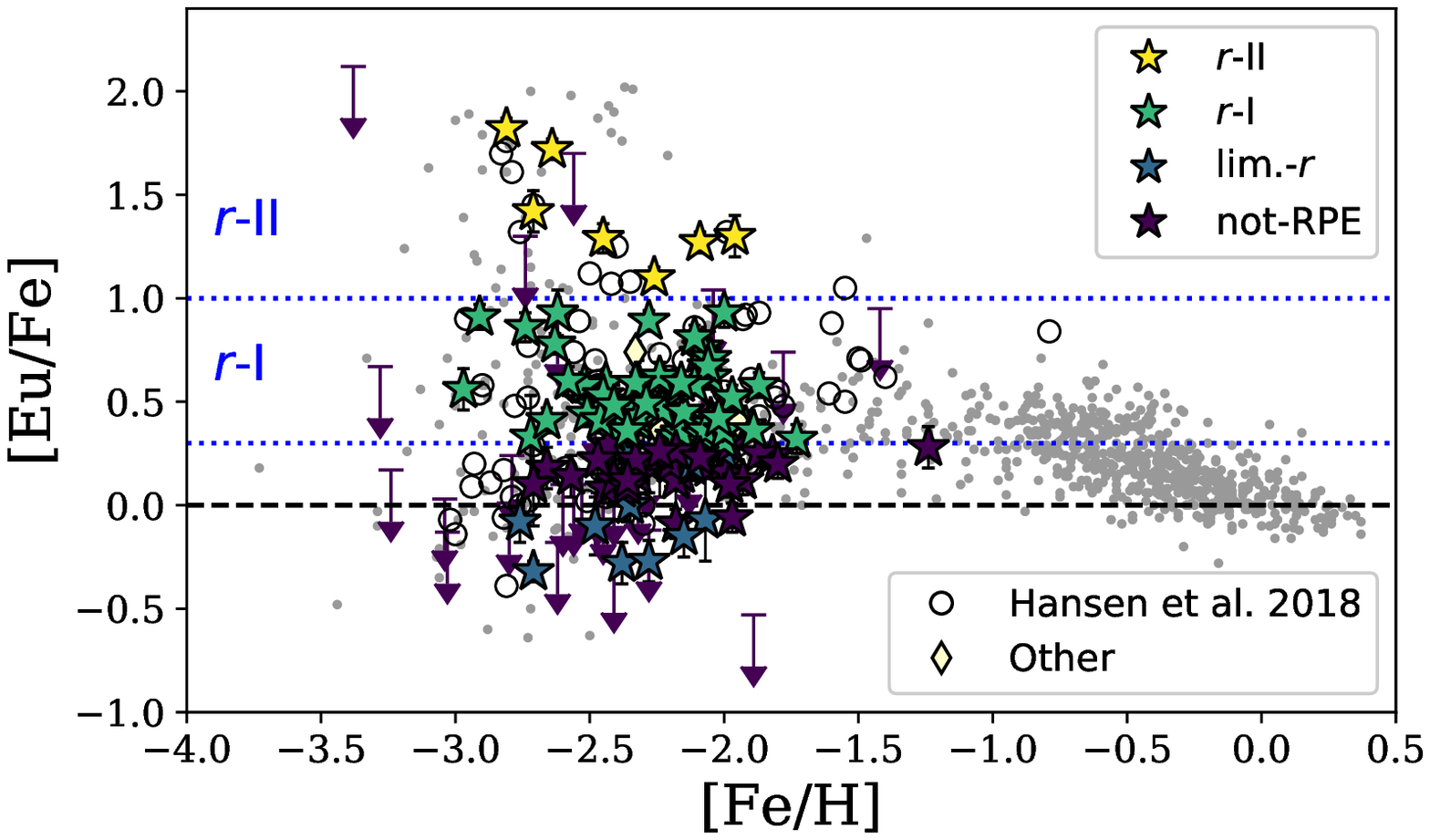}\label{fig:EuFe}}
\subfigure[]{\includegraphics[scale=0.53,trim=0.0in 0.1in 0.5in 0.45in,clip]{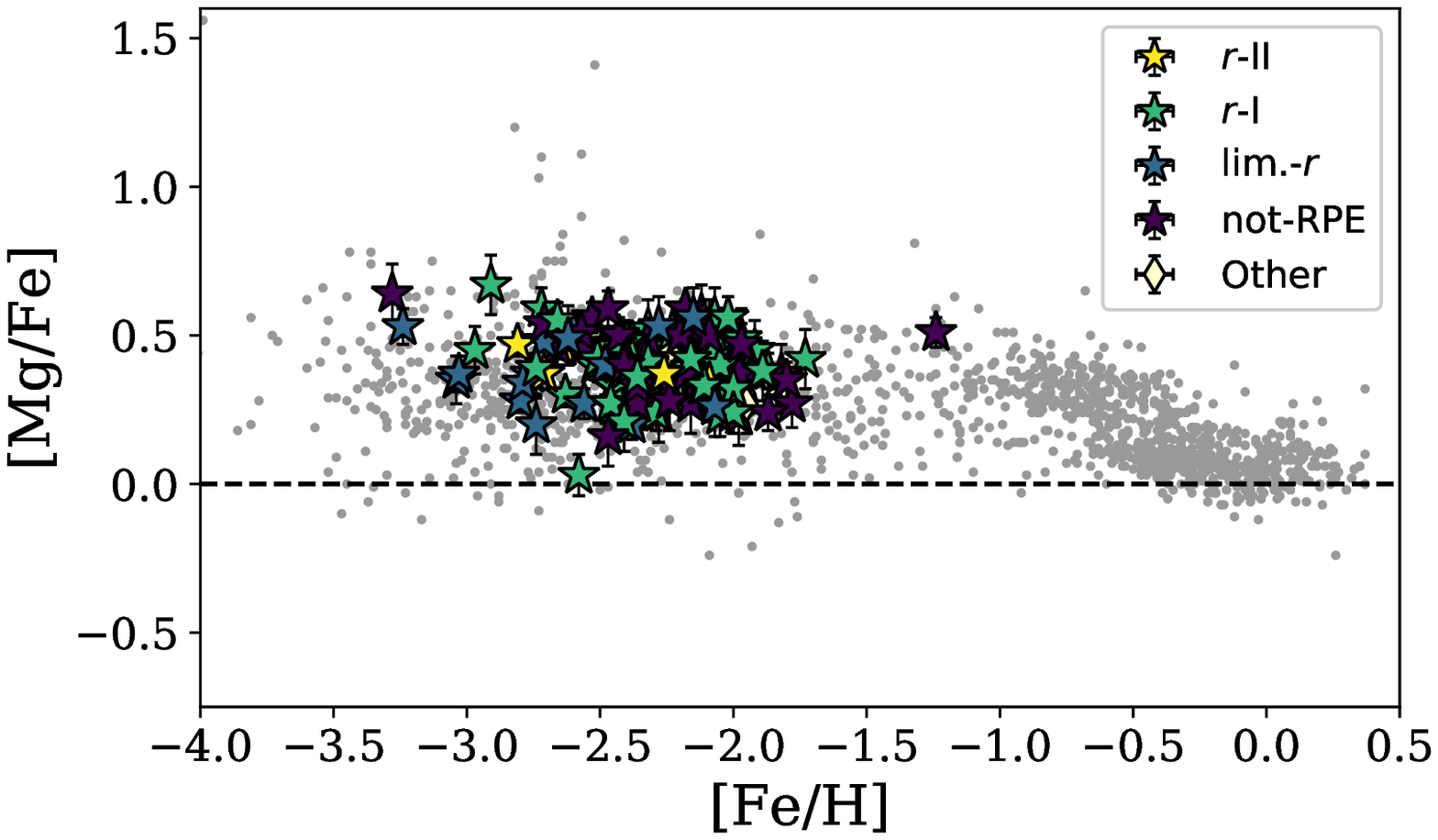}\label{fig:MgFe}}
\subfigure[]{\includegraphics[scale=0.53,trim=0.0in 0.1in 0.5in 0.45in,clip]{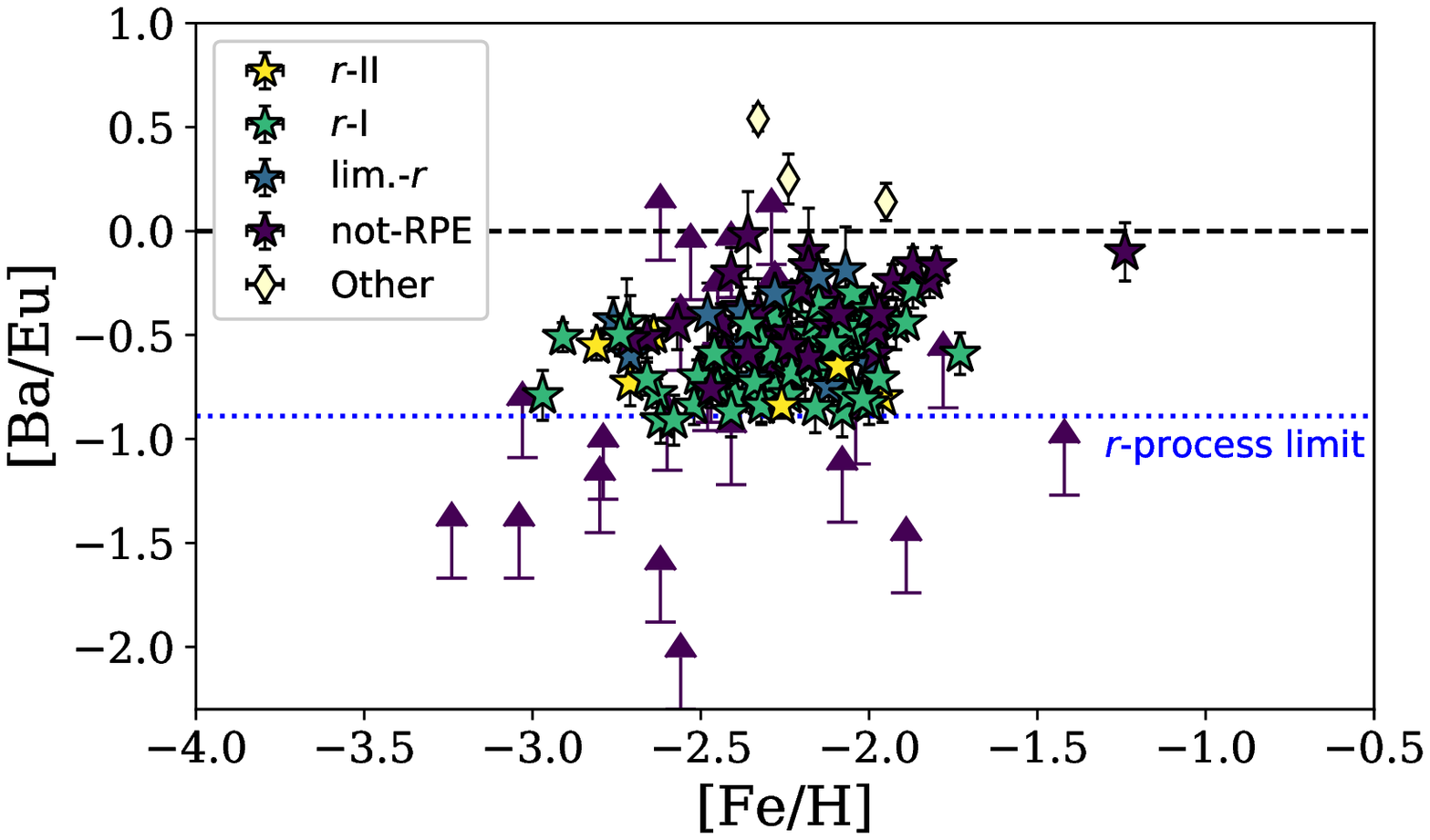}\label{fig:BaEuFe}}
\subfigure[]{\includegraphics[scale=0.53,trim=0.0in 0.1in 0.5in 0.45in,clip]{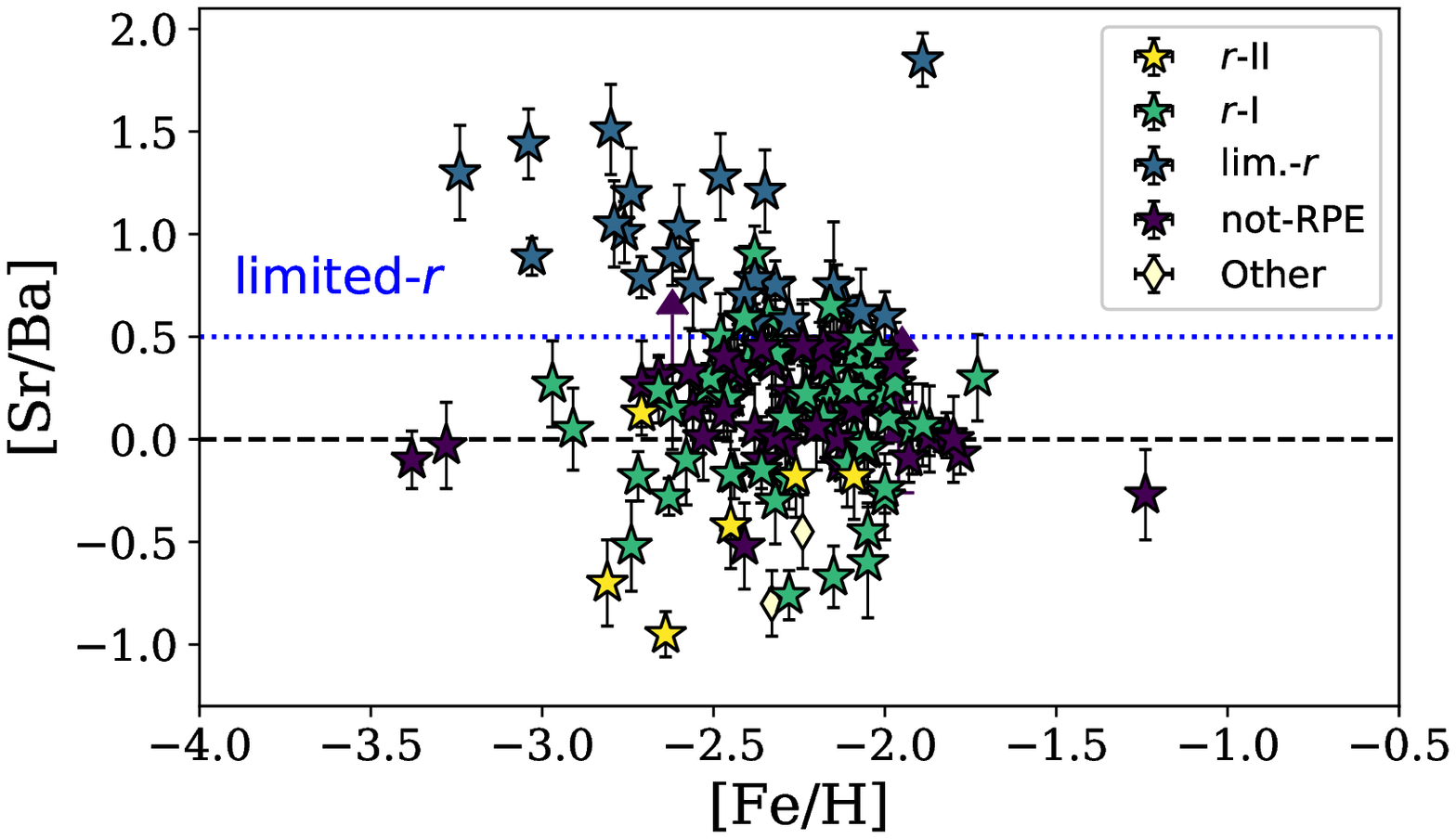}\label{fig:SrBaFe}}
\caption{Ratios of [Eu/Fe], [Mg/Fe], [Ba/Eu], and [Sr/Ba] as a
  function of [Fe/H].  The standard and target stars are grouped by
  $r$-process enhancement.  ``Other'' includes stars with $s$ and
  $r/s$ classifications. For reference, the \citet{Hansen2018} stars
  are shown as open circles, while MW field stars from
\citet{Venn2004} and \citet{Reddy2006} are shown as gray dots.  In the
top left plot, the $r$-I and $r$-II limits in [Eu/Fe] are shown with a dotted
line.  The Solar $r$-process [Ba/Eu] ratio from \citet{Burris2000} is
indicated in the bottom left plot.  Finally, in the bottom right
panel, the limit for the limited-$r$ stars, [Sr/Ba]$>+0.5$, is shown.}\label{fig:XFe}
\end{center}
\end{figure*}

Figure \ref{fig:XFe} shows [Eu/Fe], [Mg/Fe], [Ba/Eu], and [Sr/Ba] as a
function of [Fe/H], grouped by their $r$-process enhancement.  This
northern survey has discovered 4 new $r$-II stars, including
J1538$-$1804 (published by \citealt{Sakari2018}), 60 new $r$-I stars
(three of them CEMP-$r$), and 19 new limited-$r$ stars.  Combined with
the results from \citet{Hansen2018}, \citet{Placco2017},
\citet{Gull2018}, \citet{Holmbeck2018}, \citet{Cain2018},
  and \citet{Roederer2018b}, the RPA has so far identified, in total,
  18 new $r$-II, 101 new $r$-I (including 6 CEMP-$r$), 39 limited-$r$,
  and 1 $r+s$ star.  The properties of the stars from this paper are
discussed below.

\subsubsection{The Sub-populations of $r$-process-Enhanced Stars}\label{subsubsec:RPEnumbers}
\begin{figure}[h!]
\begin{center}
\centering
\subfigure[]{\includegraphics[scale=0.55,trim=0.3in 0in 0.45in 0.3in,clip]{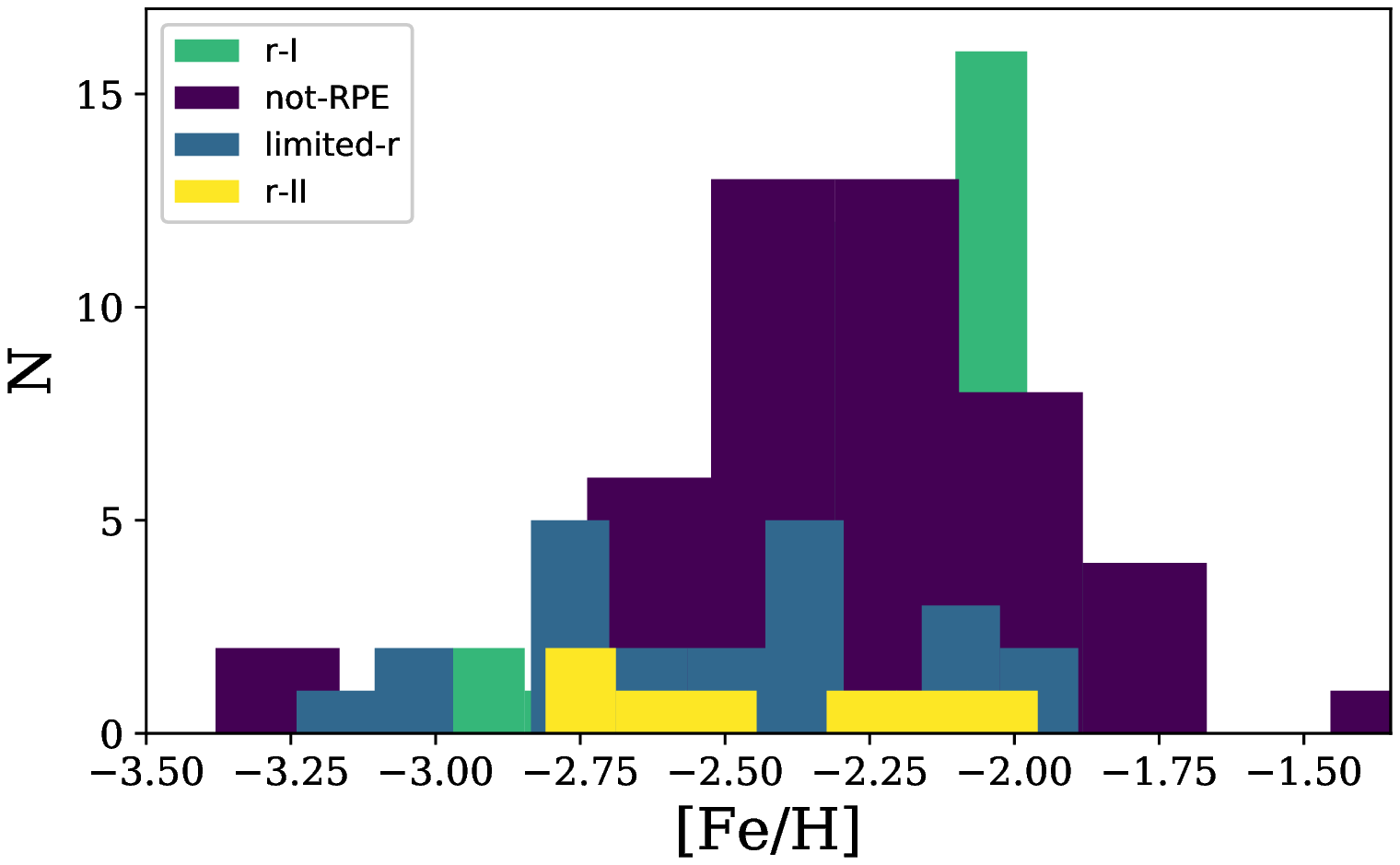}\label{fig:FeHHist}}
\subfigure[]{\includegraphics[scale=0.55,trim=0.3in 0in 0.6in 0.3in,clip]{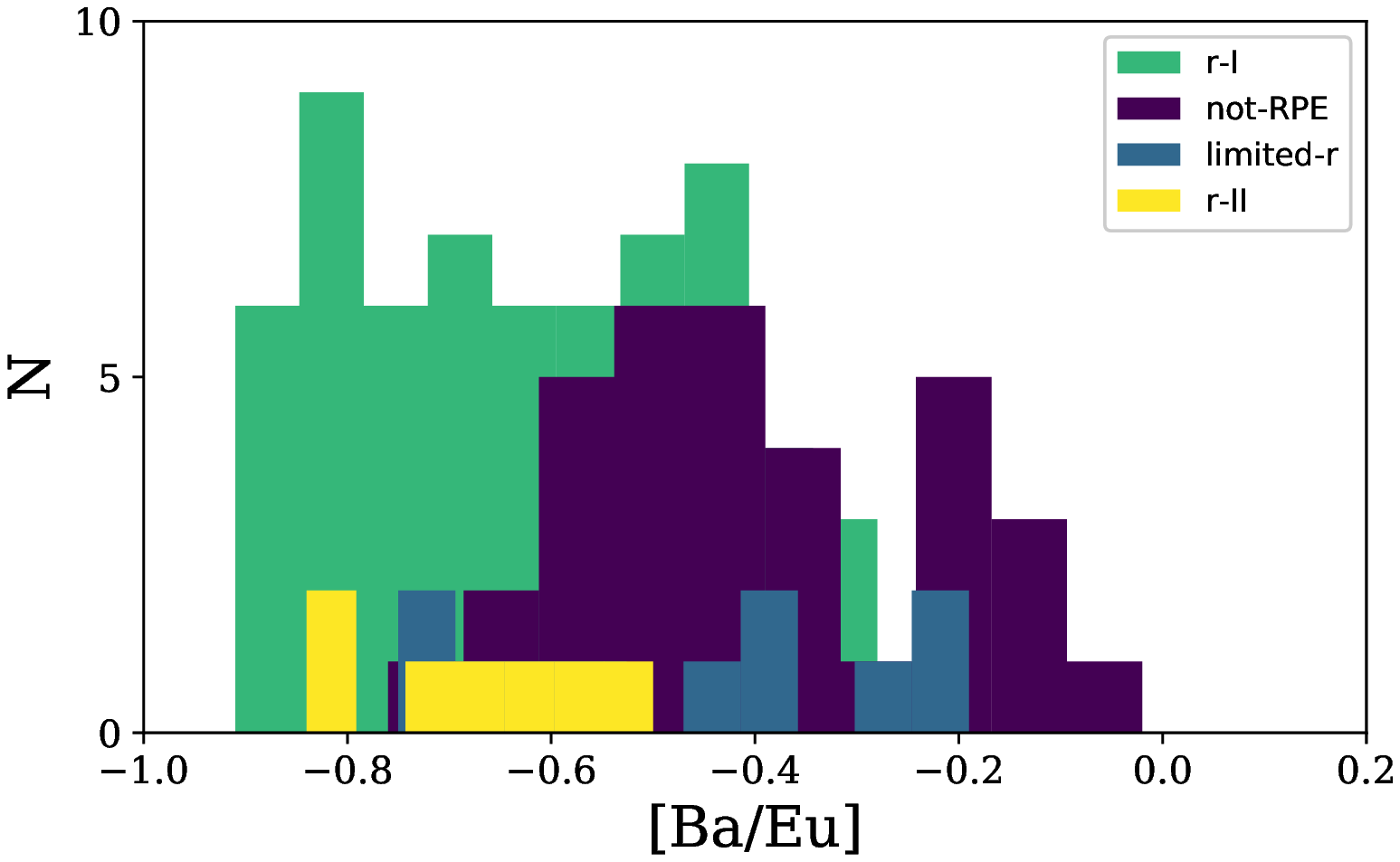}\label{fig:BaEuHist}}
\caption{Histograms showing the [Fe/H] (left) and [Ba/Eu] (right) distributions for the different groups of stars.  The stars with $s$- or $r$/$s$-process signatures have been removed.\label{fig:Hists}}
\end{center}
\end{figure}

The metallicity distribution of the different $r$-process
sub-populations is very similar to that found in \citet{Hansen2018}, as
shown in Figure \ref{fig:FeHHist}. The $r$-I and $r$-II stars are
found across the full [Fe/H] range; there is a hint that the
limited-$r$ stars are only found at lower metallicities, but more
stars are necessary to validate this.

Figure \ref{fig:BaEuHist} shows the distribution of [Ba/Eu]
values. The $r$-II stars and many of the $r$-I stars have low [Ba/Eu],
consistent with little enrichment from the main $s$-process.  The
not-RPE and limited-$r$ stars seem to extend to higher [Ba/Eu],
indicating some amount of $s$-process contamination.  Figure
\ref{fig:SrBaFe} also demonstrates that the $r$-II stars have 
low [Sr/Ba].  As in the Hansen et al. sample, some $r$-I stars are
found to have enhanced [Sr/Ba] and [Sr/Eu] ratios, similar to the
stars in the limited-$r$ class.

Note that the large spread in [Eu/Fe] at a given metallicity is not
accompanied by a similar spread in [Mg/Fe] (see Figure \ref{fig:XFe}),
which has been noted by many other authors.  With one exception, all
the target stars have light, $\alpha$, and Fe-peak abundances that are
consistent with normal MW halo stars, regardless of $r$-process
enhancement. This places important constraints on the nucleosynthetic
signature and site of the $r$-process.  For instance, the robust Mg
abundances rule out traditional core-collapse supernovae as the only
source of the heavy $r$-process elements (also see
\citealt{MaciasRamirezRuiz2016}).

\subsubsection{Kinematics}\label{subsubsec:Toomre}
\begin{figure}[h!]
\begin{center}
\centering
\includegraphics[scale=0.75,trim=0.7in 0.4in 0.8in 0.5in,clip]{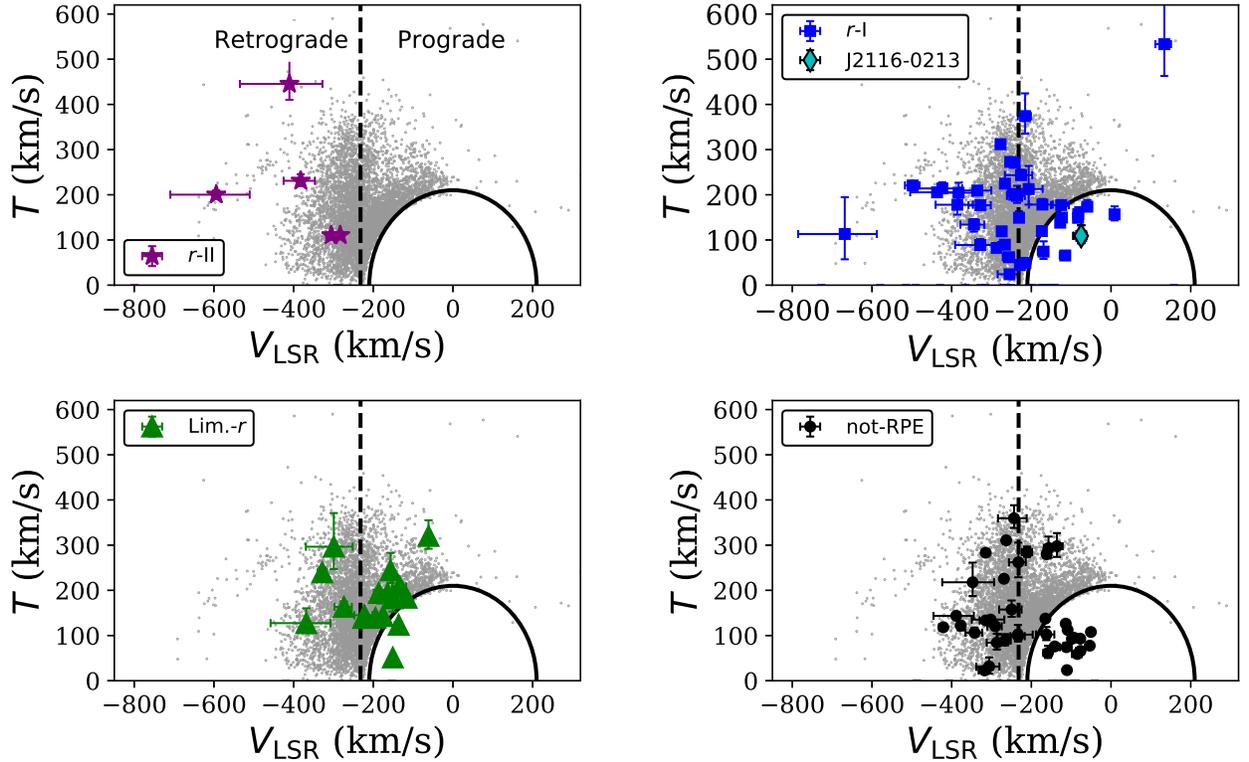}
\caption{Toomre diagrams for the four main sub-populations in this
  paper, where $T = \sqrt{U^2+W^2}$.  This plot helps distinguish halo
  stars from disk stars and retrograde halo stars from prograde halo
  stars.  The gray points show MW disk and halo stars within 1 kpc
  from \citet{Koppelman2018}---the large circle shows their criterion for
  halo membership; disk stars lie within the circle.  The colored points
  use {\it Gaia} DR2 data; when radial velocities were not available, the
  values from this paper were used.  Only stars with parallax
  uncertainties $<20$\% are shown (see \citealt{BailerJones2018}). The
  upper left panel shows the $r$-II stars, the upper right panel the
  $r$-I stars, the lower left panel the limited-$r$ stars, and the
  lower right panel the ``not-RPE'' stars.
\label{fig:Toomre}}
\end{center}
\end{figure}

All of these stars are {\it Gaia} DR2 targets; all but one have proper
motions and parallaxes, though the parallax errors are occasionally
too large to provide reliable distances \citep{BailerJones2018}.
Figure \ref{fig:Toomre} shows a Toomre diagram for stars with parallax
errors $<20$\%, generated with the \texttt{gal\_uvw}
code.\footnote{\url{https://github.com/segasai/astrolibpy/blob/master/astrolib/gal_uvw.py}}
This diagram distinguishes between disk and halo stars, and between
retrograde and prograde halo stars.  The errors in Figure
\ref{fig:Toomre} reflect the uncertainties in the parallax and proper
motion.  The velocities have been corrected for the solar motion,
according to the values from \citet{Coskunoglu2011}.

In Figure \ref{fig:Toomre} the stars are grouped by their
$r$-process-enhancement classification, and are compared with
kinematically-selected MW halo stars from \citet{Koppelman2018}.
Several of the non-RPE stars are consistent with membership in the
metal-weak thick disk \citep{Kordopatis2013}. The majority of the
$r$-process-enhanced stars are consistent with membership in the halo,
and a large number are retrograde halo stars. All of the $r$-II stars
and more than half of the $r$-I stars in this paper are
retrograde, possibly indicating they originated in a
satellite. The kinematics of three of the $r$-II stars
  from \citet{Hansen2018} are presented in \citet{Roederer2018}; only
  those three pass the stringent cut in parallax error, but note that
  two of these stars are prograde halo stars.  The kinematics of
$r$-process-enhanced stars will have important consequences for the
birth sites of these stars.  Full orbital calculations will be even
more useful \citep{Roederer2018}.

\subsubsection{Detailed $r$-process Patterns}\label{subsubsec:Patterns}
\begin{figure}[h!]
\begin{center}
\centering
\subfigure[$r$-II stars]{\includegraphics[scale=0.43,trim=0.35in 0.55in 0.9in 1.0in,clip]{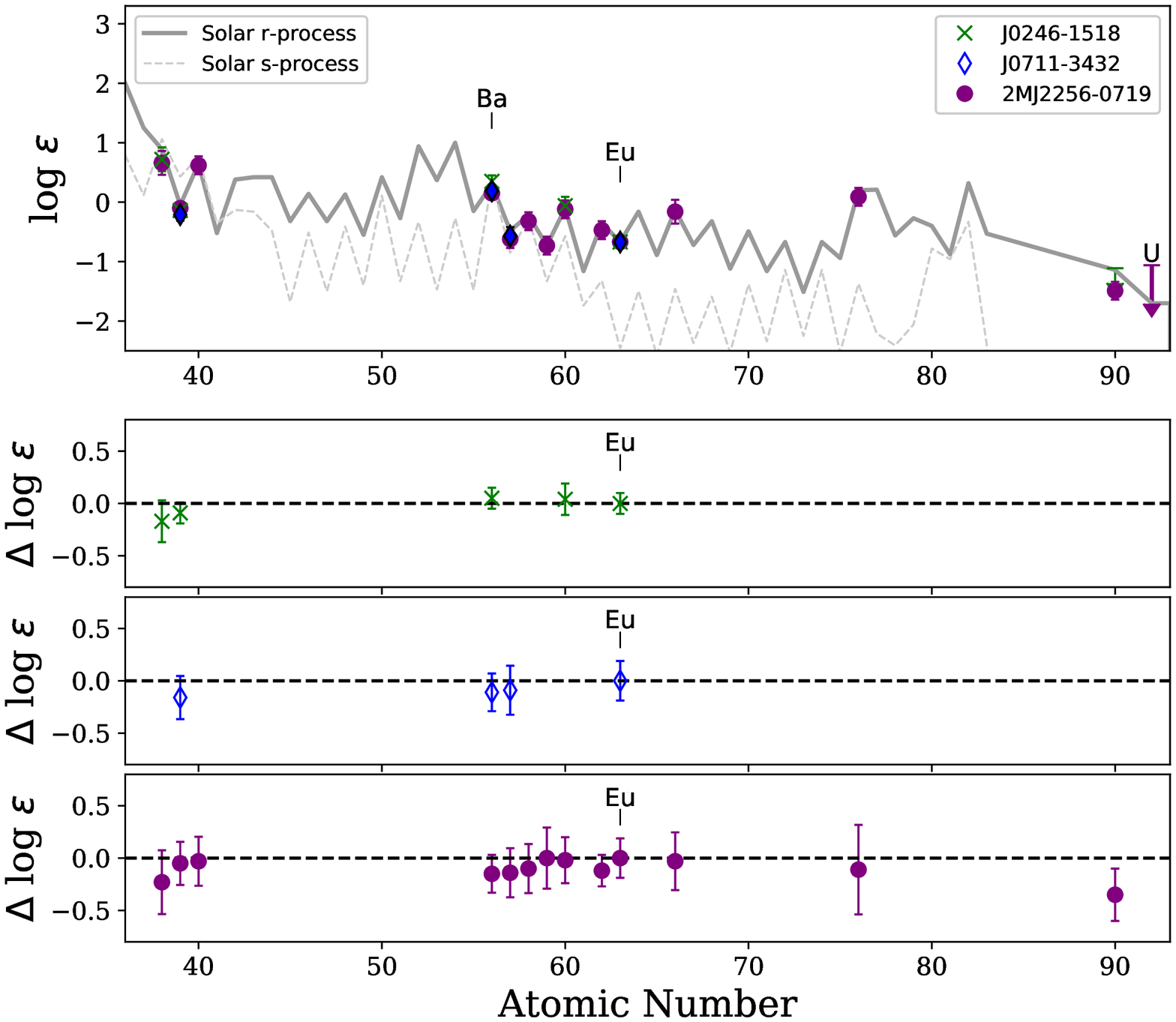}\label{fig:rIIPatterns}}
\subfigure[$r$-I stars]{\includegraphics[scale=0.43,trim=0.25in 0.55in 0.9in 1.0in,clip]{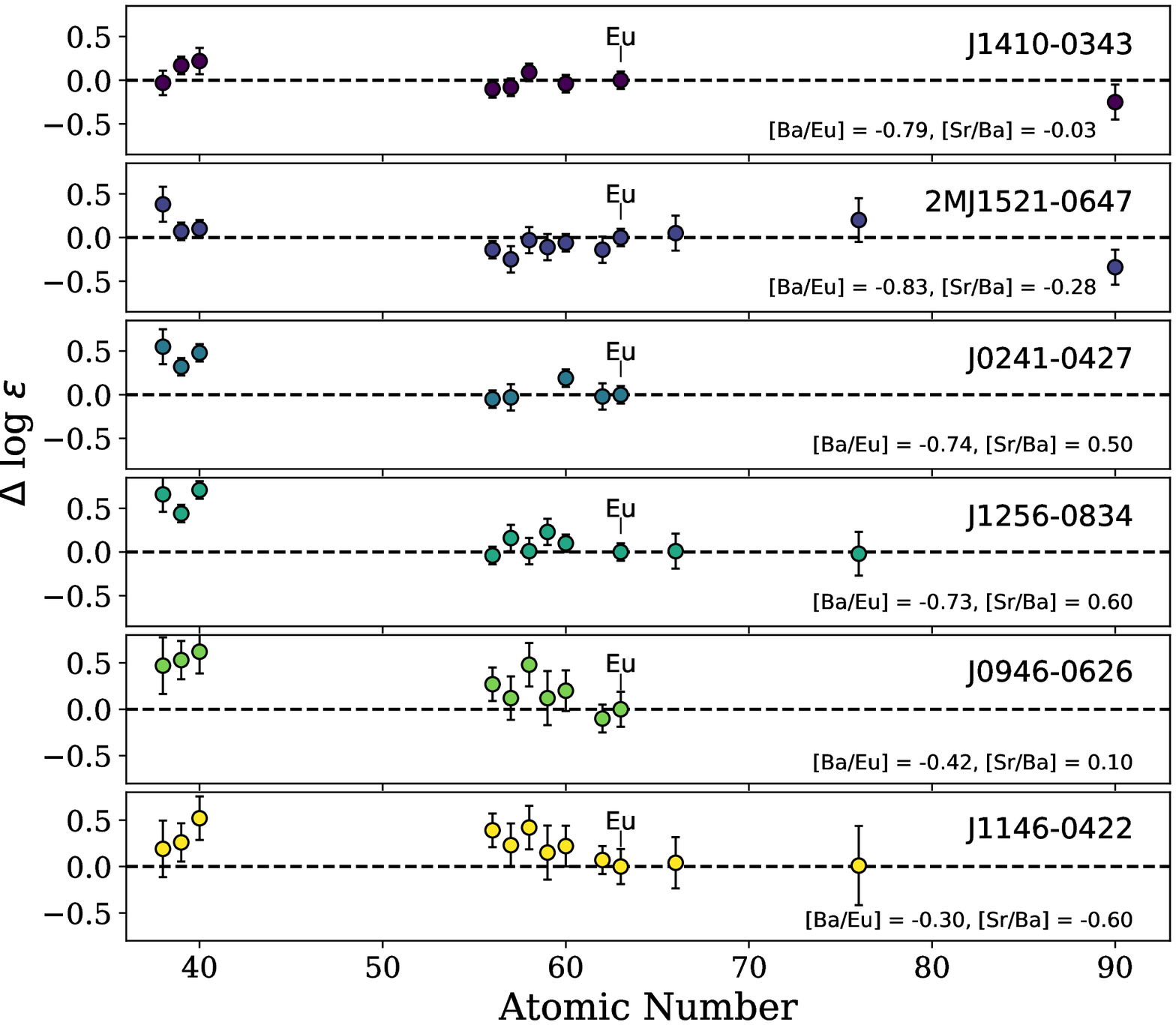}\label{fig:rIPatterns}}
\caption{Detailed $r$-process patterns for new $r$-II (left) and
  $r$-I (right) stars compared to the Solar $r$-process residual from
  \citet{Arlandini1999}.  The points have been shifted to common
  Eu abundances.  The $r$-I stars are grouped by [Ba/Eu] and [Sr/Ba],
  demonstrating pure $r$-process enrichment in the top two panels,
  limited-$r$ enhancement in Sr, Y, and Zr in the middle two panels,
  and $s$-process enhancement in the bottom two panels.
\label{fig:Patterns}}
\end{center}
\end{figure}

Figure \ref{fig:rIIPatterns} shows the detailed $r$-process patterns and
residuals with respect to the scaled-Solar $r$-process pattern in three
$r$-II stars (the pattern for J1538$-$1804 was presented in
\citealt{Sakari2018}).  As has been found in numerous other studies,
the abundance patterns are consistent with the scaled-Solar $r$-process
pattern (but see below for Th).  Figure \ref{fig:rIPatterns} shows
patterns for six of the $r$-I stars.  The top two panels show $r$-I
stars with low [Ba/Eu] and [Sr/Ba]; as expected, their abundances are
consistent with a pure $r$-process pattern.  The next two panels show
$r$-I stars with low [Ba/Eu], but elevated [Sr/Ba].  These stars have
elevated Sr, Y, and Zr compared to the scaled-Solar pattern, but the
pattern of the lanthanides is consistent.  Finally, the last two
panels show $r$-I stars with slightly sub-Solar [Ba/Eu], indicating
some $s$-process contamination.  These stars have high Sr, Y, Zr, Ba,
La, and Ce, relative to the Solar pattern.

These detailed patterns support the classifications from the more
general [Ba/Eu] and [Sr/Ba] ratios (e.g.,
\citealt{Frebel2018,Spite2018}), and will be useful in identifying the
nucleosynthetic signatures of the limited-$r$ and $r$-processes.
Follow-up of the limited-$r$ and $r$-I stars with enhanced [Sr/Ba]
will enable detailed comparisons between abundance patterns and model
predictions, particularly in the 38~$\leq Z \leq$~47 range, which
could distinguish between limited-$r$ and weak $s$-process scenarios
(e.g.,
\citealt{Chiappini2011,Frischknecht2012,Cescutti2013,Frischknecht2016}).

\subsubsection{Cosmochronometric Ages}\label{subsubsec:Ages}
The few $r$-I and $r$-II stars with Th detections enable
determinations of 1) cosmo-chronometric ages and 2) the possible
presence of an actinide boost.  Table \ref{table:Ages} shows the Th
abundances relative to Eu and ages derived from Equation 1 in
\citet{Placco2017}, using two different sets of production ratios:
the \citet{Schatz2002} values, from waiting-point calculations, and
the \citet{Hill2017} values, from a high-entropy wind.  Although the
errors in age are quite large (due to high uncertainties in the Th
abundance), all of the stars have Th/Eu ratios that are consistent
with ancient $r$-process production; none appear to exhibit an
actinide boost.  Several of the ages are quite old, comparable to the
results found for Reticulum II \citep{JiFrebel2018}.  These old ages
are consistent with recent results from simulations, which suggest
that many of the most metal-poor MW halo stars should be ancient
\citep{Starkenburg2017,ElBadry2018}.  These ages will be greatly
improved through higher precision Th abundances and U detections,
which require observations at higher resolution and higher S/N.

\startlongtable
\begin{deluxetable}{@{}lDDDD}
\tabletypesize{\scriptsize}
\tablecolumns{10}
\tablewidth{0pt}
\tablecaption{Th/Eu Abundance Ratios and Ages\label{table:Ages}}
\hspace*{-4in}
\tablehead{
           & \multicolumn{2}{c}{} & \multicolumn{2}{c}{} & \multicolumn{4}{c}{Age (Gyr)} \\
Star       & \multicolumn{2}{c}{[Fe/H]} & \multicolumn{2}{r}{$\log \epsilon$ (Th/Eu)} & \multicolumn{2}{c}{\citet{Schatz2002}} & \multicolumn{2}{c}{\citet{Hill2017}}}
\decimals
\startdata
J0053$-$0253   & -2.16\pm0.01 & -0.61\pm0.11 & 13.1\pm5.1 & 17.3\pm5.1 \\
J0246$-$1518   & -2.45\pm0.03 & <-0.70       & >17.3      & >21.5 \\ 
J0313$-$1020   & -2.05\pm0.02 & -0.58\pm0.21 & 11.2\pm9.8 & 15.9\pm9.8 \\
J0343$-$0924   & -1.92\pm0.01 & -0.58\pm0.17 & 11.2\pm7.9 & 15.9\pm7.9 \\
J1410$-$0343   & -2.06\pm0.02 & -0.65\pm0.21 & 14.9\pm9.8 & 19.1\pm9.8 \\
2MJ1521$-$0607 & -2.00\pm0.01 & -0.75\pm0.17 & 19.6\pm7.9 & 23.8\pm7.9 \\
2MJ2256$-$0719 & -2.26\pm0.01 & -0.76\pm0.20 & 20.1\pm9.3 & 24.3\pm9.3 \\
\enddata
\end{deluxetable}

\subsection{J2116$-$0213: A Globular Cluster Star?}\label{subsec:GC}

\begin{figure*}[h!]
\begin{center}
\includegraphics[scale=0.6,clip,trim=0.5in 0 0 0]{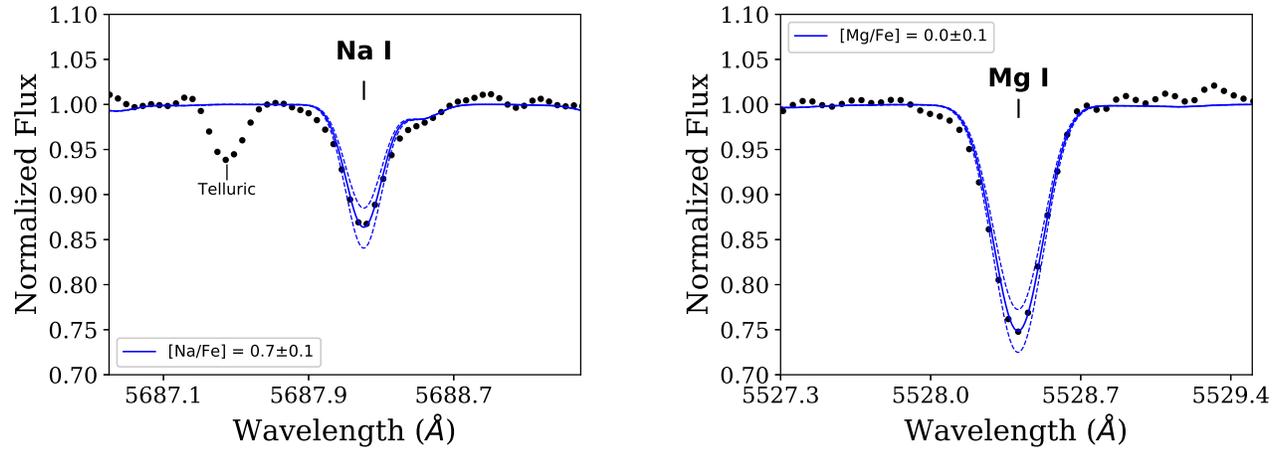}
\caption{Syntheses to the 5688 \AA \hspace{0.025in} \ion{Na}{1} and 5528
  \AA \hspace{0.025in} \ion{Mg}{1} lines in J2116$-$0213.  Uncertainties of $\pm0.1$ dex are shown.\label{fig:NaMgSynths}}
\end{center}
\end{figure*}

One of the $r$-I stars in this sample, J2116$-$0213, has elevated sodium
($[\rm{Na/Fe}] = +0.68\pm0.07$) and has low magnesium
($[\rm{Mg/Fe}]~=~+~0.03\pm0.05$; see Figure \ref{fig:NaMgSynths})
coupled with normal Si, Ca, and Ti.  The Al lines at 6696 and
6698 \AA $\;$ are too weak for a robust [Al/Fe] measurement.
These abundances are not like typical halo stars; instead, this
abundance pattern is a signature of multiple populations in globular
clusters (GCs; e.g., \citealt{Carretta2009}).  This suggests that
J2116$-$0213 may have originated in a GC and was later ejected into the
Milky Way halo. Escaped GC stars have been identified from their
unique abundance signatures in the MW halo \citep{Martell2016} and
bulge \citep{Schiavon2017}.  J2116$-$0213 is an $r$-I star with
$[\rm{Eu/Fe}]~\sim~+~0.6$---this is consistent with other metal-poor
GCs, which contain large numbers of $r$-I stars \citep{Gratton2004}.
However, J2116$-$0213 is more metal-poor ($[\rm{Fe/H}] \sim -2.6$) than
the intact MW GCs.  Note that this star's location in the Toomre
diagram is right between the thick halo/halo classification; a more
detailed orbit for this star could potentially identify its birth
environment more clearly.

\clearpage
\section{Conclusions}\label{sec:Conclusion}
This paper has presented high-resolution spectroscopic observations of
126 new metal-poor stars and 20 previously observed standards, as part
of the $R$-Process Alliance (also see \citealt{Hansen2018}).
Atmospheric parameters and metallicities were derived differentially
with respect to a set of standards, applying $<$3D$>$ NLTE
corrections.  Abundances of a wide variety of elements were then
determined.  Sr, Ba, and Eu were used to classify the stars according
to their $r$-process enhancement, using [Eu/Fe] as the indicator of
the main $r$-process, [Ba/Eu] as the indicator for the amount of
main $s$-process contamination, and [Sr/Ba] as the indicator for the
amount of limited-$r$ (or weak-$s$) contamination.  Proper motions and
parallaxes from {\it Gaia} DR2 enabled the 3D kinematics of these
stars to be probed.

Out of the 126 metal-poor targets, four were discovered to be
highly Eu-enhanced $r$-II stars.  All four are found to have
$r$-process patterns that are consistent with the scaled Solar
$r$-process residual, and all show no signs of significant
contributions from the limited-$r$ or $s$-processes.  In other words,
the $r$-II stars have retained a pure main $r$-process signature, even
though they span a large range in metallicity.  All the $r$-II stars
in this paper have retrograde halo orbits.  The 60 new $r$-I stars
show more variation; some exhibit a limited-$r$ signature and some
have contributions from the $s$-process, but many have low [Ba/Eu] and
[Sr/Ba] ratios consistent with a pure $r$-process signal.  As with the
$r$-II stars, the $r$-I stars span a wide range in [Fe/H].  The
majority of the $r$-I stars are likely halo stars, many of them with
retrograde orbits.  The smaller number of limited-$r$ stars prohibits
making firm conclusions about them as a stellar population, but the 19
in this paper are restricted to lower metallicities.

A number of interesting individual stars were identified in this
survey, most of which are being targeted for follow-up observations at
higher spectral resolution.  Nine CEMP stars were discovered:
three are $r$-I stars, four are CEMP-no, and two are CEMP-$r/s$.
Another star was found to have an $r/s$ signature, but its corrected C
abundance ratio, $[\rm{C/Fe}]~=~+~0.67$, lies slightly below the CEMP
threshold. An $r$-I star, J2116$-$0213, is also found to have high
[Na/Fe] and low [Mg/Fe], a characteristic sign
of the ``intermediate'' or ``extreme'' populations in GCs
\citep{Carretta2009}.  J2116$-$0213 may therefore have been accreted
from a very metal-poor globular cluster.

These results are part of an ongoing survey by the RPA to assess the
$r$-process-enhancement phenomenon in MW halo stars.  The first two
releases from the Northern (this paper) and Southern Hemisphere
\citep{Hansen2018} observing campaigns have significantly increased
the numbers of known $r$-I, $r$-II, and limited-$r$ stars. By
incorporating the kinematic information from {\it Gaia}, these stars
can start to be investigated as stellar populations rather than
interesting anomalies.  Future releases from the RPA will continue to
increase these numbers and identify more chemically interesting stars,
ultimately placing essential constraints on the cosmic site(s) of the
$r$-process.

\acknowledgements
The authors thank the anonymous referee for helpful comments which
improved this manuscript. The authors also thank Anish Amarsi and
Karin Lind for providing the NLTE grids and assisting with their
usage.  C.M.S. thanks Brett Morris for observing some of the stars in
this paper.  The authors thank the current and previous observing
specialists on the 3.5-m telescope at Apache Point Observatory for
their continued help and support. C.M.S. and G.W. acknowledge funding
from the Kenilworth Fund of the New York Community Trust.
V.M.P., T.C.B., A.F., E.M.H., and I.U.R. acknowledge partial support
from grants PHY08-22648 and PHY 14-30152 (Physics Frontier
Center/JINA/CEE), awarded by the US National Science Foundation.  
C.C. acknowledges support from DFG Grant CH1188/2-1 and the "ChETEC"
COST Action (CA16117), supported by COST (European Cooperation in
Science and Technology).
I.U.R. acknowledges funding from NSF grants AST~16-13536 and
AST~18-15403.
This research is based on observations obtained with the Apache Point
Observatory 3.5-meter telescope, which is owned and operated by the
Astrophysical Research Consortium.
This research has made use of the SIMBAD database, operated at CDS,
Strasbourg, France. This work has also made use of data from the
European Space Agency (ESA) mission {\it Gaia}
(\url{http://www.cosmos.esa.int/gaia}),  processed by the {\it Gaia}
Data Processing and Analysis Consortium  (DPAC,
\url{http://www.cosmos.esa.int/web/gaia/dpac/consortium}).  Funding
for the DPAC has been provided by national institutions, in particular
the institutions participating in the {\it Gaia} Multilateral
Agreement.
Funding for RAVE has been provided by: the Leibniz-Institut fuer
Astrophysik Potsdam (AIP); the Australian Astronomical Observatory;
the Australian National University; the Australian Research Council;
the French National Research Agency; the German Research Foundation
(SPP 1177 and SFB 881); the European Research Council (ERC-StG 240271
Galactica); the Istituto Nazionale di Astrofisica at Padova; The Johns
Hopkins University; the National Science Foundation of the USA
(AST-0908326); the W. M. Keck foundation; the Macquarie University;
the Netherlands Research School for Astronomy; the Natural Sciences
and Engineering Research Council of Canada; the Slovenian Research
Agency; the Swiss National Science Foundation; the Science \&
Technology Facilities Council of the UK; Opticon; Strasbourg
Observatory; and the Universities of Groningen, Heidelberg and
Sydney. 
The RAVE web site is at \url{https://www.rave-survey.org}.

\software{IRAF (Tody 1986, Tody 1993), DAOSPEC
  \citep{DAOSPECref}, MOOG (v2017; \citealt{Sneden,Sobeck2011}),
  MULTI2.3 \citep{Carlsson1986,Carlsson1992}, linemake
  (\url{https://github.com/vmplacco/linemake}), MARCS
  \citep{Gustafsson2008}, gal\_uvw (\url{https://github.com/segasai/astrolibpy/blob/master/astrolib/gal_uvw.py})}

\facilities{ARC 3.5m (ARCES), \textit{Gaia}}

\footnotesize{

}

\clearpage
\normalsize

\appendix

\section{Comparisons of Atmospheric Parameters with Independent Methods}\label{appendix:Comp}

\subsection{Comparison of Spectroscopic vs. Photometric Temperatures}\label{subsec:CompRM05}
\begin{figure}[h!]
\begin{center}
\centering
\subfigure{\includegraphics[scale=0.55,trim=0.1in 0.1in 0.7in 0.4in,clip]{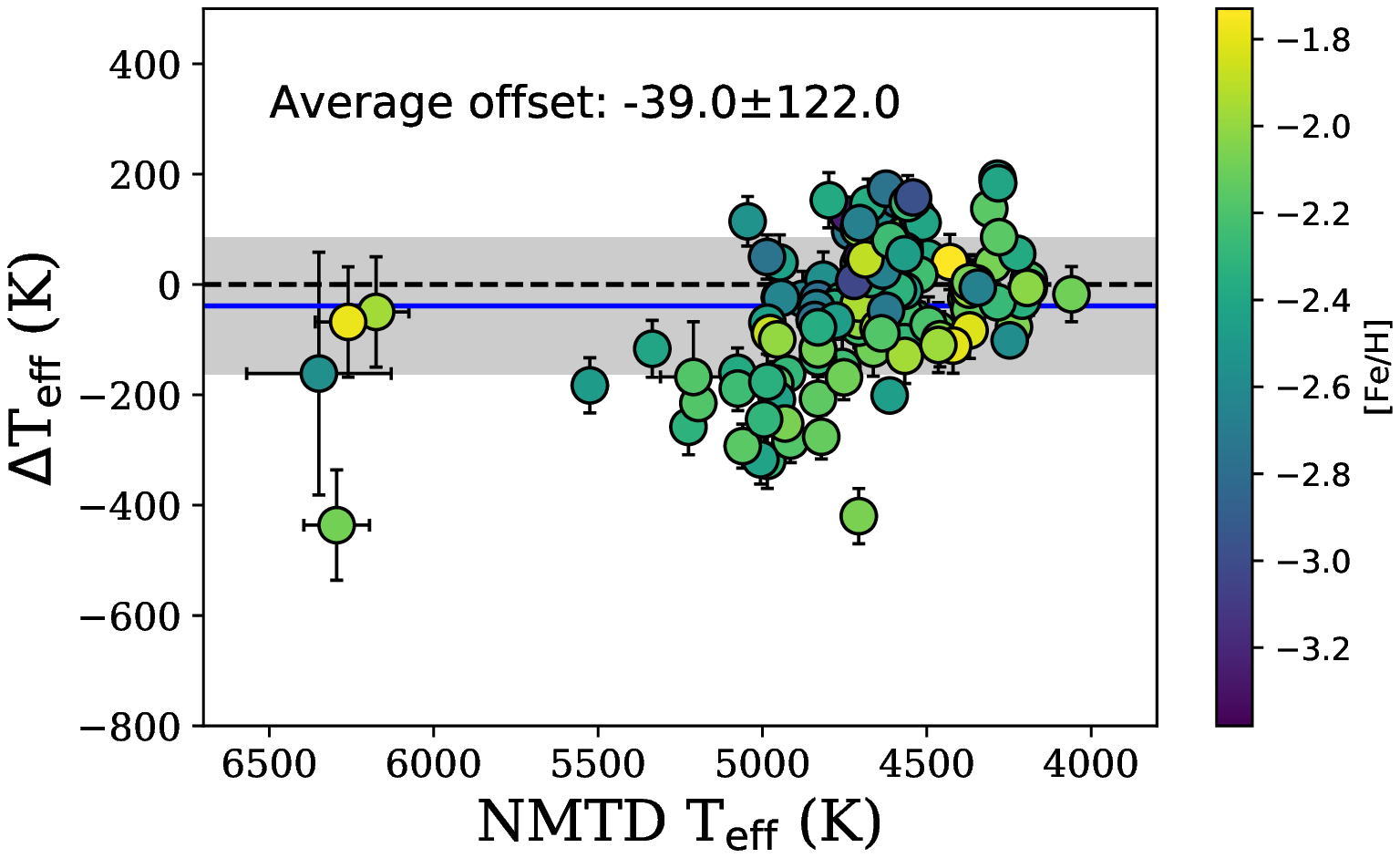}}
\subfigure{\includegraphics[scale=0.55,trim=0.1in 0.1in 0.7in 0.4in,clip]{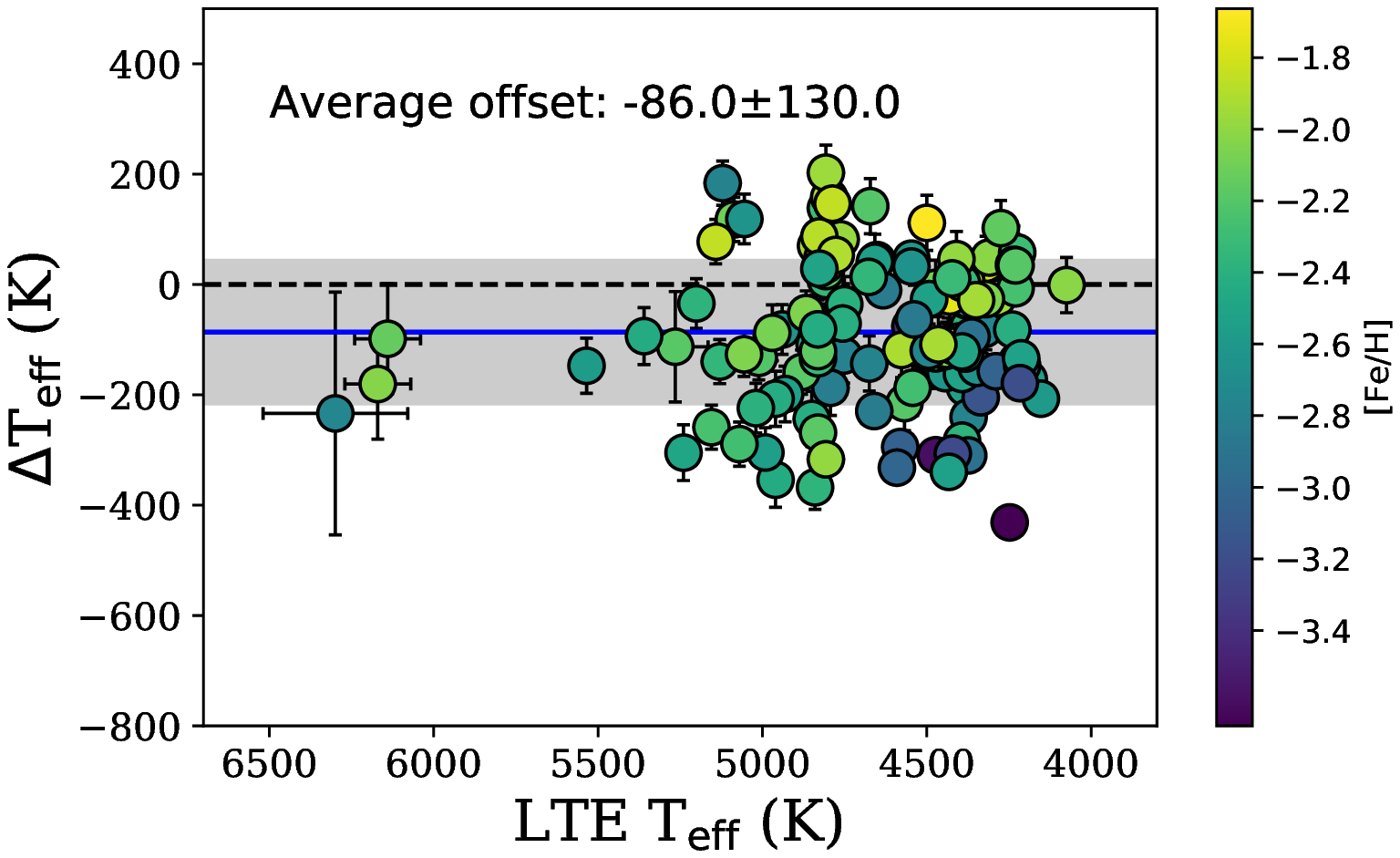}}
\caption{Offsets in effective temperature (spectroscopic $-$
  photometric) for the $<$3D$>$, NLTE temperatures (left) and the LTE
  temperatures (right).  The points are color-coded by [Fe/H].
  Average offsets are shown with a solid line, while the
  $1\sigma$ dispersion is shown with a gray band.\label{fig:CompRM05}}
\end{center}
\end{figure}

Stellar temperatures can be predicted from their colors with
1) empirically calibrated relationships between color, $T_{\rm{eff}}$,
and metallicity (for dwarfs and giants), 2) accurate photometry, 3)
estimates of the reddening, and 4) appropriate reddening laws.  To
compare with the spectroscopic temperatures, photometric temperatures
have been derived from the $(V-K)$ colors and the
\citet{RamirezMelendez2005} color-$T_{\rm{eff}}$ relation, using the
Johnson $V$ and 2MASS $K$ magnitudes from SIMBAD. Estimates of the
reddening have been derived from the \citet{SchlaflyFinkbeiner2011}
extinction
maps,\footnote{\url{http://irsa.ipac.caltech.edu/applications/DUST/}}
and have been converted to E(V-K) with the reddening law from
\citet{McCall2004}. Comparisons of the photometric and spectroscopic
temperatures (NLTE and LTE) are shown in Figure \ref{fig:CompRM05}.
With some exceptions, the spectroscopic temperatures
of the giants agree with the photometric temperatures within 200 K.
On average, the NLTE temperatures are in slightly better agreement
than the LTE temperatures, but there is a scatter of $\sim 150$ K.
The points that lie below the average offset (with lower spectroscopic
temperatures) may be due to uncertainties in the reddening.  The
\citet{SchlaflyFinkbeiner2011} $E(B-V)$ values are determined from
dust maps, and could be higher than the actual foreground
reddening---a higher reddening would lead to a higher photometric
temperature.  The offsets with the dwarfs could be due to issues with
reddening, or could reflect insufficient NLTE corrections or problems
in the adopted color-temperature relations at low metallicity.  Note
that this offset is seen in the dwarfs regardless of whether the
\citet{RamirezMelendez2005} or \citet{Casagrande2010} relation is
used.

\subsection{Comparisons of $\log g$ with {\it Gaia} DR2 Results}\label{subsec:CompGAIA}
All of the target stars have parallax measurements from {\it Gaia}
DR2, though the errors are quite large in some cases. These
parallax-based distances, combined with $V$ magnitudes and $E(B-V)$
reddenings, give absolute $V$ magnitudes, $M_V$.  Only parallaxes with
errors $<20\%$ are utilized to derive distances (see
\citealt{BailerJones2018}).

\begin{figure}[h!]
\begin{center}
\centering
\subfigure[NMTD]{\includegraphics[scale=0.55,trim=0.1in 0.1in 0.7in 0.4in,clip]{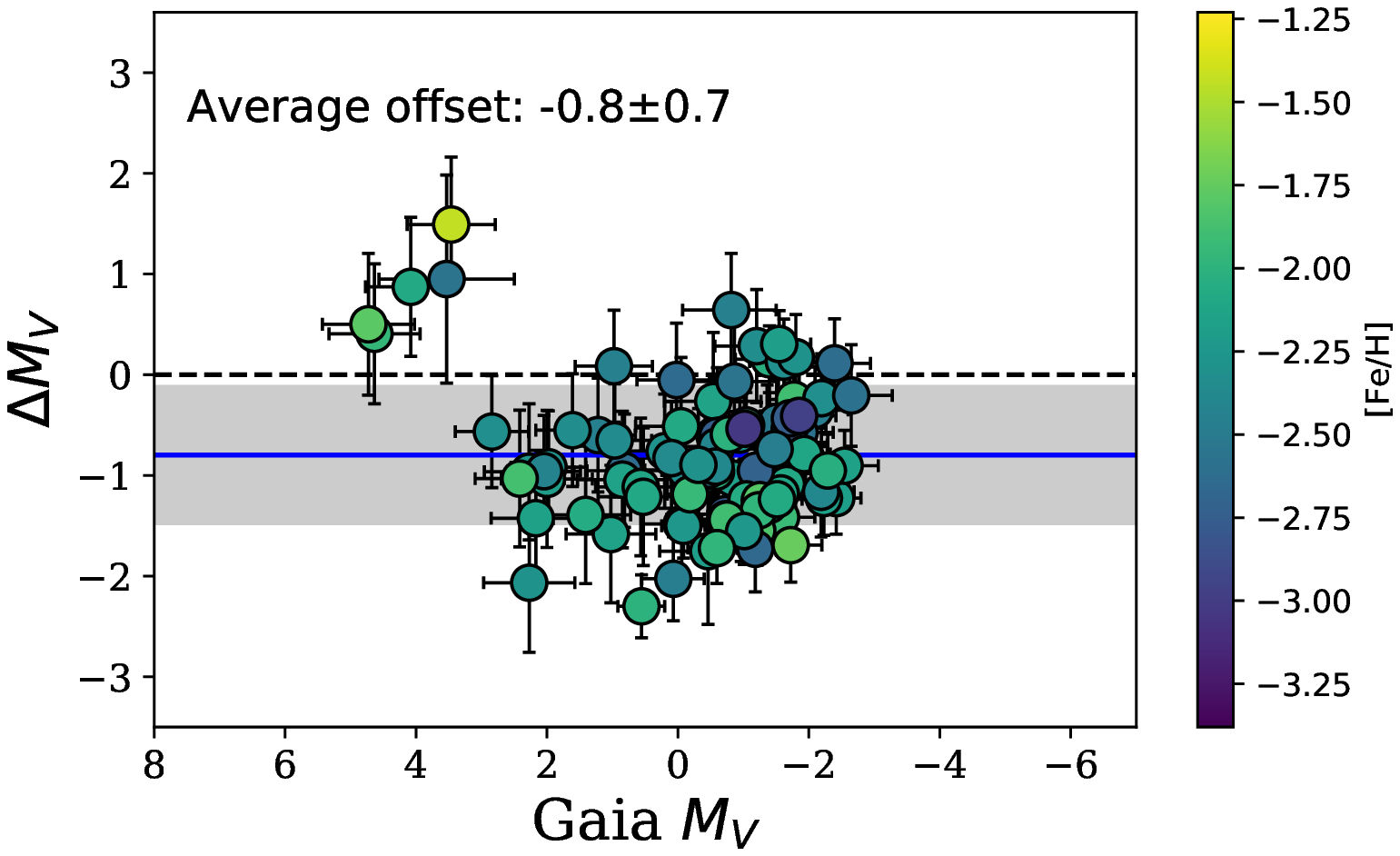}}
\subfigure[LTE]{\includegraphics[scale=0.55,trim=0.1in 0.1in 0.7in 0.4in,clip]{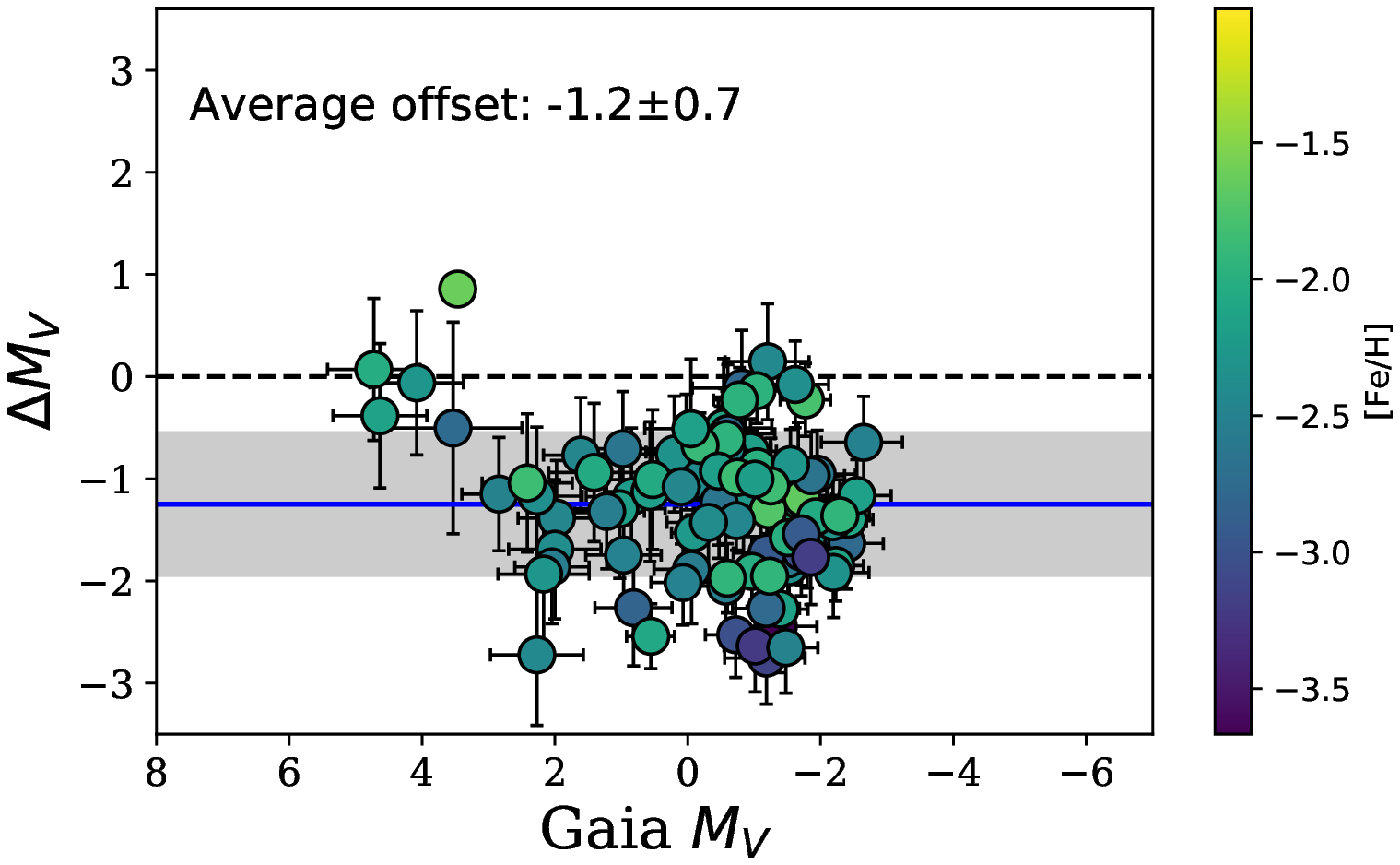}}
\caption{Offsets in $M_V$ (the spectroscopic value derived from
  $\log g$ and bolometric corrections minus the parallax-based
  photometric value) for the $<$3D$>$, NLTE temperatures (left) and
  the LTE temperatures (right).  The points are color-coded by
  [Fe/H]. Average offsets are shown with a solid line, while the
  $1\sigma$ dispersion is shown with a gray band.\label{fig:Mv}}
\end{center}
\end{figure}

Absolute visual magnitudes can also be calculated from the
spectroscopic surface gravities.  The spectroscopic surface gravities
are converted into luminosities and bolometric absolute magnitudes via
Equations 3 and 4 of \citet{MB08}.  These bolometric magnitudes are
then converted into absolute $V$ magnitudes with the bolometric
corrections from the Kurucz database, adopting the $T_{\rm{eff}}$,
$\log g$, and [Fe/H] interpolation scheme from \citet{MB08}.  Figure
\ref{fig:Mv} shows the differences between the spectroscopic (NLTE and
LTE) and photometric absolute magnitudes for the subset of stars with
sufficiently small errors in the parallax.  Both the NLTE
  and LTE values lead to lower predicted $M_{\rm{V}}$ magnitudes, on
  average, than predicted by {\it Gaia}; in other words, the
  spectroscopic surface gravities indicate that the stars are slightly
  brighter than predicted by {\it Gaia}, though the average offset and
  dispersion are smaller when the NLTE corrections are
  utilized. Although this also may reflect problems with the adopted
  bolometric corrections, the assumed stellar mass, or the adopted
  temperature, it may also indicate that additional NLTE corrections
  are necessary.

\clearpage
\section{LTE Abundances and Atmospheric Parameters for the Target Stars}\label{appendix:LTE}
Table \ref{table:LTEAtmParams} shows the spectroscopic parameters for
the target stars if non-LTE corrections are not applied.

\startlongtable
\begin{deluxetable}{@{}ccccDDD}
\tabletypesize{\scriptsize}
\tablecolumns{8}
\tablewidth{0pt}
\tablecaption{LTE Atmospheric Parameters: Target Stars\tablenotemark{a}\label{table:LTEAtmParams}}
\hspace*{-4in}
\tablehead{
Star & $T_{\rm{eff}}$ (K) & $\log g$ & $\xi$ (km/s) & \multicolumn{2}{c}{[\ion{Fe}{1}/H]} & \multicolumn{2}{c}{[\ion{Fe}{2}/H]}
}
\decimals
\startdata
J001236.5$-$181631 & 4985 & 2.44 & 1.49 & -2.28\pm0.01 & -2.42\pm0.015 \\ 
J000738.2$-$034551 & 4663 & 1.48 & 2.32 & -2.09\pm0.01 & -2.23\pm0.031 \\ 
J002244.9$-$172429 & 4718 & 1.11 & 1.95 & -3.38\pm0.03 & -3.88\pm0.11 \\ 
J003052.7$-$100704 & 4831 & 1.48 & 2.2 & -2.35\pm0.02 & -2.42\pm0.061 \\ 
J005327.8$-$025317 & 4370 & 0.56 & 1.95 & -2.16\pm0.01 & -2.19\pm0.036 \\ 
J005419.7$-$061155 & 4707 & 1.03 & 2.01 & -2.32\pm0.02 & -2.36\pm0.075 \\ 
J010727.4$-$052401 & 5225 & 3.03 & 1.43 & -2.32\pm0.01 & -2.51\pm0.025 \\ 
J014519.5$-$280058 & 4582 & 0.69 & 2.09 & -2.60\pm0.02 & -2.55\pm0.049 \\ 
J015656.3$-$140211 & 4622 & 1.09 & 2.4 & -2.08\pm0.02 & -2.09\pm0.041 \\ 
2MJ02134021$-$0005183 & 6175 & 4.47 & 2.6 & -1.96\pm0.03 & -2.17\pm0.08 \\ 
\enddata
\tablenotetext{a}{Only a portion of this table is shown here to
  demonstrate its form and content. A machine-readable version of the
  full table is available.}
\end{deluxetable}

\section{Systematic Errors}\label{appendix:Errors}
The systematic errors in the abundances are quantified according to
the uncertainties in the atmospheric parameters, using the techniques
outlined in \citet{McWilliam2013} and \citet{Sakari2017}.  First, the
variances and covariances in the atmospheric parameters were
estimated, as shown in Table \ref{table:AppendixErrors}.  For the
temperature and microturbulence, the uncertainties were determined
based on the errors in the slopes of Fe abundance vs. EP and REW,
respectively.  The uncertainty in gravity was based on the random error
in the \ion{Fe}{2} abundance, while the uncertainty in the metallicity
was based on the random error in the \ion{Fe}{1} abundance.  The
covariances were calculated according to Equation A6 in
\citet{McWilliam2013}.

\begin{deluxetable}{@{}lccccccccc}
\tabletypesize{\scriptsize}
\tablecolumns{10}
\tablewidth{0pt}
\tablecaption{Variances and Covariances in Atmospheric Parameters for
  Several Standard Stars\label{table:AppendixErrors}}
\hspace*{-4in}
\tablehead{
 & \multicolumn{1}{l}{[Fe/H] $\sim-3$} & & \multicolumn{3}{l}{[Fe/H] $\sim-2.5$} & & \multicolumn{3}{l}{[Fe/H] $\sim-2$} \\
 & $\log g \sim 1$ & & $\log g \sim 1$ & $\log g \sim 2$ & $\log g \sim 4$ & & $\log g \sim 1$ & $\log g \sim 2$ & $\log g \sim 4$\\
 & HE 1523$-$0901    & & J2038$-$0023      & CS 31082-001   & BD$-$13 3442      & & TYC 6535-3183-1 & TYC 4924-33-1   & TYC 5911-452-1
}
\startdata
$\sigma_T$          & 30     & & 25   & 30   & 220    & & 55 & 40 & 95 \\
$\sigma_g$          & 0.25   & & 0.05 & 0.15 & 0.30   & & 0.05 & 0.20 & 0.20\\
$\sigma_\xi$         & 0.16   & & 0.20 & 0.14 & 0.35   & & 0.15 & 0.13 & 0.28\\
$\sigma_{\rm{[M/H]}}$ & 0.02    & & 0.02 & 0.02 & 0.02   & & 0.02 & 0.01 & 0.02\\
& & & & & & & & & \\
$\sigma_{T\xi}$       & 0.08   & & 0.03  & 0.03  & 0.10   & & 0.0 & 2.4 & 8.07 \\
$\sigma_{Tg}$        & 0.0     & & 0     & 0.0  & 0.0     & & 0.0 & 0.0 & 0.0\\
$\sigma_{g\xi}$       & 0.08   & & 0.003 & 0.03  & -0.090 & & 0.0 & -0.012 & -0.015\\
$\sigma_{T\rm{[M/H]}}$ & 0      & & 0.32  &       & 0.0     & & 0.0 & 0.48 & 0.90 \\
\enddata
\end{deluxetable}

The uncertainties in the [Fe/H] and [X/Fe] abundance ratios were then
calculated using Equation A1 in \citet{Sakari2017} and Equations A4
and A5 in \citet{McWilliam2013}.  Table \ref{table:AppendixAbundErrors}
shows the total errors (systematic and random) in the abundance ratios
for the six representative standard stars.  Only the uncertainties in
[X/Fe] are shown; note that the errors in the [X/Fe] ratios are often
lower than the errors in the absolute $\log \epsilon$ abundances,
since the abundances change together as the atmospheric parameters are
varied.

\begin{deluxetable}{@{}lccccccccc}
\tabletypesize{\scriptsize}
\tablecolumns{10}
\tablewidth{0pt}
\tablecaption{Total Errors (Systematic and Random) in the Abundance Ratios for Several Standard Stars\label{table:AppendixAbundErrors}}
\hspace*{-4in}
\tablehead{
 & \multicolumn{1}{l}{[Fe/H] $\sim-3$} & & \multicolumn{3}{l}{[Fe/H] $\sim-2.5$} & & \multicolumn{3}{l}{[Fe/H] $\sim-2$} \\
 & $\log g \sim 1$ & & $\log g \sim 1$ & $\log g \sim 2$ & $\log g \sim 4$ & & $\log g \sim 1$ & $\log g \sim 2$ & $\log g \sim 4$\\
 & HE 1523$-$0901    & & J2038$-$0023      & CS 31082-001 & BD$-$13 3442      & & TYC 6535-3183-1 & TYC 4924-33-1   & TYC 5911-452-1
}
\startdata
$\sigma$[\ion{Fe}{1}/H]  & 0.03 & & 0.02 & 0.02 & 0.03 & & 0.02 & 0.02 & 0.03 \\
$\sigma$[\ion{Fe}{2}/H]  & 0.07 & & 0.08 & 0.07 & 0.10 & & 0.10 & 0.07 & 0.09 \\
$\sigma$[\ion{Li}{1}/Fe] &      & &      & 0.11 & 0.12 & & 0.09 & 0.11 & 0.11 \\
$\sigma$[\ion{O}{1}/Fe]  &      & &      &      &      & & 0.12 &      &  \\
$\sigma$[\ion{Na}{1}/Fe] &      & &      &      & 0.14 & & 0.16 &      &  \\
$\sigma$[\ion{Mg}{1}/Fe] & 0.01 & & 0.06 & 0.05 & 0.12 & & 0.11 & 0.03 & 0.04 \\
$\sigma$[\ion{Si}{1}/Fe] &      & &      &      &      & & 0.08 & 0.12 & 0.11 \\
$\sigma$[\ion{K}{1}/Fe]  & 0.13 & & 0.11 & 0.11 &      & & 0.11 & 0.10 & 0.10 \\
$\sigma$[\ion{Ca}{1}/Fe] & 0.02 & & 0.04 & 0.03 & 0.08 & & 0.04 & 0.02 & 0.03 \\
$\sigma$[\ion{Sc}{2}/Fe] & 0.06 & & 0.07 & 0.07 & 0.14 & & 0.07 & 0.04 & 0.08 \\
$\sigma$[\ion{Ti}{1}/Fe] & 0.02 & & 0.04 & 0.04 &      & & 0.04 & 0.03 & 0.05 \\
$\sigma$[\ion{Ti}{2}/Fe] & 0.05 & & 0.06 & 0.08 & 0.07 & & 0.08 & 0.03 & 0.09 \\
$\sigma$[\ion{V}{1}/Fe]  & 0.34 & & 0.18 & 0.11 &      & & 0.21 & 0.18 &  \\
$\sigma$[\ion{Cr}{2}/Fe] & 0.30 & & 0.18 & 0.08 &      & & 0.12 & 0.08 &  \\
$\sigma$[\ion{Mn}{1}/Fe] & 0.08 & & 0.11 & 0.11 &      & & 0.05 & 0.03 &  \\
$\sigma$[\ion{Co}{1}/Fe] & 0.40 & & 0.19 & 0.11 &      & & 0.10 & 0.07 &  \\
$\sigma$[\ion{Ni}{1}/Fe] & 0.05 & & 0.05 & 0.03 &      & & 0.02 & 0.05 &  \\
$\sigma$[\ion{Cu}{1}/Fe] &      & &      &      &      & & 0.17 &      &  \\
$\sigma$[\ion{Zn}{1}/Fe] & 0.15 & & 0.11 & 0.06 &      & & 0.12 & 0.15 &  \\
$\sigma$[\ion{Sr}{2}/Fe] & 0.13 & & 0.12 & 0.10 & 0.24 & & 0.11 & 0.11 & 0.18 \\
$\sigma$[\ion{Y}{2}/Fe]  & 0.06 & & 0.09 & 0.06 &      & & 0.06 & 0.08 &  \\
$\sigma$[\ion{Zr}{2}/Fe] & 0.09 & & 0.13 & 0.12 &      & & 0.14 & 0.11 &  \\
$\sigma$[\ion{Ba}{2}/Fe] & 0.18 & & 0.14 & 0.06 & 0.39 & & 0.12 & 0.05 & 0.14 \\
$\sigma$[\ion{La}{2}/Fe] & 0.06 & & 0.08 & 0.08 &      & & 0.11 & 0.02 &  \\
$\sigma$[\ion{Ce}{2}/Fe] & 0.05 & & 0.07 & 0.07 &      & & 0.06 &      &  \\
$\sigma$[\ion{Pr}{2}/Fe] & 0.09 & & 0.08 & 0.12 &      & & 0.06 & 0.13 &  \\
$\sigma$[\ion{Nd}{2}/Fe] & 0.06 & & 0.09 & 0.06 &      & & 0.07 & 0.12 &  \\
$\sigma$[\ion{Sm}{2}/Fe] & 0.07 & & 0.13 & 0.12 &      & & 0.07 &      &  \\
$\sigma$[\ion{Eu}{2}/Fe] & 0.11 & & 0.12 & 0.11 & 0.20 & & 0.18 & 0.06 & 0.13 \\
$\sigma$[\ion{Dy}{2}/Fe] & 0.12 & &      & 0.12 &      & &      &      &  \\
$\sigma$[\ion{Os}{1}/Fe] & 0.17 & &      & 0.12 &      & & 0.11 &      &  \\
$\sigma$[\ion{Th}{2}/Fe] &      & & 0.12 & 0.12 &      & &      &      &  \\
\enddata
\end{deluxetable}

\section{Equivalent Widths and Line Abundances}\label{appendix:EWs}
Tables \ref{table:EWs} and \ref{table:SSs} show the EW measurements
and abundances for the lines that were determined via EW techniques
and spectrum syntheses, respectively.

\begin{deluxetable}{@{}ccccccccc}
\tabletypesize{\scriptsize}
\tablecolumns{9}
\tablewidth{0pt}
\tablecaption{Equivalent Widths\tablenotemark{a}\label{table:EWs}}
\hspace*{-4in}
\tablehead{
Element & Wavelength & EP & $\log gf$ & J0007$-$0345 EW & J0012$-$1816 EW & J0022$-$1724 EW & J0030$-$1007 EW & J0053$-$0253 EW \\
 & (\AA) &  (eV) &  & (m\AA) & (m\AA) & (m\AA) & (m\AA) & (m\AA) 
}
\startdata
\ion{Fe}{1} & 4383.54 & 1.48 &  0.208 &  ---  &  ---  &  ---  &  ---  &  ---  \\
\ion{Fe}{1} & 4401.44 & 2.83 & -1.650 &  85.1 &  39.2 &  ---  &  ---  &  ---  \\
\ion{Fe}{1} & 4404.75 & 1.56 & -0.147 &  ---  &  ---  &  ---  &  ---  &  ---  \\
\ion{Fe}{1} & 4408.42 & 2.20 & -1.775 &  ---  &  49.6 &  ---  &  ---  &  ---  \\
\ion{Fe}{1} & 4415.12 & 1.61 & -0.621 &  ---  &  ---  &  ---  &  ---  &  ---  \\
\ion{Fe}{1} & 4430.61 & 2.22 & -1.728 &  ---  &  50.5 &  20.8 &  43.3 &  ---  \\
\ion{Fe}{1} & 4442.34 & 2.22 & -1.228 &  ---  &  62.4 &  ---  &  69.1 &  ---  \\
\ion{Fe}{1} & 4443.19 & 2.86 & -1.043 &  76.9 &  38.4 &  ---  &  ---  &  87.5 \\
\ion{Fe}{1} & 4447.72 & 2.22 & -1.339 &  ---  &  ---  &  ---  &  64.7 &  ---  \\
\ion{Fe}{1} & 4466.55 & 2.83 & -0.600 &  ---  &  67.3 &  26.8 &  84.8 &  ---  \\
\enddata
\tablenotetext{a}{Only a portion of this table is shown here to
  demonstrate its form and content. A machine-readable version of the
  full table is available.}
\end{deluxetable}

\begin{deluxetable}{@{}ccccccccc}
\tabletypesize{\scriptsize}
\tablecolumns{9}
\tablewidth{0pt}
\tablecaption{Abundances from Synthesized Lines\tablenotemark{a}\label{table:SSs}}
\hspace*{-4in}
\tablehead{
Element & Wavelength & EP & $\log gf$ & J0007$-$0345 & J0012$-$1816 & J0022$-$1724 & J0030$-$1007 & J0053$-$0253 \\
 & (\AA) &  (eV) &  & $\log \epsilon$ & $\log \epsilon$ & $\log \epsilon$ & $\log \epsilon$ & $\log \epsilon$ 
}
\startdata
\ion{Li}{1} & 6707.3\tablenotemark{b} & 0.000 & 0.18 &  ---  &  ---  &  ---  &  --- &  ---  \\
\ion{O}{1}  & 6300.304 & 0.000 & -9.82  &  --- &  ---  &  ---  &  --- &  ---  \\
\ion{O}{1}  & 6363.776 & 0.020 & -10.30 &  --- &  ---  &  ---  &  --- &  ---  \\
\ion{Na}{1} & 5682.633 & 2.101 & -0.70  & 4.25 &  ---  &  ---  &  --- & 4.03 \\
\ion{Na}{1} & 5688.205 & 2.103 & -0.45  & 4.15 &  ---  &  ---  &  --- & 4.08 \\
\ion{Na}{1} & 5889.951 & 0.000 &  0.12  &  ---  &  ---  &  ---  &  --- &  ---  \\
\ion{Na}{1} & 5895.924 & 0.000 & -0.18  &  ---  &  ---  &  ---  &  --- &  ---  \\
\ion{Cu}{1} & 5105.5\tablenotemark{b} & 1.388 & -1.52 & 1.43 &  --- &  ---  &  --- & 1.08 \\
\ion{Cu}{1} & 5782.1\tablenotemark{b} & 1.641 & -1.72 &  --- &  --- &  ---  &  --- &  ---  \\
\ion{Zn}{1} & 4722.153 & 4.027 & -0.340 & 2.57 & 2.68 &  ---  & 2.69 & 2.50 \\
\ion{Zn}{1} & 4810.528 & 4.075 & -0.140 & 2.60 & 2.33 &  ---  &  --- & 2.45 \\
\enddata
\tablenotetext{a}{Only a portion of this table is shown here to
  demonstrate its form and content. A machine-readable version of the
  full table is available.}
\tablenotetext{b}{This line has HFS and/or isotopic splitting.}
\end{deluxetable}

\end{document}